\RequirePackage{fix-cm}
\documentclass[twocolumn,epjc3]{svjour3}
\smartqed  % flush right qed marks, e.g. at end of proof
\RequirePackage{graphicx}
\RequirePackage{placeins}
\RequirePackage{amsmath}
\RequirePackage{subfig}
\RequirePackage[hyphens]{url}
\journalname{Eur. Phys. J. C}
\begin{document}

\title{Study of energy response and resolution of the ATLAS Tile Calorimeter to hadrons of energies from 16 to 30 GeV
}


\author{%\thanksref{e1,addr1}
Jalal Abdallah\thanksref{addr1}
\and
Stylianos  Angelidakis\thanksref{addr2}
\and
Giorgi Arabidze\thanksref{addr3}
\and
Nikolay Atanov\thanksref{addr4}
\and
Johannes Bernhard\thanksref{addr5}
\and
Roméo  Bonnefoy\thanksref{addr6}
\and
Jonathan Bossio\thanksref{addr7}
\and
Ryan  Bouabid\thanksref{addr8}
\and
Fernando Carrio\thanksref{addr9}
\and
Tomas  Davidek\thanksref{addr10}
\and
Michal  Dubovsky\thanksref{addr11}
\and
Luca  Fiorini\thanksref{addr9}
\and
Francisco Brandan  Garcia Aparisi\thanksref{addr9}
\and
Tancredi Carli\thanksref{addr5}
\and
Alexander  Gerbershagen\thanksref{addr5}
\and
Hazal Goksu\thanksref{addr8}
\and
Haleh Hadavand\thanksref{addr1}
\and
Siarhei Harkusha\thanksref{addr12}
\and
Dingane  Hlaluku\thanksref{addr13}
\and
Michael James Hibbard\thanksref{addr1}
\and
Kevin   Hildebrand\thanksref{addr8}
\and
Juansher  Jejelava\thanksref{addr14}
\and
Andrey Kamenshchikov\thanksref{addr15}
\and
Stergios  Kazakos\thanksref{addr16}
\and
Tomas  Kello\thanksref{addr10}
\and
Ilya Korolkov\thanksref{addr16}
\and
Yuri Kulchitsky\thanksref{addr12}
\and
Hadar  Lazar\thanksref{addr8}
\and
Nthabiseng Lekalakala\thanksref{addr13}
\and
Jared  Little\thanksref{addr1}
\and
Romain Madar\thanksref{addr6}
\and
Samuel Manen\thanksref{addr6}
\and
Filipe Martins\thanksref{addr17}
\and
Thabo Masuku\thanksref{addr13}
\and
Irakli Minashvili\thanksref{addr4}
\and
\underline{Tigran Mkrtchyan}\thanksref{e1,addr18}
\and
Michaela  Mlynarikova\thanksref{addr19}
\and
Seyedali Moayedi\thanksref{addr1}
\and
Stanislav Nemecek\thanksref{addr20}
\and
Lawrence  Nodulman\thanksref{addr21}
\and
Robert Oganezov\thanksref{addr22}
\and
Mats Joakim Robert Olsson\thanksref{addr23}
\and
Mark Oreglia\thanksref{addr8}
\and
Priscilla  Pani\thanksref{addr24}
\and
Alexander  Paramonov\thanksref{addr21}
\and
Ruth Pottgen\thanksref{addr25}
\and
Tres Reid\thanksref{addr8}
\and
Sergi  Rodriguez Bosca\thanksref{addr9}
\and
Andrea  Rodriguez Perez\thanksref{addr16}
\and
Rachel Christine  Rosten\thanksref{addr26}
\and
Puja  Saha\thanksref{addr19}
\and
\underline{Claudio  Santoni}\thanksref{e2,addr6}
\and
Laura Sargsyan\thanksref{addr22}
\and
Douglas Michael  Schaefer\thanksref{addr8}
\and
Nikolay Shalanda\thanksref{addr15}
\and
Andrew Caldon   Smith\thanksref{addr27}
\and
Alexander  Solodkov\thanksref{addr15}
\and
Oleg Solovyanov\thanksref{addr15}
\and
Pavel Starovoitov\thanksref{addr18}
\and
Evgeny Starchenko\thanksref{addr15}
\and
Petr  Tas\thanksref{addr10}
\and
Viacheslav Tereshchenko\thanksref{addr4}
\and
Sijiye Humphry  Tlou\thanksref{addr13}
\and
Michael Ughetto\thanksref{addr28}
\and
Lea   Uhliarova\thanksref{addr10}
\and
Giulio  Usai\thanksref{addr1}
\and
Eduardo Valdes Santurio\thanksref{addr28}
\and
Alberto Valero Biot\thanksref{addr9}
\and
Guido Volpi\thanksref{addr16}
\and
Tamar Zakareishvili\thanksref{addr29}
\and
Pedro Diego Zuccarello\thanksref{addr9}
}

\thankstext{e1}{Corresponding author e-mail: tigran.mkrtchyan@cern.ch}
\thankstext{e2}{Corresponding author e-mail: claudio.santoni@cern.ch}


\institute{%First address \label{addr1}
Department of Physics, University of Texas at Arlington, Arlington TX \label{addr1}
\and
Physics Department, National and Kapodistrian University of Athens, Athens \label{addr2}
\and
Department of Physics and Astronomy, Michigan State University, East Lansing MI \label{addr3}
\and
Joint Institute for Nuclear Research, Dubna \label{addr4}
\and
CERN, Geneva \label{addr5}
\and
LPC, Universit\'e Clermont Auvergne, CNRS/IN2P3, Clermont-Ferrand \label{addr6}
\and
Department of Physics, McGill University, Montreal QC \label{addr7}
\and
Enrico Fermi Institute, University of Chicago, Chicago IL \label{addr8}
\and
Instituto de F\'isica Corpuscular (IFIC), Centro Mixto Universidad de Valencia - CSIC, Valencia \label{addr9}
\and
Charles University, Faculty of Mathematics and Physics, Prague \label{addr10}
\and
Faculty of Mathematics, Physics and Informatics, Comenius University, Bratislava \label{addr11}
\and
B.I. Stepanov Institute of Physics, National Academy of Sciences of Belarus, Minsk \label{addr12}
\and
School of Physics, University of the Witwatersrand, Johannesburg \label{addr13}
\and
E. Andronikashvili Institute of Physics, Iv. Javakhishvili Tbilisi State University, Tbilisi \label{addr14}
\and
Institute for High Energy Physics of the National Research Centre Kurchatov Institute, Protvino \label{addr15}
\and
Institut de F\'isica d'Altes Energies (IFAE), Barcelona Institute of Science and Technology, Barcelona \label{addr16}
\and
Laborat\'orio de Instrumenta\c{c}\~ao e F\'isica Experimental de Part\'iculas - LIP, Lisboa \label{addr17}
\and
Kirchhoff-Institut f\"{u}r Physik, Ruprecht-Karls-Universit\"{a}t Heidelberg, Heidelberg \label{addr18}
\and
Department of Physics, Northern Illinois University, DeKalb IL \label{addr19}
\and
Institute of Physics of the Czech Academy of Sciences, Prague \label{addr20}
\and
High Energy Physics Division, Argonne National Laboratory, Argonne IL \label{addr21}
\and
Alikhanyan National Science Laboratory, Yerevan \label{addr22}
\and
Department of Physics and Astronomy, University of California Irvine, Irvine CA \label{addr23}
\and
Deutsches Elektronen-Synchrotron DESY, Hamburg and Zeuthen \label{addr24}
\and
Fysiska institutionen, Lunds universitet, Lund \label{addr25}
\and
Ohio State University, Columbus OH \label{addr26}
\and
Nevis Laboratory, Columbia University, Irvington NY \label{addr27}
\and
Oskar Klein Centre, Stockholm \label{addr28}
\and
High Energy Physics Institute, Tbilisi State University, Tbilisi \label{addr29}
       %    \emph{Present Address:} if needed\label{addr3}
}

\date{Received: date / Accepted: date}

\maketitle

\begin{abstract}
\hfill \break
Three spare modules of the ATLAS Tile Calorimeter were exposed to test beams from the Super Proton Synchrotron accelerator at CERN in 2017. The measurements of the energy response and resolution of the detector to positive pions and kaons and protons with energy in the range 16 to 30 GeV are reported. The results have uncertainties of few percent. They were compared to the predictions of the Geant4-based simulation program used in ATLAS to estimate the response of the detector to proton-proton events at Large Hadron Collider. The determinations obtained using experimental and simulated data agree within the uncertainties.
\end{abstract}
%-------------------------------------------------------------------------------
\section{Introduction}
\label{sec:introduction}
%-------------------------------------------------------------------------------
Three spare modules of the Tile Calorimeter (TileCal) of the ATLAS experiment~\cite{PERF-2007-01},
two long-barrels and one extended-barrel, were exposed to muons, electrons, pions, kaons and protons with different energies and incident angles at test beams (TBs) in 2017~\cite{H8website}. 
%The modules were equipped with different upgraded front-end electronics systems proposed for the ATLAS LHC Phase-II operations~\cite{Phase_II_Upgrade}. 
The role of the hadron calorimetry in ATLAS is to measure the energy and the angle of isolated hadrons and jets. To achieve good performance, the study of the sub-detector response to isolated hadrons is important. In this paper, the measurements of the calorimeter response and resolution to positive pions and kaons and protons, with energies in the range 16-30 GeV are presented. The results are compared with the ones obtained analyzing simulated data produced using the ATLAS Geant4 toolkit~\cite{Agostinelli:2002hh}, \cite{Allison:2006ve} and \cite{Costanzo:916030}. The experimental setup including the beam line counters and the detector is described in Section~\ref{sec:exp_setup}. The data sets, the event selections and the reconstruction of the particle energies in the case of experimental and simulated data are presented in Sections~\ref{sec:data_selection} and~\ref{sec:MC_selection}, respectively. The determinations of the calorimeter responses and resolutions are discussed in %Section~\ref{sec:energy_reconstruction} and
Section~\ref{sec:calorimeter_response}. The results are compared with hadronic cascade model predictions in Section~\ref{sec:cascade_models}. The conclusions are stated in Section~\ref{sec:conclusion}.
%\newpage
\section{The experimental setup}
\label{sec:exp_setup}
\subsection{The beam line}
\label{subsec:beam_line}
The measurements discussed in this paper were performed using tertiary particle beams at the H8 line in the North Area of CERN~\cite{H8website}. Secondary beams are produced
%directly
by targeting 400 GeV protons, from the Super Proton Synchrotron (SPS) accelerator, on a 100 mm thick T4 target made of beryllium (primary target). Using Secondary Targets located at about 130 m downstream of the T4 target, tertiary beams can be produced. A large spectrometer constructed of four Main Bend North Area  dipole magnets is used for the momentum definition. Beam particles can have energies from 10 to 350 GeV. Beam intensity decreases dramatically at the low energies. To have mixed hadron enriched tertiary beams, the Secondary Target is made of copper and has a thickness of 300 mm. Additionally, a lead absorber (6~mm) is moved into the beam about 270~m downstream of the target. It absorbs the electrons, while the hadrons mostly pass through it. For electron enriched tertiary beams, the Secondary Target is made by aluminum and has a thickness of 400 mm. It is immediately followed by 6~mm of lead. The lead absorber further downstream is moved out of the beam trajectory. 

\begin{figure}[]
\centering
\includegraphics[width=8.4cm,clip]{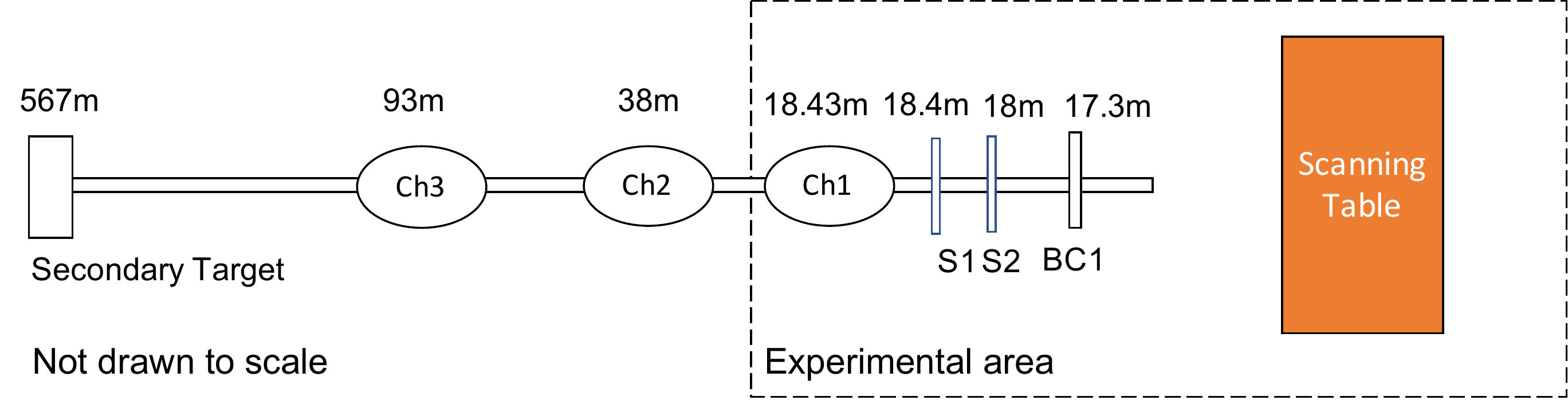}
\caption{Schematic layout of the H8 beam line detectors. The distances of the beam line components and of the Secondary Target from the Scanning Table (on which was placed the TB calorimeter) setup are shown.
%\textcolor {red} {Take out BC2 and 4.4 m}
}
\label{fig:beam_line}       % Give a unique label
\end{figure}

The layout of the beam line detectors is shown in Figure~\ref{fig:beam_line}.
The transverse beam profile was monitored by the wire chamber BC1~\cite{Beam_chambers}.
%by two wire chambers (BC1 and BC2)~\cite{Beam_chambers}. 
Two scintillating counters, S1 and S2 with an active surface of $5\times5$ cm$^2$~\cite{TB_instrumentation}, were used in coincidence to trigger the data acquisition (Physics Trigger) and to provide the trigger timing. These two detectors were also used to reject beam particles interacting upstream of the detector. The Cherenkov counters Ch1, Ch2 and Ch3 allowed identification of beam particles. The counters Ch1 and Ch3 distinguish electrons and pions from kaons and protons. They were filled with CO$_2$ and He, respectively. The pressure values set for the different beam energies are reported in Table~\ref{tab:Chs_pressure}. The Cherenkov counter Ch2 was also filled with CO$_2$. The higher pressure in Ch2 
%value with respect to Ch2, %see Table~\ref{tab:Chs_pressure}, it 
allows for separation of kaons from protons. More details can be found in Ref.~\cite{TB_instrumentation}.
% ----------
%\FloatBarrier
% ----------
% ----------
\begin{table}[]
\centering
\caption{Cherenkov radiator materials and corresponding gas pressure values set for the different beam energies.
%\textcolor {red} {Update the numbers} 
}
\label{tab:Chs_pressure}       % Give a unique label
% For LaTeX tables you can use 
\begin{tabular}{	|c|c|c|c|} 
\hline 
Cherenkov Counter& Ch1 & Ch2 & Ch3 \\ \hline \hline
Radiator Material & CO$_2$ & CO$_2$ & He  \\ \hline
$E_\text{beam}$ [GeV] & \multicolumn{3}{c|}{Pressure [bar]}\\ \hline
16 & 0.19 & 0.75 & 2.6 \\ 
%\hline
18 & 0.19 & 0.75 & 2.6 \\ 
%\hline
20 & 0.19 & 0.75 & 2.6 \\ 
%\hline
30 & 0.3 & 2.0 & 2.6 \\ \hline
\end{tabular}
%\end{center}
\end{table}
%\FloatBarrier
% ----------
\subsection{The detector}
\label{subsec:detector}
% -----------
\begin{figure}[]
\centering
\includegraphics[width=8.4cm,clip]{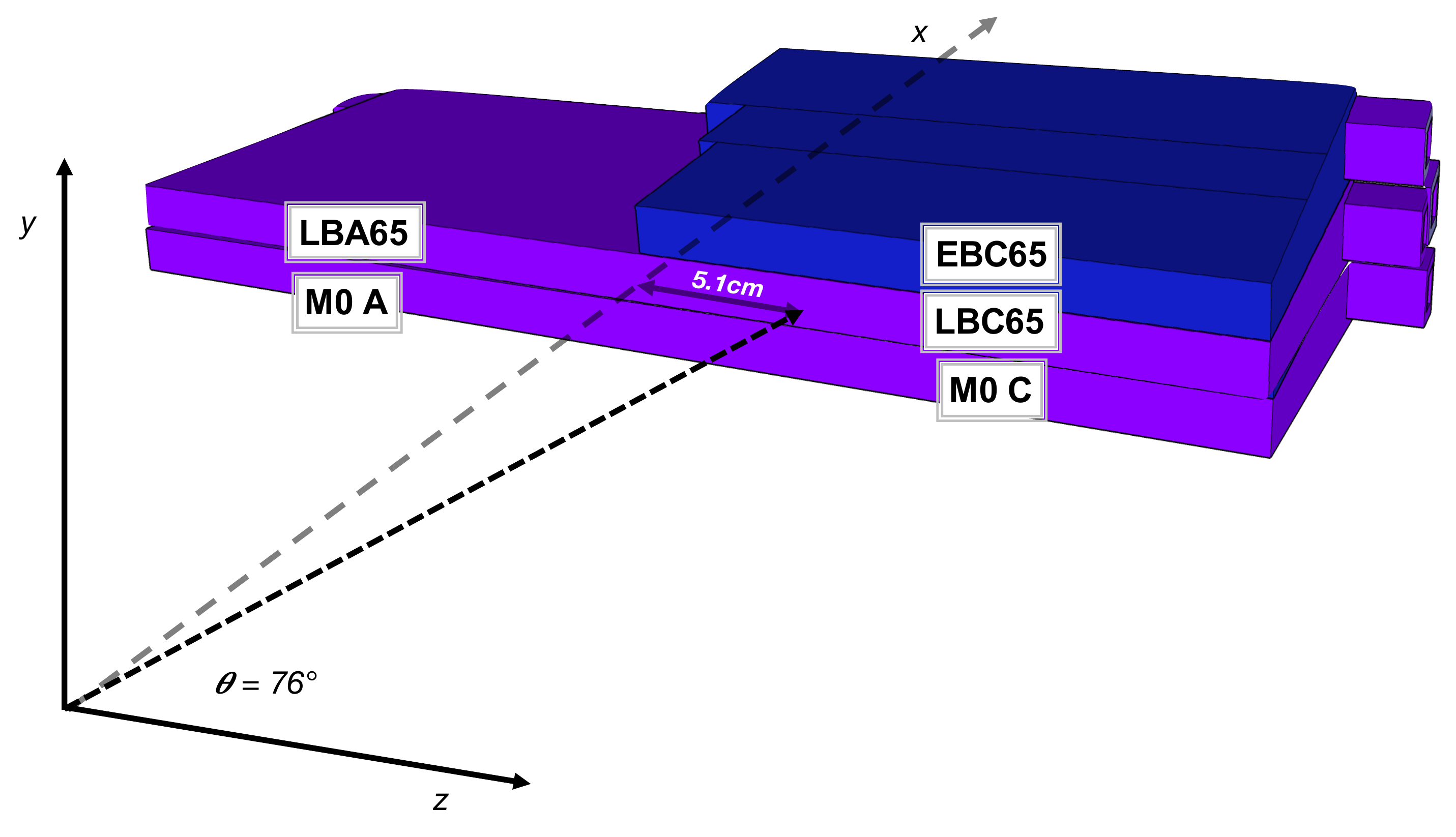}
\caption{Schematic view of the TileCal modules as stacked on the scanning table at the H8 beam line. The names of the super-drawers
%(white fields) and the corresponding front-end electronics are indicated 
and the direction and the interaction point of the particle beams in the detector are shown.
%(see the text for the details).
%\textcolor {red} {The arrow of the beam direction must hit vertically at the centre of LBC65 super drawer} 
% The modules electronics are discussed in ref.~\cite{Phase_II_Upgrade}.
}
%\label{fig:TB_setup}       % Give a unique label
\label{fig:TB_setup}       % Give a unique label
\end{figure}
\begin{figure}[ht]
\centering
\includegraphics[width=8.4cm,clip]{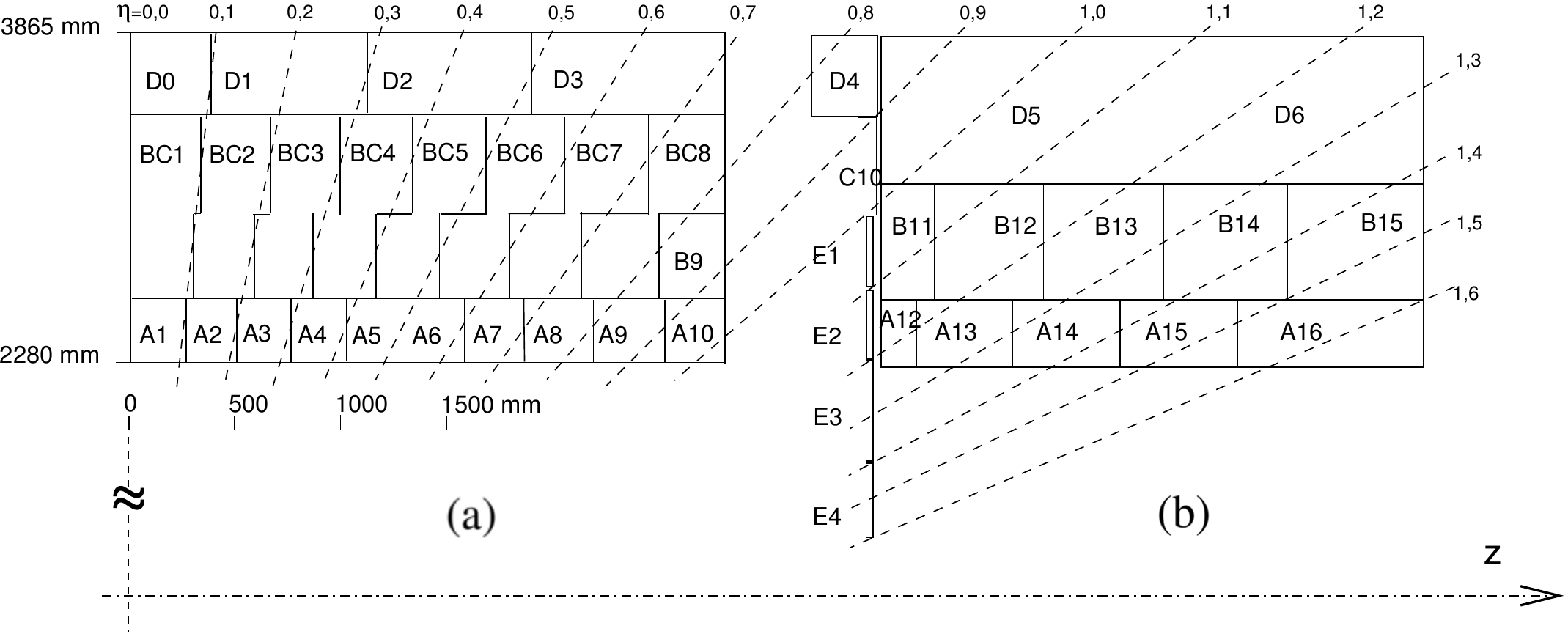}
\caption{Cell structure in half long-barrel (a) and extended-barrel (b) modules of the calorimeter. Solid lines show the cell boundaries formed by grouping optical fibers from the tiles for read out by separate photomultipliers. Also shown are dashed lines of fixed pseudo-rapidity $\eta$~\cite{PERF-2007-01}.}
\label{fig:module_scheme}       % Give a unique label
\end{figure}

%\FloatBarrier
% ----------
The TB setup, shown in Figure~\ref{fig:TB_setup}, consists of three spare ATLAS modules~\cite{PERF-2007-01} 
of TileCal,
%mechanically equal to those installed in ATLAS
two long-barrels and one extended-barrel, stacked on a scanning table (see Figure~\ref{fig:beam_line}) that is capable of placing modules at different position and angle with respect to the incoming beam particles. 
%The mechanical differences between the TB modules and the ones of the modules installed in ATLAS are discussed in Ref.~\cite{Phase_II_Upgrade}. An extended-barrel (long-barrel) consists of one (two) super-drawer(s). The super-drawer of the long barrel LBC65 is equipped with the upgraded 3-in-1 front-end electronics cards. This super-drawer is named DEMONSTRATOR and provides also all the functionality of the electronics installed at the present in ATLAS (LEGACY SD)~\cite{Phase_II_Upgrade}. The other option, alternative to the upgrated 3-in-1 system, based on Application Specific Integrated Circuits (ASICs), the so-called FATALIC, has been mounted in the super-drawer LBA65.  The super-drawers M0 C and EBC65 were equipped with the LEGACY SD electronics. Multi-Anodes  Photo-Multipliers (MA PMT) were installed in the M0 A super-drawer. Details can be found in Ref.~\cite{Phase_II_Upgrade}.
An extended-barrel (long-barrel) consists of one (two) super-drawer(s). In the figure they are named M0A and M0C (module at the bottom), LBA 65 and LBC 65 (module in the middle) and EBC 65 (module at the top). Some of the super-drawers were equipped with different upgraded front-end electronics systems proposed for the ATLAS LHC Phase-II operations
%. In Figure~\ref{fig:TB_setup} they are named: FATALIC, MA PM and DEMONSTRATOR
~\cite{Phase_II_Upgrade}. 
The super-drawers EBC 65 and M0 C were equipped with the electronics installed currently in ATLAS~\cite{PERF-2007-01}.
%(LEGACY SD)
% ----------
\begin{figure}[ht]
\centering
\includegraphics[width=4.2cm,clip]{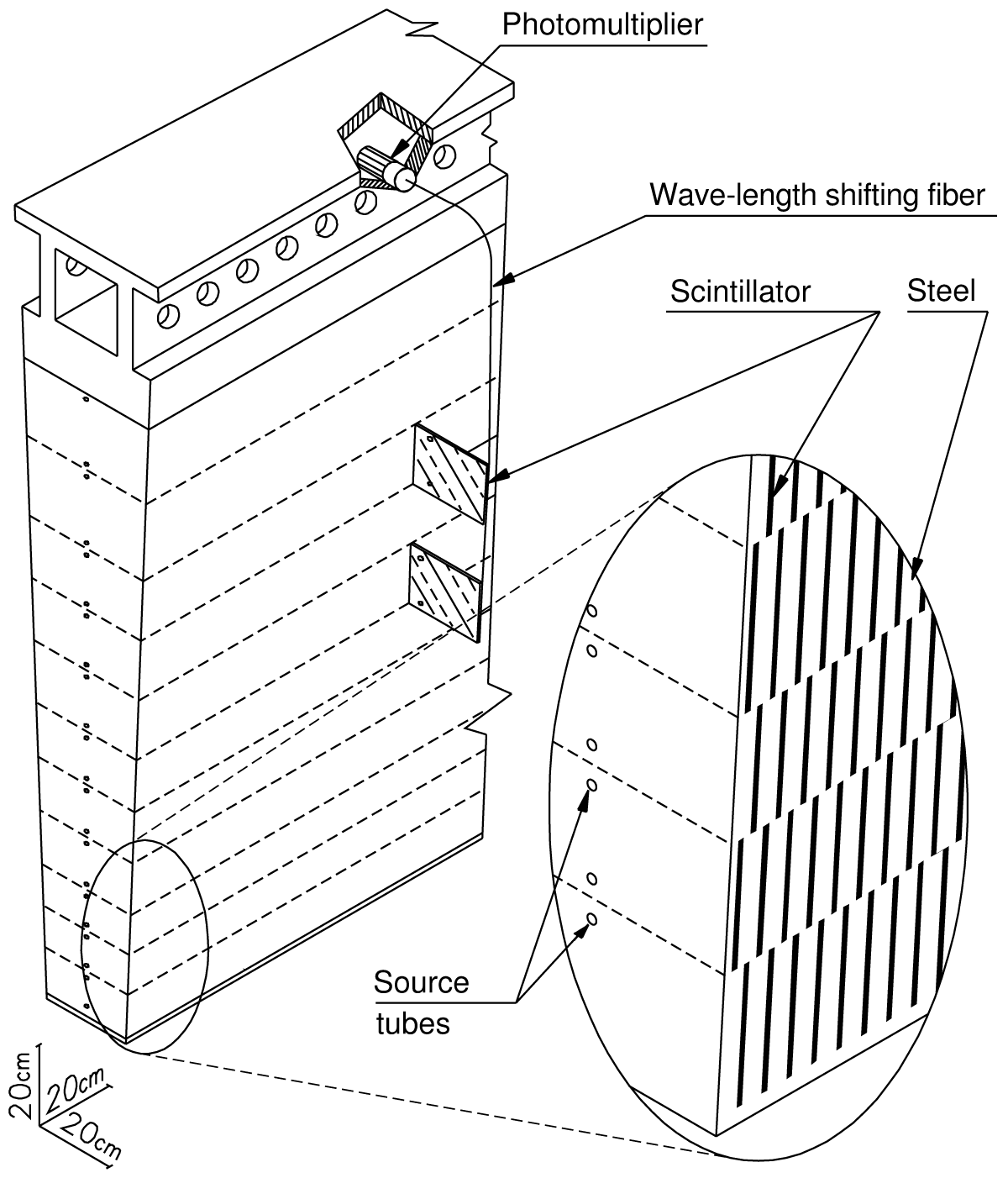}
\caption{Mechanical structure of a TileCal module, showing the slots in the steel for scintillating tiles and the method of light collection by wavelength-shifting fibres to PMs. The holes for radioactive source tubes that traverse the module perpendicularly to the iron plates and scintillating tiles are also shown.
%(see Section~\ref{subsec:detector}).
}
\label{fig:module}       % Give a unique label
\end{figure}
%\FloatBarrier
% ----------
As shown in Figure~\ref{fig:module}, the modules have a periodic structure of steel plates and scintillating tiles perpendicular to the $z$ axis.
%in Fig.~\ref{fig:TB_setup}.
Wavelength-shifting fibres transmit light produced in the tiles to the PMs~\cite{PMTs}. In each module a three-dimensional cell structure is defined by grouping optical fibres connected to the same PM~\cite{Optics}. In general two PMs read-out a cell and the signals are summed up to provide the cell response. A structure of three cell layers parallel to the $z$ axis is obtained. The cell layers A, BC and D in half long-barrel and A, B and D in extended barrel are shown in Figure~\ref{fig:module_scheme}.

As in the ATLAS detector at LHC, the energy deposited in a cell of the TB detector, $E_\text{c}^\text{raw}$, was determined making use of the Optimal Fit method~\cite{fit_method}. The linearity of the ADC's is determined using the Charge Injection System (CIS)~\cite{production_modules}. The inter-calibration of the different calorimeter cells was obtained by equalizing the PM current induced by movable radioactive $^{137}$Cs sources that cross every row of scintillating tiles near the edges (see Figure~\ref{fig:module}). Since the scintillating tile response depends on the impact point position of the particle in the tile and on the tile size, correction factors were applied for each layer of the calorimeter. Those values were determined from 1990’s Test Beam data, which measured the response to muons impinging on the calorimeter with a direction parallel to the $z$ axis (see Figure~\ref{fig:TB_setup}), and from the measurements obtained using a Sr source~\cite{production_modules}.
The scale of the reconstructed cell energy, $C_\text{c}^\text{EM}$ = 1.05 pC⁄GeV, was obtained using electron beams incident at the centre of each cell with an angle of $20^\circ$ with respect to the cell surface normal. The estimated uncertainty is $\Delta C_\text{c}^\text{EM}$ = 2.4\% ~\cite{production_modules}. The analysis of the muon and electron test beam data collected in the 2017 Test Beam~\cite{Phase_II_Upgrade} produced performance results that agree with the ones obtained using previous TBs~\cite{production_modules} and with in-situ measurements in ATLAS~\cite{Aaboud2018}.

To be consistent, the Optimal Fit method~\cite{fit_method} was applied also to reconstruct the energy deposited in the cells in the case of simulated events. The scale of the cell energy measurements was obtained %making use of the response of simulated electrons.
%the cell energy measurement used was obtained
using the response to simulated electrons

The energy deposited by the beam particles incident the detector, $E^\text{raw}$, was determined as the sum of the energy measured in the calorimeter cells.
% ----------
%\FloatBarrier
% ----------
\section{Analysis of experimental data}
\label{sec:data_selection}
% ----------
The results discussed in this paper were obtained exposing the TB calorimeter setup to enriched tertiary positive hadron beams with energy, $E_\text{beam}$, equal to 16, 18, 20 and 30 GeV. As shown in Figure~\ref{fig:TB_setup}, the beams hit at the middle of the cell A3 of the super-drawer LBC65 with an azimuth angle $\phi$ = 0 and polar angle $\theta$ of about 76°, corresponding to a pseudo-rapidity values $\eta$~=~0.25~\cite{PERF-2007-01}.
(see Figure~\ref{fig:module_scheme}). The angle from the calorimeter module normal is equal to 14 degrees. The numbers of events collected during the data taking period are reported in Table~\ref{tab:selection chain} (Physics Trigger).
\subsection{Collimated single-particle events}
\label{subsec:single_particles}
Collimated single-particle events were first selected using beam detectors upstream of the TB calorimeter setup. The selection criteria on the beam line scintillating counters signals, $E_\text{S1}$  and $E_\text{S2}$, were established making use of the responses of S1 and S2 to muons. Muon events were recognized by requiring an energy deposited in 
%each of the three layers of 
the module LBC65 compatible with the one deposited by a minimum ionizing particle. The retained events satisfy the criteria:
\begin{equation}
%E_\text{S1} < 2\times peak^\text{muon}_\text{S1}
E_\text{S1} < 2\times E^\text{m.p. }_\text{S1}(\mu)
\label{eq:S1_cut}
\end{equation}
and
\begin{equation}
E_\text{S2} < 2\times E^\text{m.p.}_\text{S2}(\mu) 
\label{eq:S2_cut}
\end{equation}
where the quantities $E^\text{m.p. }_\text{S1}(\mu)$ and $E^\text{m.p. }_\text{S2}(\mu)$ are the most probable (m. p.) values of the S1 and S2 muon signal distributions respectively. The selection criteria, especially useful for electron studies, remove particles that initiated a shower upstream of the calorimeter, as well as multi-particle beam events. The number of events retained after the application of the criterion are reported in Table~\ref{tab:selection chain} (Selection 1.).
% ----------
\begin{table}[ht]
\centering
\caption{Numbers of experimental data events collected and retained in the analysis selection steps for each of the four beam energies.
The number of events identified as electrons, pions, kaons and protons is reported. Selection criteria and determination of statistical uncertainties on the number of electrons and pions are discussed in the text.}
\label{tab:selection chain}       
% Give a unique label
% For LaTeX tables you can use 
%\begin{tabular}{	|l|c|c|c|c|} 
%\hline 
%$E_\text{beam}$ [GeV] & 16 & 18 & %20 & 30 \\ \hline
%%\hline
%\multicolumn{5}{|l|}{
%Experimental data & & & &  \\ \hline
%\multicolumn{5}{|l|}{Experimental data}\\ \hline 
%Physics Trigger & 694658 &  944460 &  1226756 &  1297099 \\ 
%\hline
%Selection~1.  & 656262 & 895863 & 1155580 & 1230470 \\ 
%Cut 2: Beam line chambers impact point &  &  &  &  \\ \hline
%Selection~2.  & 552179 & 771513 & 935131 & 1069709 \\ 
%Selection~3.  & 501013 & 700590 & 777386  & 983892 \\
%~~~~~~~~~~~events rejection & & & &%\multicolumn{4}{|c|}{}
%\hline
%$e/\pi$ & 385718 & 556782 & 611687  & 723286 \\
%$K/p$ & 86635 & 133071 & 154181  & 137119 \\
%Electrons & 67647$\pm$9198 & 70834$\pm$3665 & 62137$\pm$3548 & 28288$\pm$2481 \\ 
%Pions & 318071$\mp$9198 & 485948$\mp$3665 & 549550$\mp$3548 & 694998$\mp$2481  \\ 
%Kaons & 2372 & 4674 & 6782 & 11296  \\ 
%Protons & 84263 & 128397 & 147399 & 125823 \\ \hline
%\end{tabular}
\begin{tabular}{	|l|c|c|}
%begin{tabular}{	|l|c|c|c|c|}
\hline 
$E_\text{beam}$ [GeV] & 16 & 18 \\ \hline
Physics Trigger & 694658 &  944460 \\ 
%\hline
Selection~1.  & 656262 & 895863 \\ 
Selection~2.  & 552179 & 771513 \\ 
Selection~3.  & 501013 & 700590 \\
\hline
$e/\pi$ & 385718 & 556782 \\
$K/p$ & 86635 & 133071 \\
Electrons & 67647$\pm$9198 & 70834$\pm$3665 \\ 
Pions & 318071$\mp$9198 & 485948$\mp$3665 \\ 
Kaons & 2372 & 4674 \\ 
Protons & 84263 & 128397 \\ 
\hline
\hline
$E_\text{beam}$ [GeV] & 20 & 30 \\ \hline
Physics Trigger &  1226756 &  1297099 \\ 
%\hline
Selection~1.  & 1155580 & 1230470 \\ 
Selection~2.  & 935131 & 1069709 \\ 
Selection~3.  & 777386  & 983892 \\
\hline
$e/\pi$ & 611687  & 723286 \\
$K/p$ & 154181  & 137119 \\
Electrons & 62137$\pm$3548 & 28288$\pm$2481 \\ 
Pions & 549550$\mp$3548 & 694998$\mp$2481  \\ 
Kaons & 6782 & 11296  \\ 
Protons & 147399 & 125823 \\ \hline
\end{tabular}

%\end{center}
\end{table}
%\FloatBarrier
% ----------
Events with a beam trajectory far away from the beam axis were rejected because the beam particles might have scattered upstream and therefore be off-energy. 
The beam chamber BC1 allows a determination of the transverse beam impact point coordinates, $x_\text{BC1}$ and $y_\text{BC1}$. Gaussian functions were fitted to the distributions of each data set to determine the peak values $x_\text{BC1}^\text{peak}$ and $y_\text{BC1}^\text{peak}$ respectively. 
%The sigma values ($\sigma$'s) are about 30 mm and differ slightly from run to run. 
The accepted events have the beam impact point coordinates inside the square surface of the trigger scintillating counters:
%verifying the conditions:
\begin{equation}
|x_\text{BC1}-x_\text{BC1}^\text{peak}|< 2.5~\text{cm}
\label{eq:yBC1cut}
\end{equation}
and
\begin{equation}
|y_\text{BC1}-y_\text{BC1}^\text{peak}|< 2.5~\text{cm} .
\label{eq:yBC2cut}
\end{equation}
%where $x_\text{BC1}^\text{peak}$ and $y_\text{BC1}^\text{peak}$ are the peaks values of the distributions.
The numbers of events retained after the application of this criterion are reported in Table~\ref{tab:selection chain} (Selection~2.).
\subsection{Identification of muons and electrons}
\label{subsec:pure_hadrons}
\subsubsection{Muon rejection}
\label{subsubsec:muon_identification}
\begin{figure}[ht]
\centering
\subfloat[]{\includegraphics[width=4.3cm,clip]{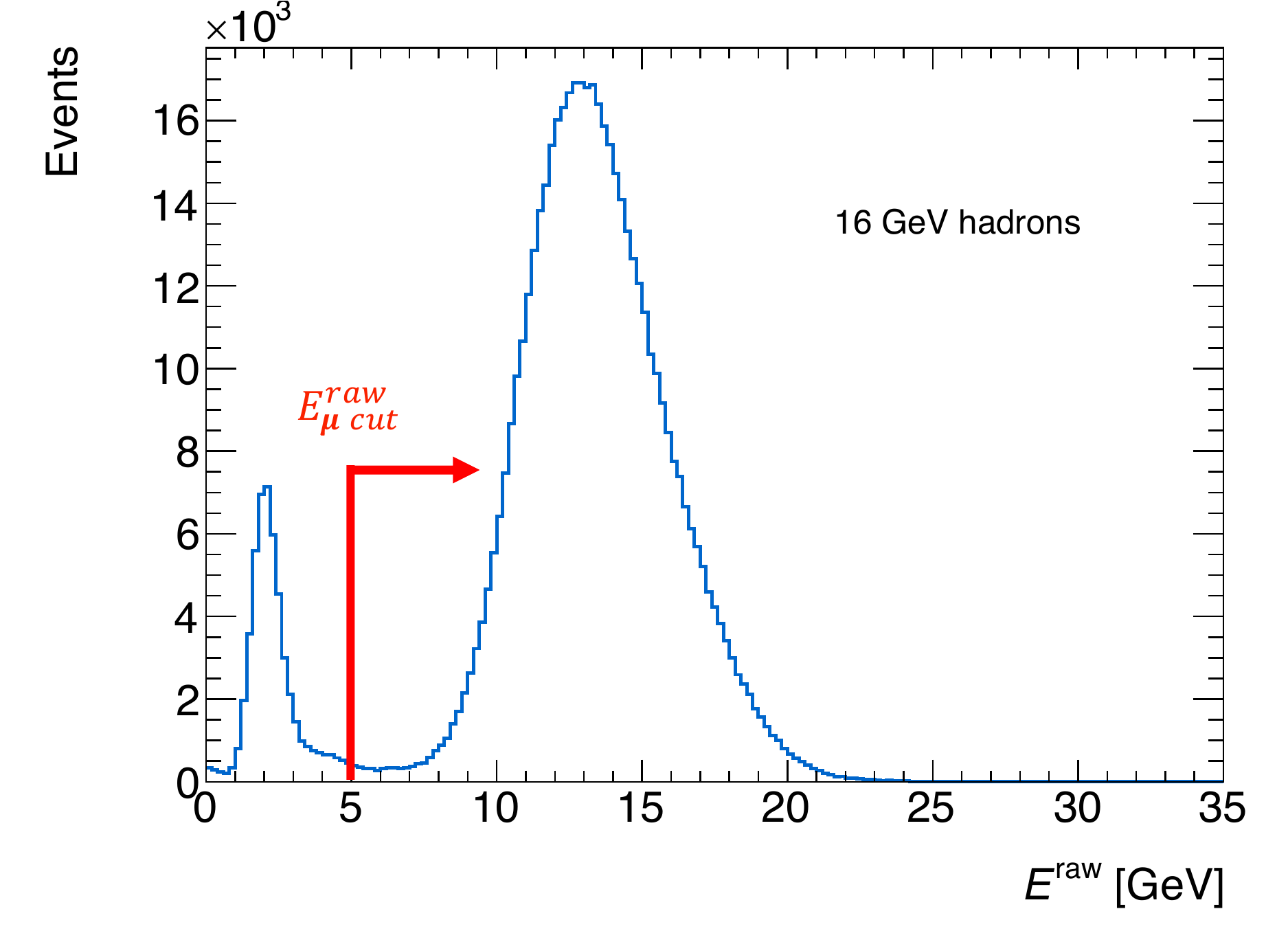}}
\subfloat[]{\includegraphics[width=4.3cm,clip]{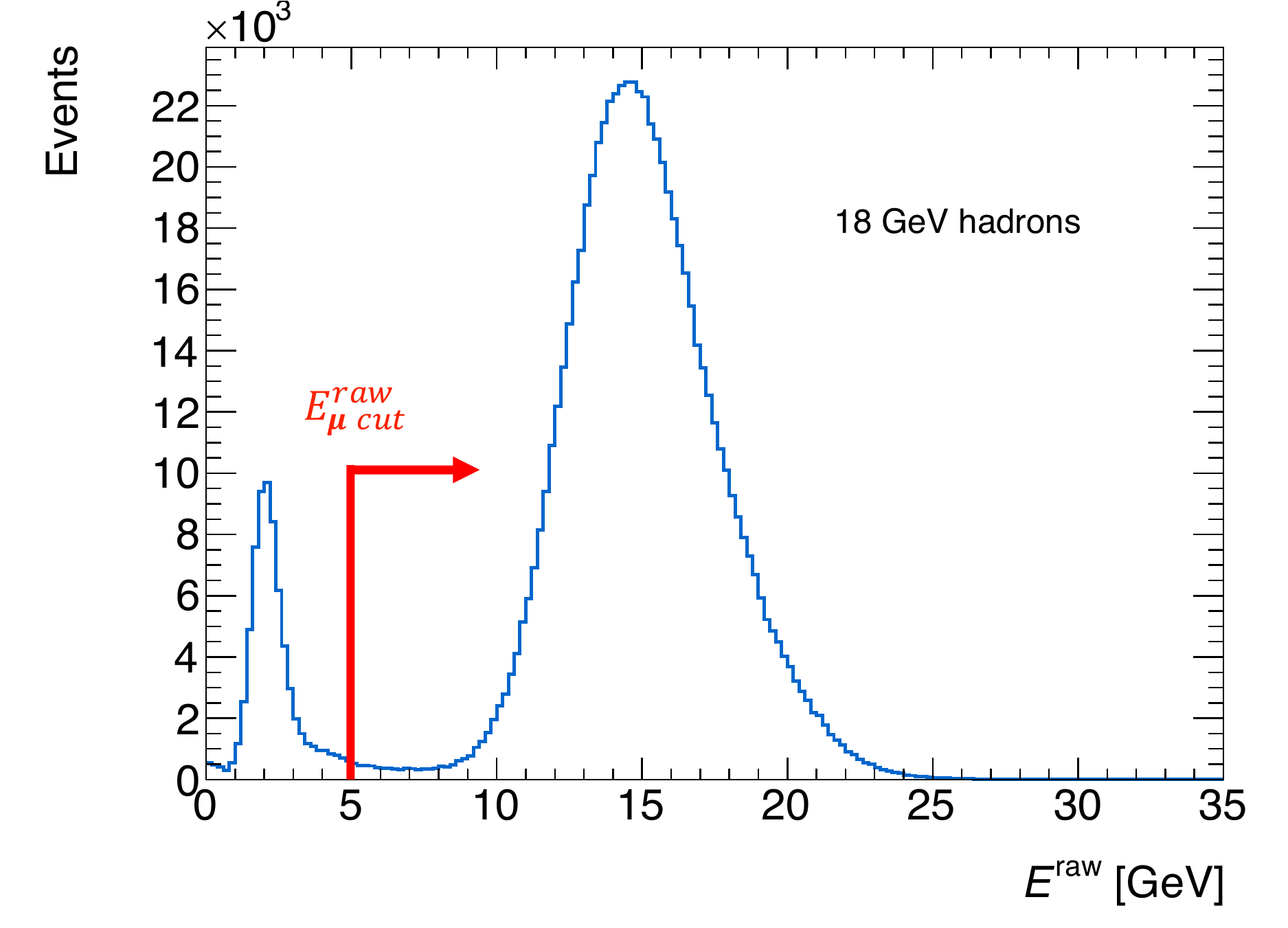}}
\caption{Distributions of the energy $E^\text{raw}$ in GeV measured in the calorimeter modules in the case of particle beam energies equal to 16 GeV (a) and 18 GeV (b). The events were selected applying the selection criteria up to Selection~2. (see Table~\ref{tab:selection chain}). The muons and spurious events were rejected in the analysis requiring $E^\text{raw}$ larger than $E^\text{raw}_{\mu~cut}$ = 5 GeV, as shown in the histograms.
%\textcolor {red} {Put an horizontal arrow from the left to the right starting on the top of the vertical red segment}
}
\label{fig:muon energy}       % Give a unique label
\end{figure}

The second set of criteria allows identifying pure samples of hadrons. As already mentioned, at the  considered beam energies, muons are minimum ionizing particles and deposit in the scintillating tiles energy much smaller than electrons and hadrons (see Figure~\ref{fig:muon energy}). The muon rejection was obtained requiring a reconstructed energy in the detector (see Section~\ref{subsec:detector})
%(see Section~\ref{subsec:shower_energy})
$E^\text{raw}$ $\rangle$ $E^\text{raw}_{\mu~cut}$ = 5 GeV. 
The selection criterion allows also a rejection of spurious trigger events. The retained events are reported in Table~\ref{tab:selection chain} (Selection~3.).
% ---------
%\FloatBarrier
% ----------
\subsubsection{Electron identification}
\label{subsubsec:electron_identification}
\begin{figure}[ht]
\centering
\subfloat[]{\includegraphics[width=4.3cm,clip]{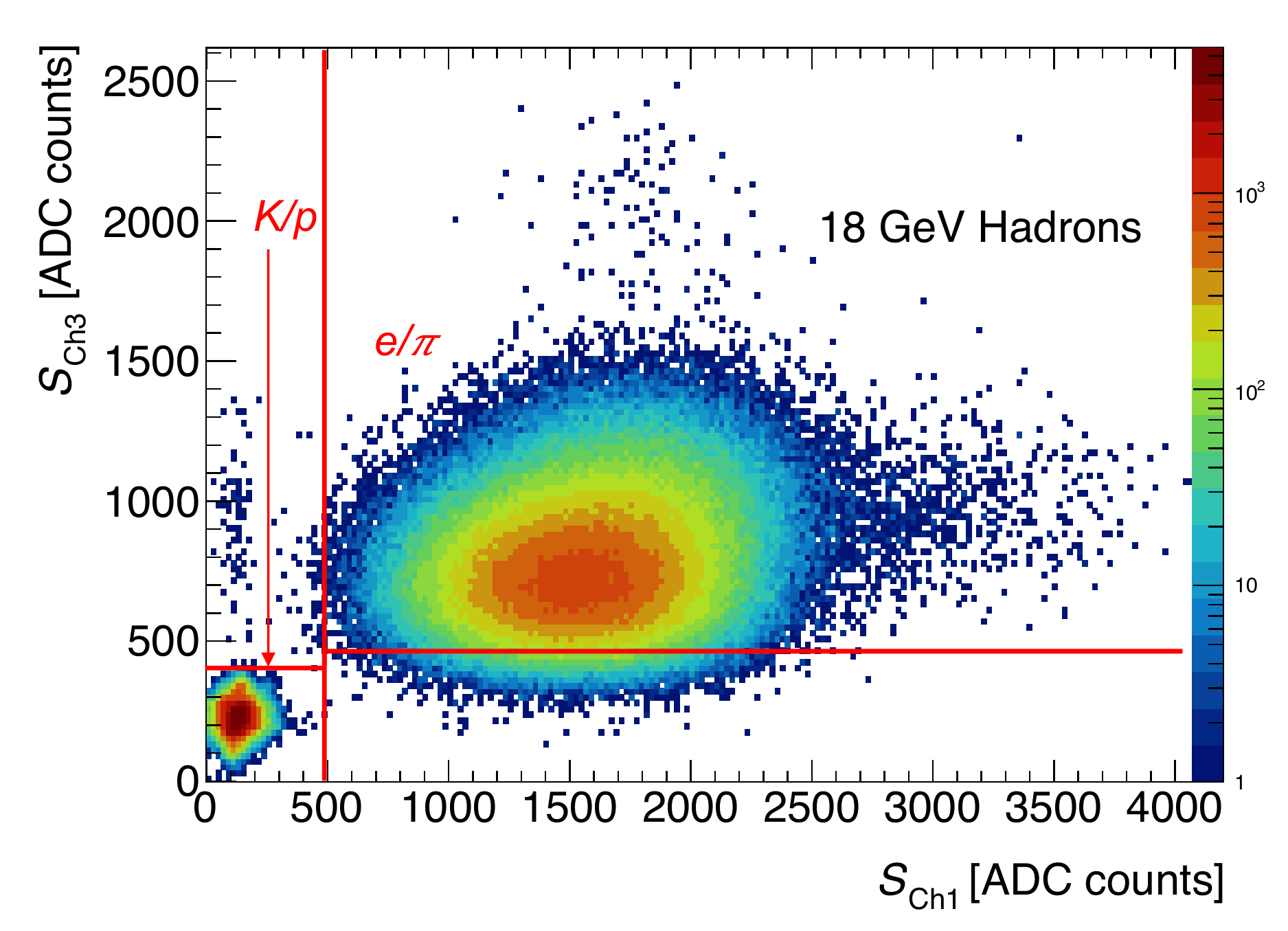}} 
%\\
\subfloat[]{\includegraphics[width=4.3cm,clip]{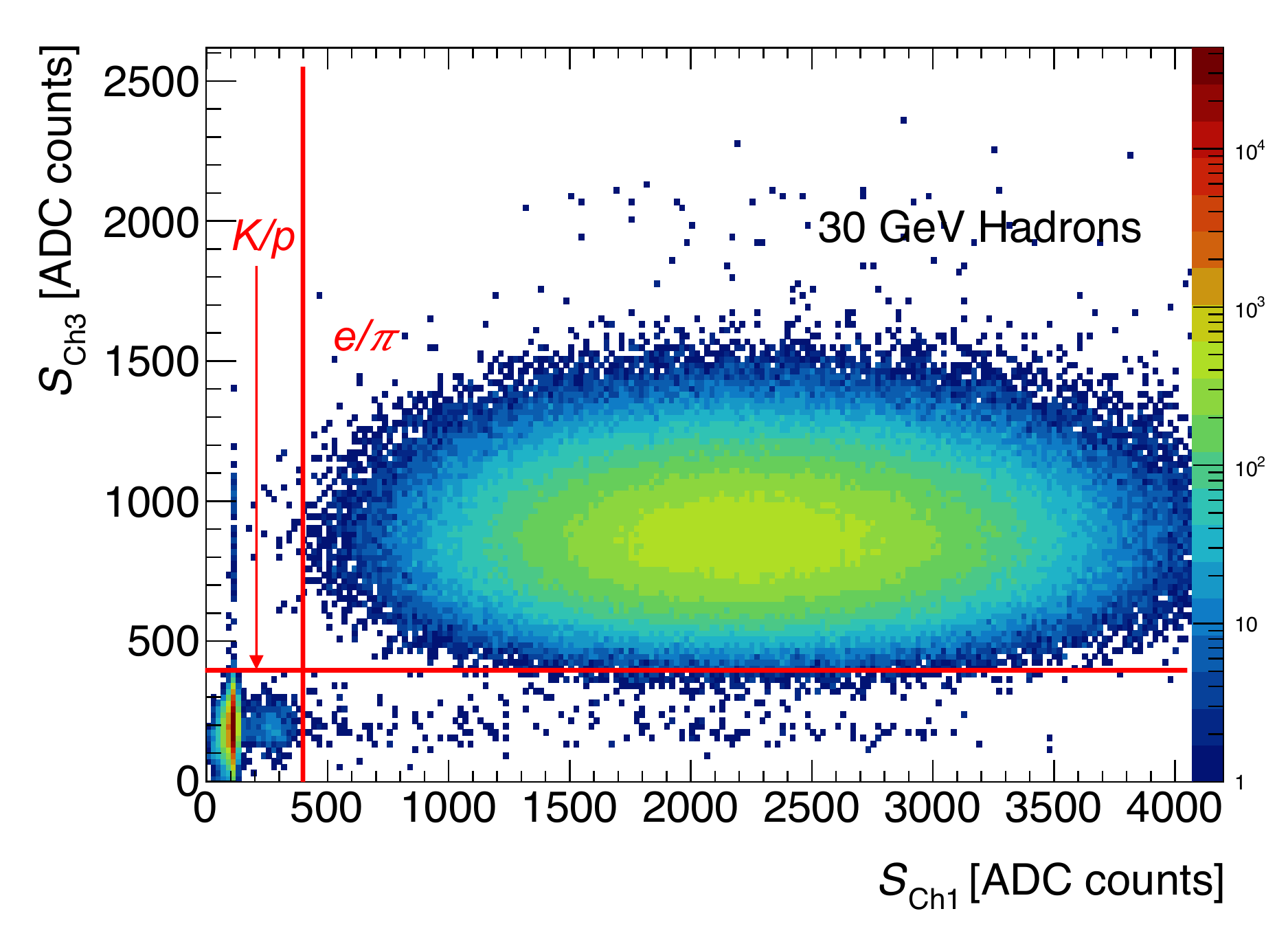}}
\caption{Scatter plots of the signals measured in the Cherenkov counter Ch3, $S_\text{Ch3}$, as a function of the signals measured in the Cherenkov counter Ch1, $S_\text{Ch1}$, in ADC counts. The histograms were obtained analysing data with beam energies equal to 18 GeV (a) and 30 GeV (b). The events were selected applying selection criteria summarized in Table~\ref{tab:selection chain}, up to Selection~3.  The cut values used to select kaon and proton, $K/p$, (left/bottom) and electron and pion, $e/\pi$, (right/top) events are shown. Colors are used in the plots to show the cell contents.
%\textcolor {red} {The regions do not correspond to the values of Table 3. Change the sens of the vectors}
}
\label{fig:Ch3 vs Ch1 18 30 GeV}       % Give a unique label
\end{figure}

As shown in Figure~\ref{fig:Ch3 vs Ch1 18 30 GeV}, the signals measured in Cherenkov counters Ch1 and Ch3, $S_\text{Ch1}$ and $S_\text{Ch3}$, respectively, allow a separation of pions and electrons ($e/\pi$) from kaons $K$ and protons $p$ ($K/p$).  
The selection criteria in ADC counts applied on the signals are reported in Table~\ref{tab:Cs cuts}. The numbers of the identified events are reported in Table~\ref{tab:selection chain}. As discussed in Section~\ref{sec:hadrons_identification}, the Ch2 measurements allow separating kaons and protons.
% ------------
\begin{table}[ht]
\centering
\caption{Selection criteria in $S_\text{Ch1}$, $S_\text{Ch2}$ and $S_\text{Ch3}$ signals applied to identify $e/\pi$, $K$ and $p$ event samples for the four particle beam energy data sets.  The Cherenkov signals are measured in ADC counts}
\label{tab:Cs cuts}       % Give a unique label
% For LaTeX tables you can use 
\begin{tabular}{	|l|c|c|c|} 
\hline 
$E_\text{beam}$  & $e/\pi$ & $K$ & $p$ \\ 
~[GeV]&  &  &  \\ \hline
16 & $S_\text{Ch1} \geq 500$ & $S_\text{Ch1} < 500$ & $S_\text{Ch1} < 500$\\
  & $S_\text{Ch3}\geq470$ & $S_\text{Ch3}\leq300$ & $S_\text{Ch3}\leq300$\\
  &  & $S_\text{Ch2}\geq1500$ & $S_\text{Ch3}\leq400$\\  
      \hline
18 & $S_\text{Ch1} \geq 500$ & $S_\text{Ch1} \leq 500$ & $S_\text{Ch1} \leq 500$\\
  & $S_\text{Ch3}\geq450$ & $S_\text{Ch3}\leq400$ & $S_\text{Ch3}\leq400$\\
  &  & $S_\text{Ch2}\geq2000$ & $S_\text{Ch3}\leq400$\\  
      \hline
20 & $S_\text{Ch1} \geq 500$ & $S_\text{Ch1} \leq 300$ & $S_\text{Ch1} \leq 300$\\
  & $S_\text{Ch3}\geq450$ & $S_\text{Ch3}\leq400$ & $S_\text{Ch3}\leq400$\\
  &  & $S_\text{Ch2}\geq2000$ & $S_\text{Ch3}\leq500$\\  
      \hline
30 & $S_\text{Ch1} \geq 500$ & $S_\text{Ch1} \leq 400$ & $S_\text{Ch1} \leq 400$\\
  & $S_\text{Ch3}\geq400$ & $S_\text{Ch3} < 400$ & $S_\text{Ch3} < 400$\\
  &  & $S_\text{Ch2}\geq1100$ & $S_\text{Ch3}\leq200$\\  
      \hline
\end{tabular}
%\begin{tabular}{	|l|c|c|c|c|} 
%\hline 
%$E_\text{beam}$ [GeV] & 16 & 18 & 20 & 30 \\ \hline
%$e/\pi$ samples &  $S_\text{Ch1} \geq 500$   & $S_\text{Ch1} \geq  500$  & $S_\text{Ch1} \geq 500$ & $S_\text{Ch1} \geq 500 $ \\
%&  $S_\text{Ch3}\geq470$   & $S_\text{Ch3}\geq450$   & $S_\text{Ch3}\geq450 $ & $S_\text{Ch3}\geq400 $ \\   \hline
%Kaons  &  $S_\text{Ch1} \leq 500$   & $S_\text{Ch1} \leq 500$   & $S_\text{Ch1} \leq 300$  & $S_\text{Ch1} \leq 400 $ \\ 
%                 &  $S_\text{Ch3} \leq 300$   & $S_\text{Ch3} \leq 400$   & $S_\text{Ch3} \leq 400$  & $S_\text{Ch3} \leq 400 $ \\   
 %                &  $S_\text{Ch2} \geq 1500$   & $S_\text{Ch2} \geq 2000$   & $S_\text{Ch2} \geq 2000$  & $S_\text{Ch2} \geq 1100 $ \\   \hline             
%Protons          &  $S_\text{Ch1} \leq 500$   & $S_\text{Ch1} \leq 500$   & $S_\text{Ch1} \leq 300$  & $S_\text{Ch1} \leq 400 $ \\ 
%                 &  $S_\text{Ch3} \leq 300$   & $S_\text{Ch3} \leq 400$   & $S_\text{Ch3} \leq 400$  & $S_\text{Ch3} \leq 400 $ \\   
%                 &  $S_\text{Ch2} \leq 400$   & $S_\text{Ch2} \leq 400$   & $S_\text{Ch2} \leq 500$  & $S_\text{Ch2} \leq 200 $ \\   \hline               
%\end{tabular}
%\end{center}
\end{table}
%\FloatBarrier
% --------------------------------------------------
%\FloatBarrier

%\FloatBarrier
% ---------------------------------------------------------
\begin{figure}[ht]
\centering
\subfloat[]{\includegraphics[width=4.3cm,clip]{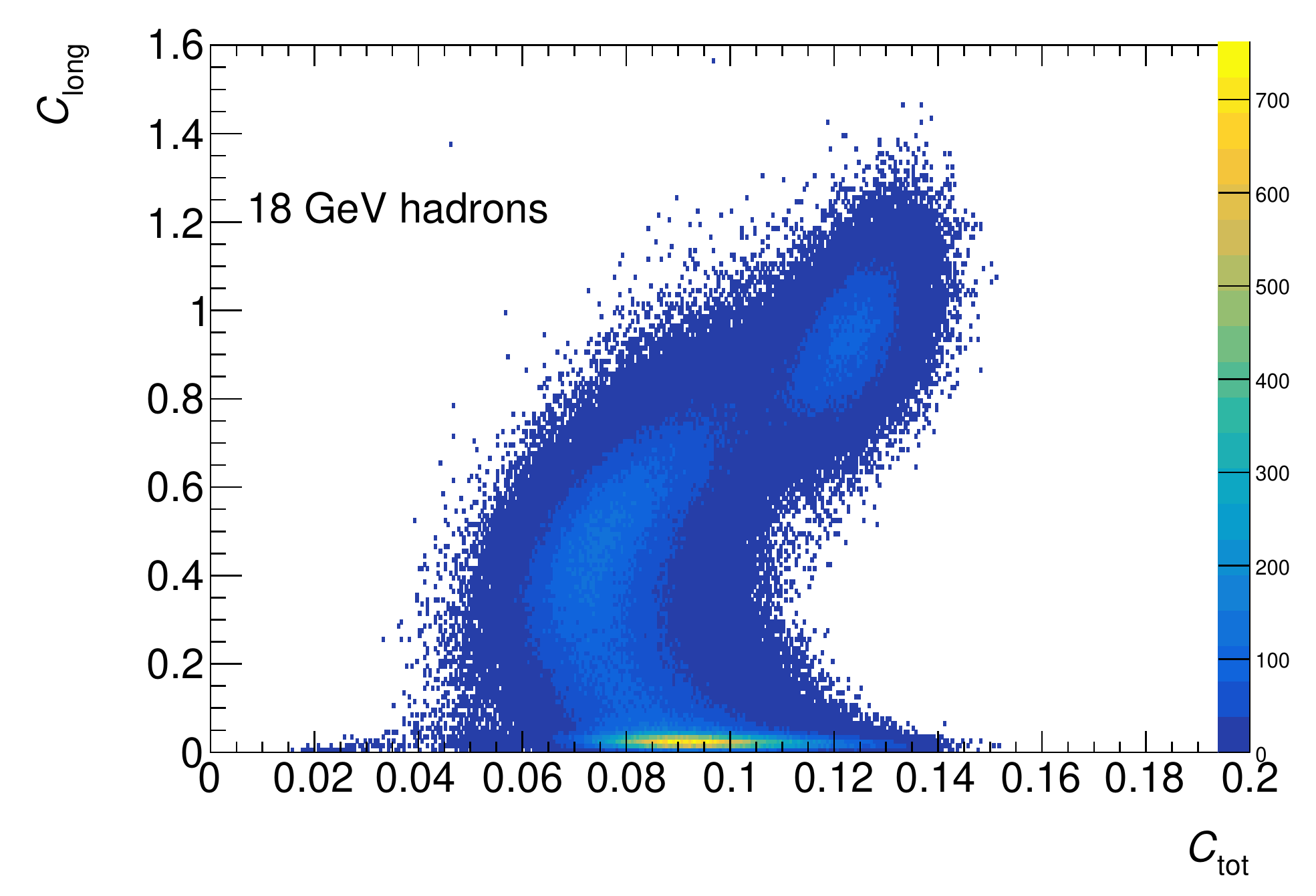}}
\subfloat[]{\includegraphics[width=4.3cm,clip]{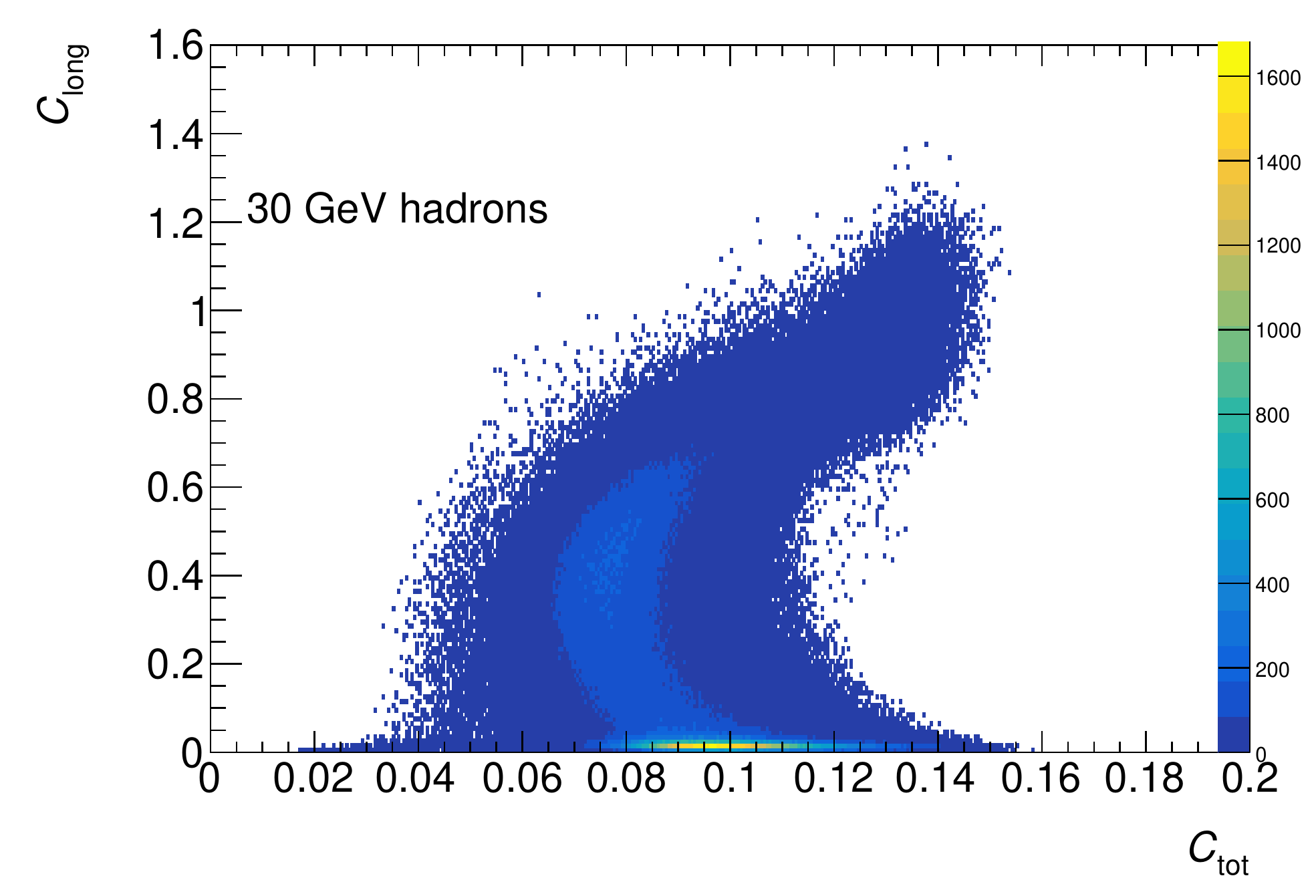}}
\caption{Scatter plot $C_\text{long}$ vs $C_\text{tot}$  of $e/\pi$ sample events produced by beams of particles with energies equal to 18 GeV (a) and 30 GeV (b). Colors are used in the plots to show the cell contents.}
\label{fig:Clong vs C_tot 18 30}       % Give a unique label
\end{figure}
%\FloatBarrier
% ---------------------------------------------------------
The electron components in $e/\pi$ samples were determined statistically exploiting the difference of electromagnetic and hadronic shower profiles in the calorimeter modules~\cite{production_modules}. Two separators, $C_\text{long}$ and $C_\text{tot}$, were used:
\begin{enumerate}
\item 
The shower profile parameter $C_\text{long}$ represents the fraction of the beam energy, $E_\text{beam}$, deposited in the layers A of the modules (see Figure~\ref{fig:module_scheme}) :
\begin{equation}
 C_\text{long}=\frac{\sum_{i=1}^3 \sum_{j=1}^3 (E_\text{c}^\text{raw})_{i,j}} {E_\text{beam}}
\label{eq:C_long}
\end{equation}
where $i$ = 1, 2 and 3 indicate the super-drawers M0 C, LBC65 and EBC65 respectively. The parameter $j$ runs over 3 contiguous cells of the three layers A around the cell hit by the beam and $E_\text{c}^\text{raw}$ stands for the energy measured in a cell (see Section~\ref{subsec:detector}).
\item
The separator $C_\text{tot}$ measures the spread of the energy E$_\text{c}^\text{raw}$ deposited in the cells of the modules:
\begin{equation}
\begin{split}
& C_\text{tot}=\frac{1}{\sum_{i=1}^{N_\text{cell}}[(E_\text{c}^\text{raw})_i]^\alpha} \times \\
& \sqrt{\frac{1}{N_\text{cell}} \sum_{i=1}^{N_\text{cell}}\Big([(E_\text{c}^\text{raw})_i]^\alpha-\frac{1}{N_\text{cell}}\sum_{i=1}^{N_\text{cell}}[(E_\text{c}^\text{raw})_i]^\alpha\Big)^2}
\label{eq:C_tot}
\end{split}
\end{equation}
where $N_\text{cell}$ = 24 stands for the total number of contiguous cells, around the hit cell, considered for the shower profile estimate and the exponent $\alpha$ = 0.6 was tuned using a Monte Carlo (MC) simulation program to achieve maximum electron pion separation~\cite{production_modules}.
\end{enumerate}
%Scatter plots, $C_\text{long}$ vs $C_\text{tot}$, of $e/\pi$ sample events obtained using beams of particles with $E_\text{beam}$ equal to 18 and 30 GeV are shown in Figure~\ref{fig:Clong vs C_tot 18 30}. They can be compared with the ones in Figure~\ref{fig:Clong vs C_tot 18 30 MC} obtained using simulated electrons and pions events with the same beam energies. In general the pions have small values of $C_\text{long}$ and C$_\text{tot}$, while in the case of electrons, the parameters have larger values localized in narrower regions. Pion events with large $C_\text{long}$ and C$_\text{tot}$ values are due to showers with large electromagnetic component.
%\FloatBarrier
% ----------------------------------------------------------
\begin{figure}[ht]
\centering
%\includegraphics[width=10cm,clip]{fig_04.pdf}
%\subfloat[]{\includegraphics[width=8cm,clip]{Scat_clongctot_mc_elec_18gev.pdf}}
\subfloat[]{\includegraphics[width=4.3cm,clip]{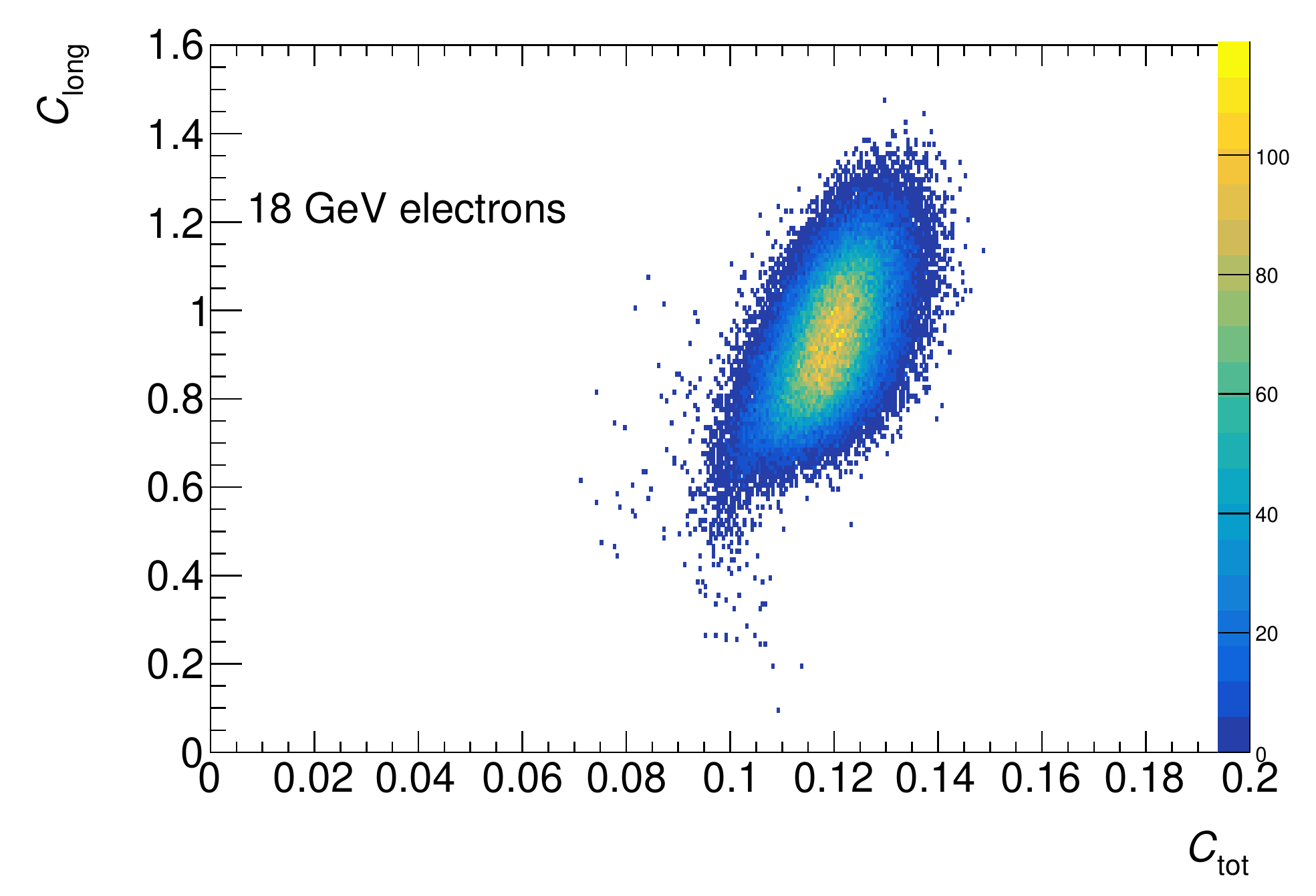}}
\subfloat[]{\includegraphics[width=4.3cm,clip]{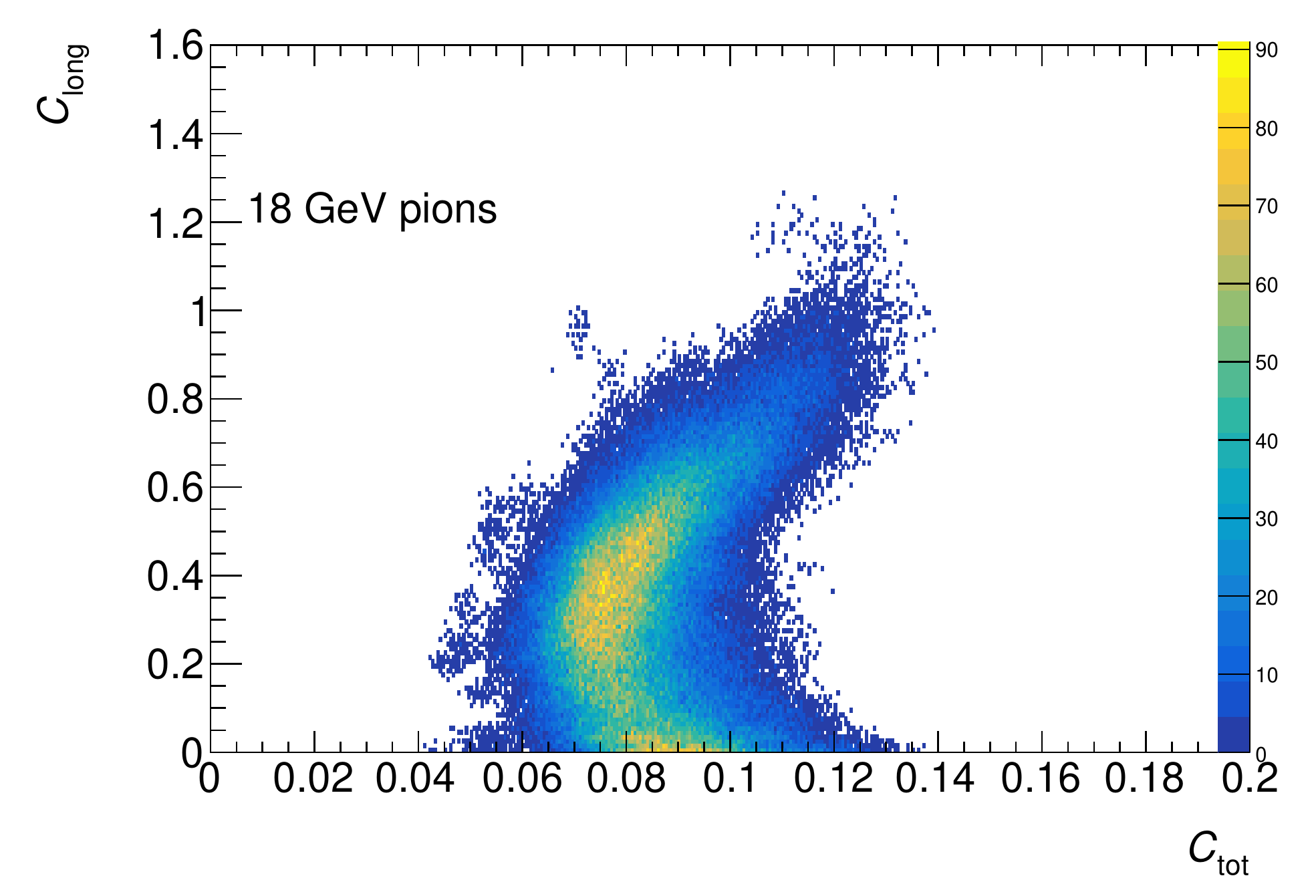}} \\
\subfloat[]{\includegraphics[width=4.3cm,clip]{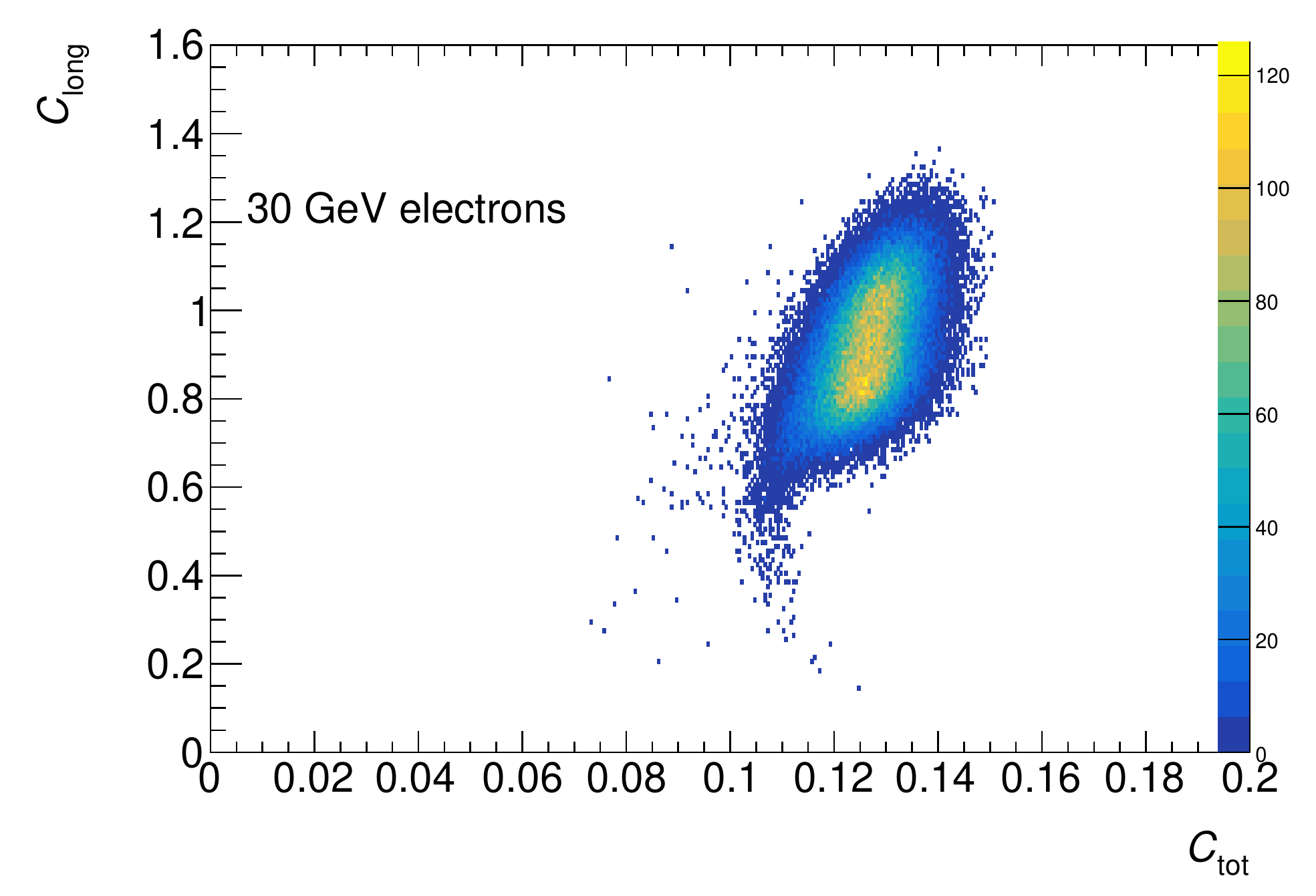}}
\subfloat[]{\includegraphics[width=4.3cm,clip]{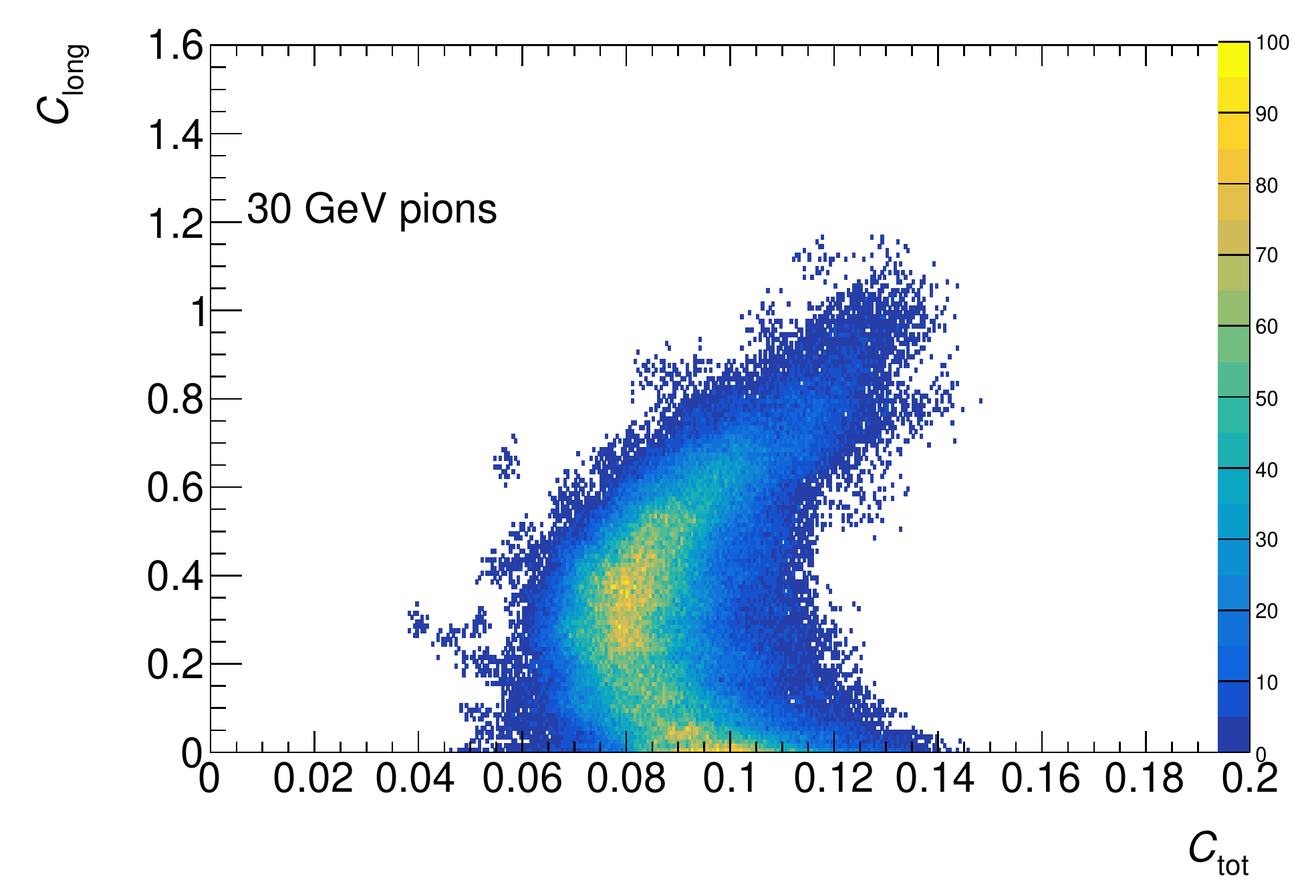}}
\caption{Scatter plot $C_\text{long}$ vs $C_\text{tot}$  obtained using simulated 18 GeV electrons (a), 18 GeV pions (b), 30 GeV electrons (c) and 30 GeV pions (d) beam. The color bands represent the number of events in the bins.}
\label{fig:Clong vs C_tot 18 30 MC}       % Give a unique label
\end{figure}
Scatter plots, $C_\text{long}$ vs $C_\text{tot}$, of $e/\pi$ sample events obtained using beams of particles with $E_\text{beam}$ equal to 18 and 30 GeV are shown in Figure~\ref{fig:Clong vs C_tot 18 30}. They can be compared with the ones in Figure~\ref{fig:Clong vs C_tot 18 30 MC} obtained using simulated electrons and pions events with the same beam energies. In general the pions have small values of $C_\text{long}$ and C$_\text{tot}$, while in the case of electrons, the parameters have larger values localized in narrower regions. Pion events with large $C_\text{long}$ and C$_\text{tot}$ values are due to showers with large electromagnetic component.

\begin {figure} [t]
\centering
\subfloat[]{\includegraphics[width=4.3cm,clip]{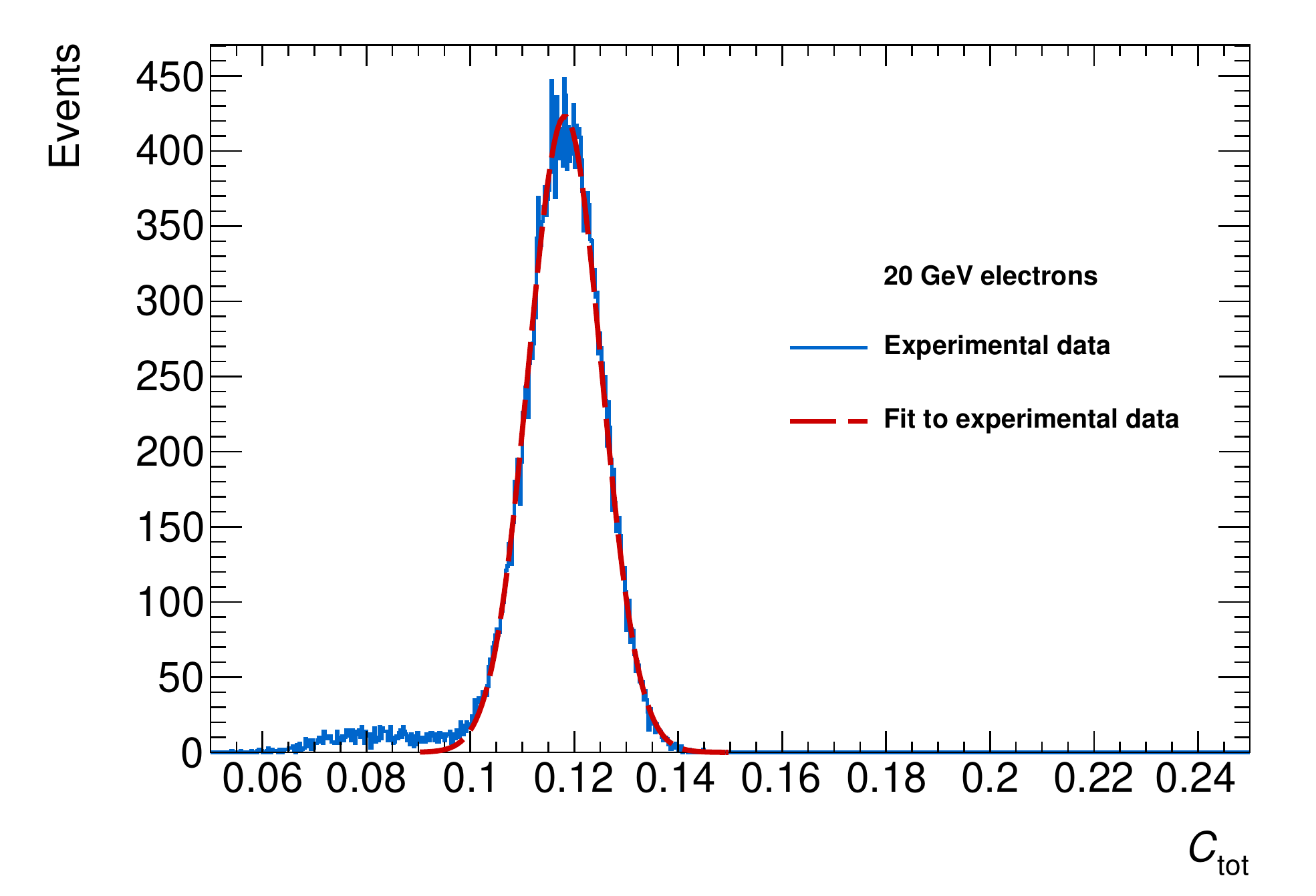}} \\
\subfloat[]{\includegraphics[width=4.3cm,clip]{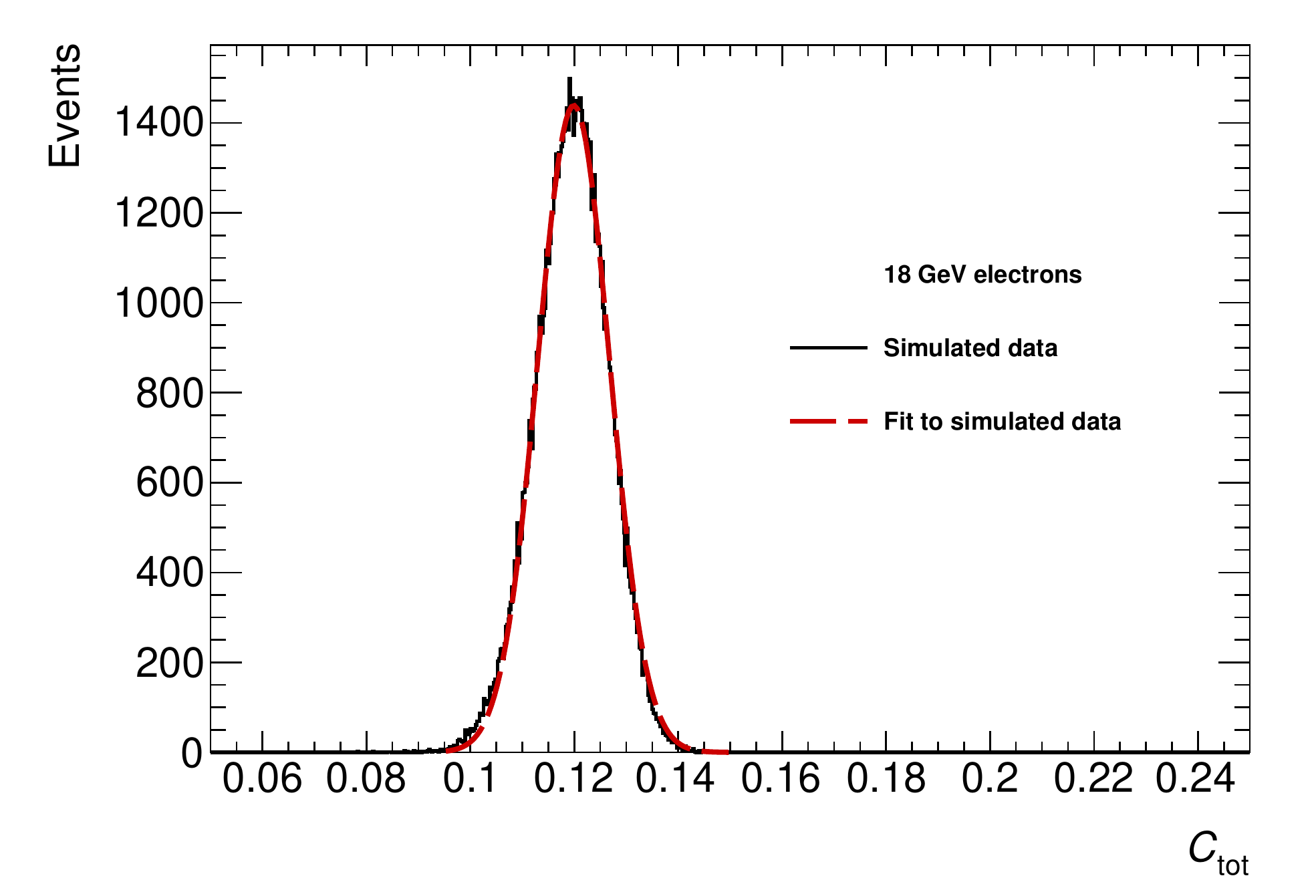}} 
\subfloat[]{\includegraphics[width=4.3cm,clip]{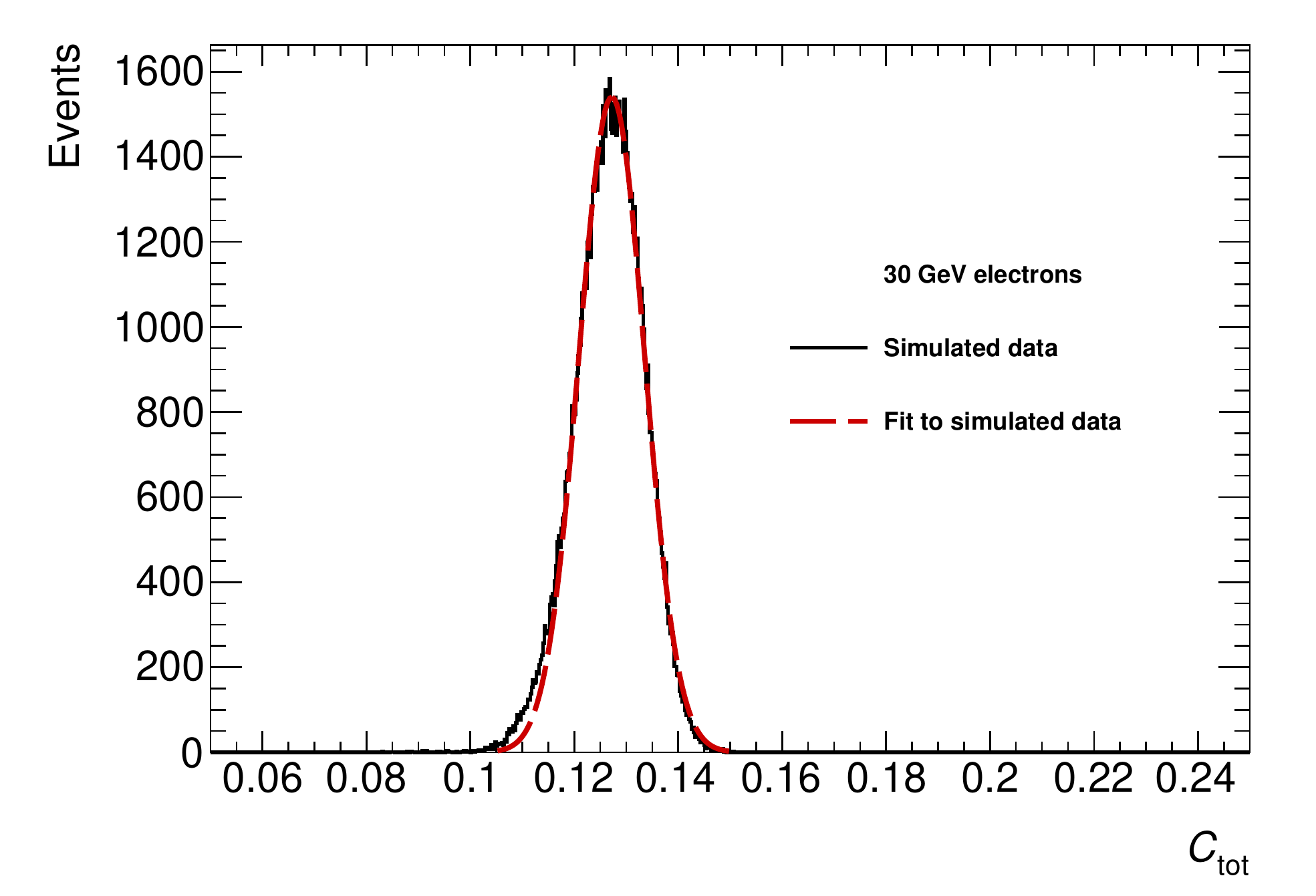}}
\caption{
%The black histograms show 
Distributions of $C_\text{tot}$ obtained using experimental electron-enriched beams of particles with an energy equal to 20 GeV (a), and simulated electrons with $E_\text{beam}$ equal to 18 GeV (b) and 30 GeV (c). Fit Gaussian functions obtained using the method of the least squares are superimposed in red on the distributions.}
\label{fig:e C_tot 18 30 MC 20 Exp}       % Give a unique label
\end{figure}

\begin{figure}[ht]
\centering
\subfloat[]{\includegraphics[width=4.3cm,clip]{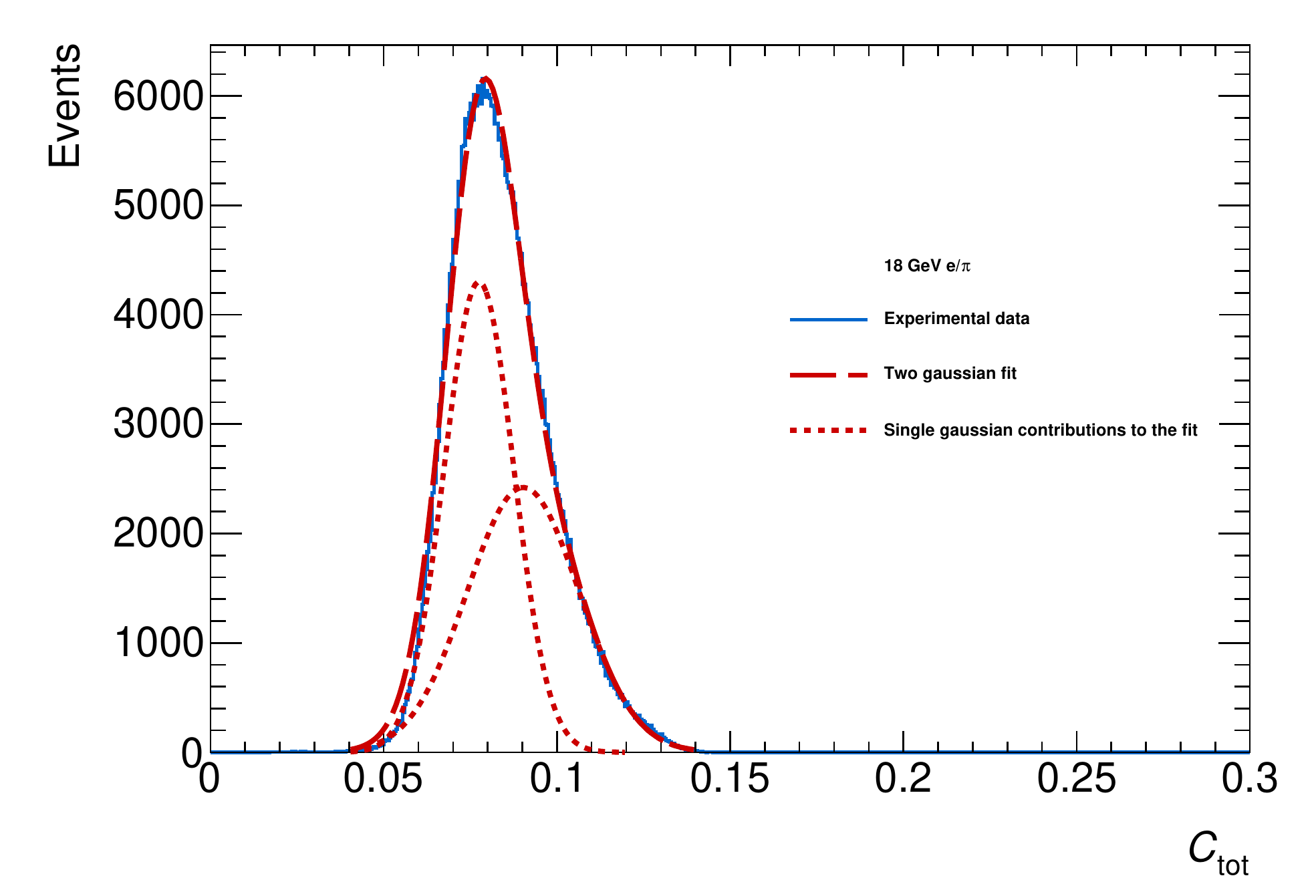}}
\subfloat[]{\includegraphics[width=4.3cm,clip]{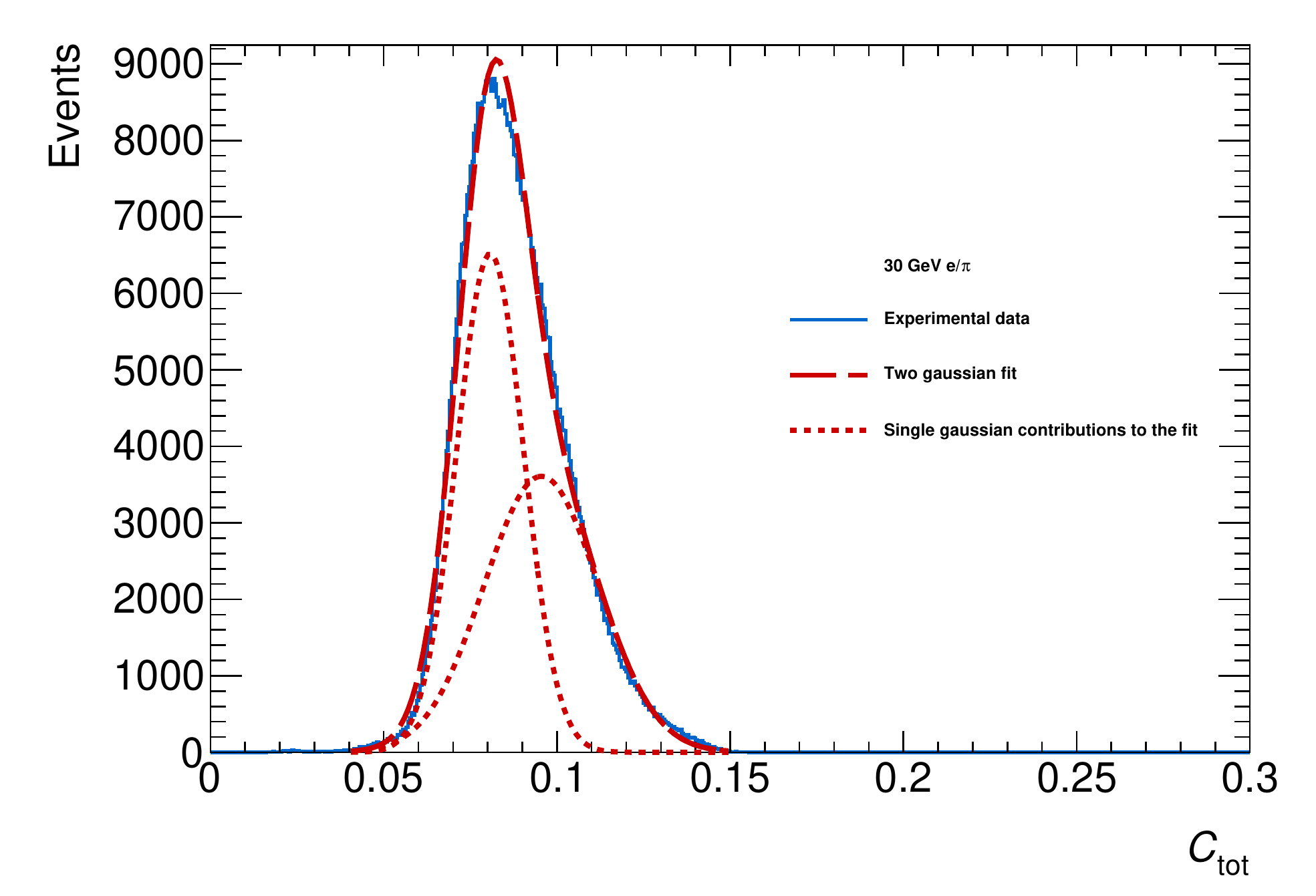}} \\
\subfloat[]{\includegraphics[width=4.3cm,clip]{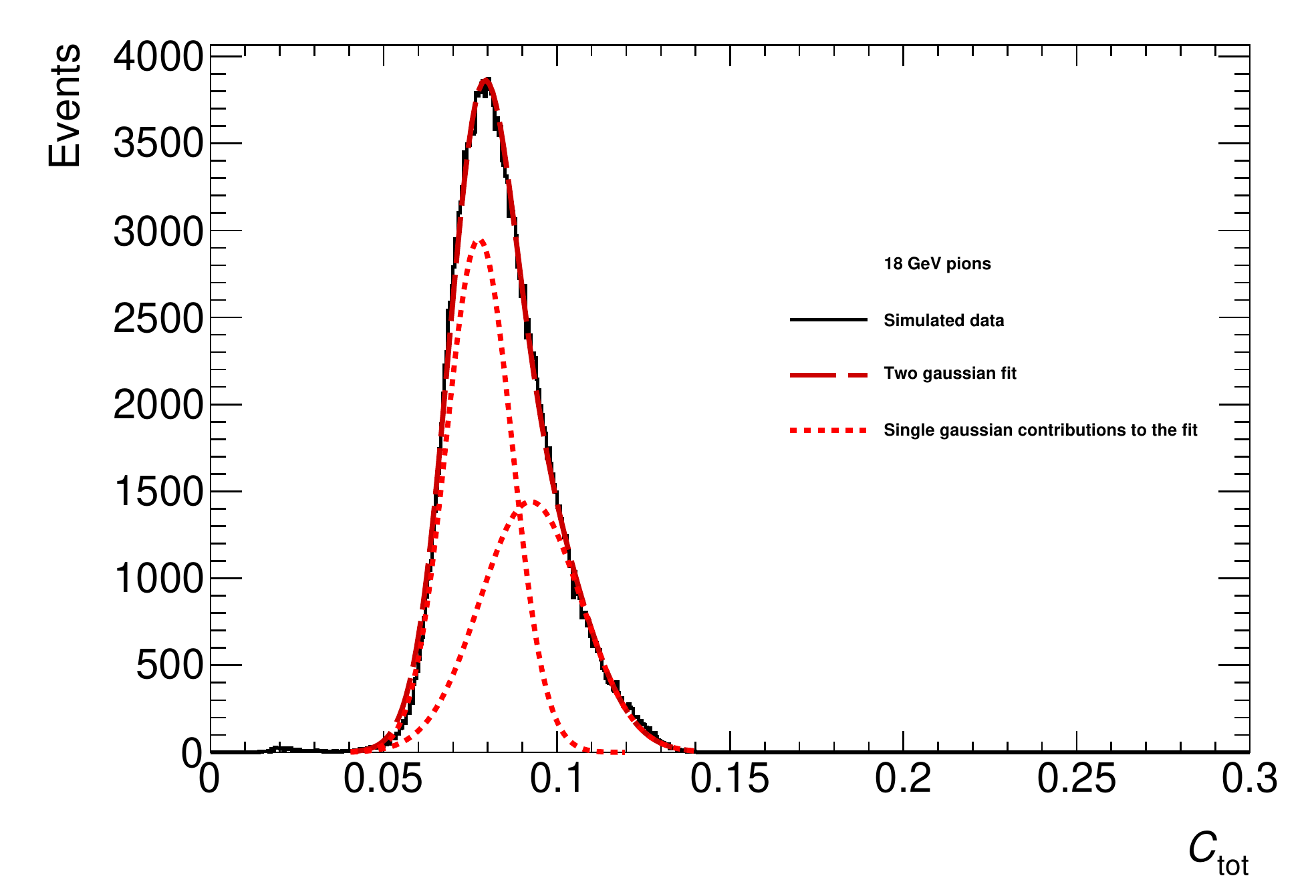}}
\subfloat[]{\includegraphics[width=4.3cm,clip]{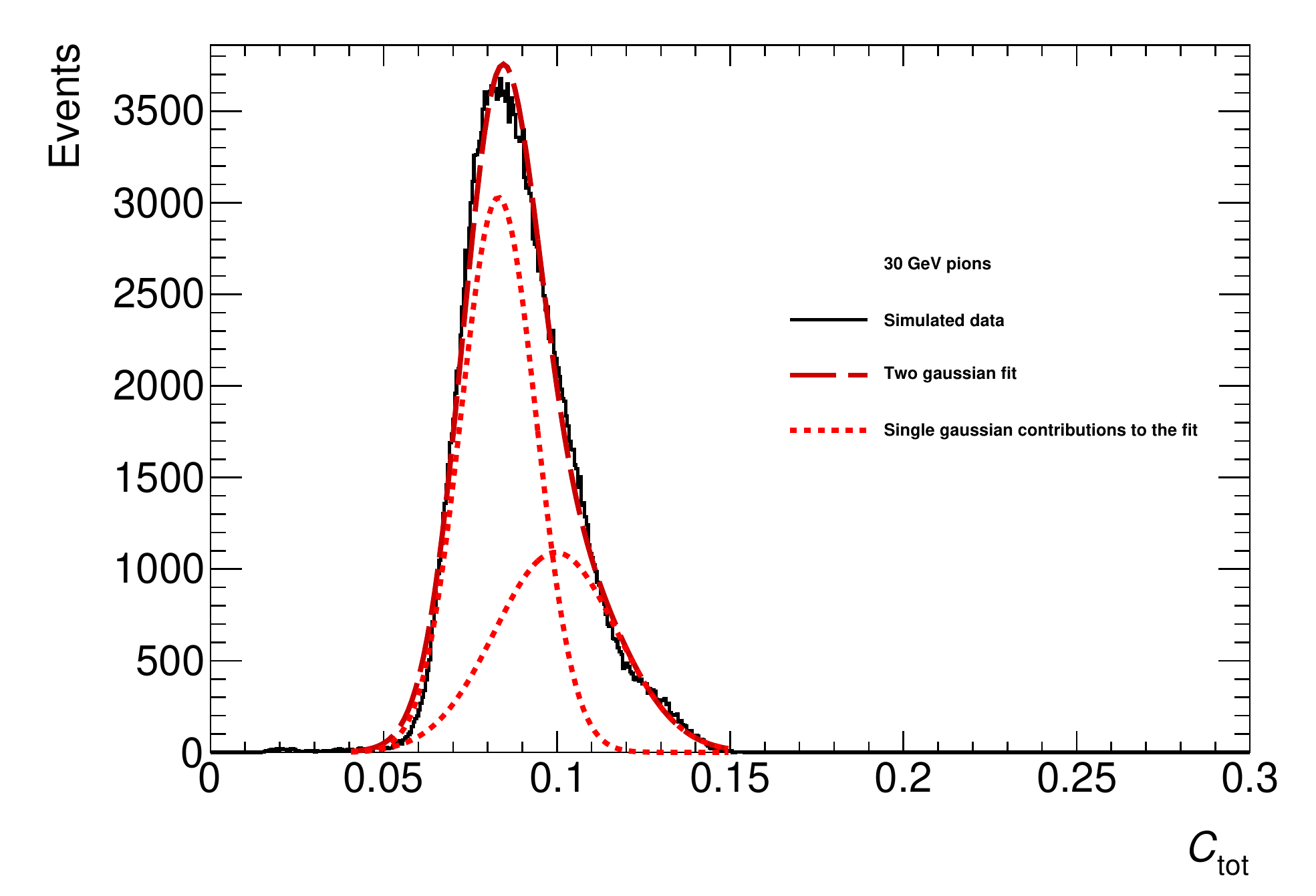}}
\caption{The blue histograms in (a) and (b) show $C_\text{tot}$ distributions of experimental $e/\pi$ sample events with E$_\text{beam}$ equal to 18 GeV and 30 GeV, respectively. Samples of pion events were selected requiring C$_\text{long}$ $<$ 0.6. The black histograms (c) and (d) show the same distributions for simulated pion events selected applying the same selection criteria and with E$_\text{beam}$ equal to 18 GeV and 30 GeV, respectively. Two Gaussian functions fits, obtained using the method of the least squares, are overlapped to the data (red dashed curves). Red dotted curves show the individual Gaussian contributions.}
\label{fig:C_tot 18 30 C_long < 0.6}       % Give a unique label
\end{figure}

\begin{figure}[ht]
\centering
\subfloat[]{\includegraphics[width=4.3cm,clip]{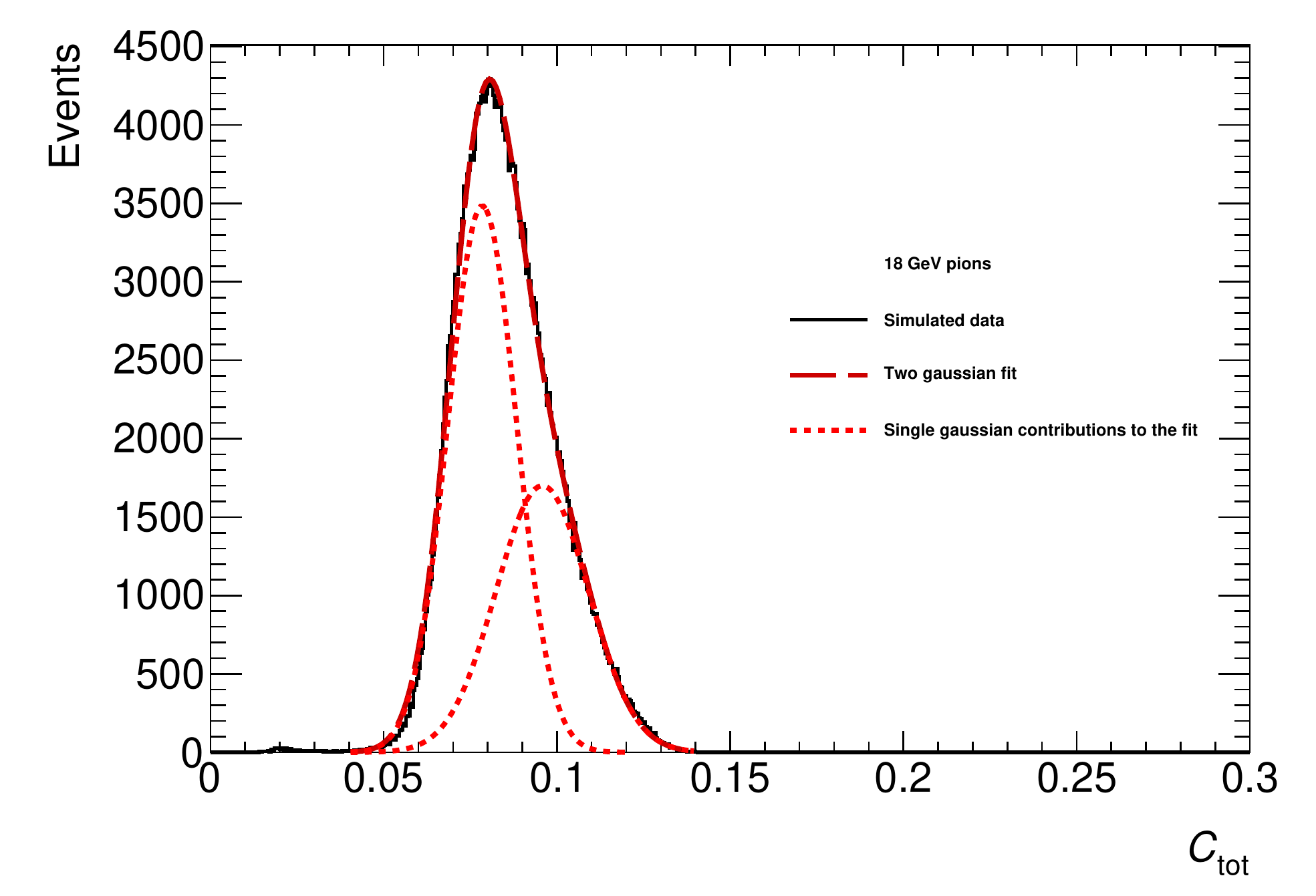}}
\subfloat[]{\includegraphics[width=4.3cm,clip]{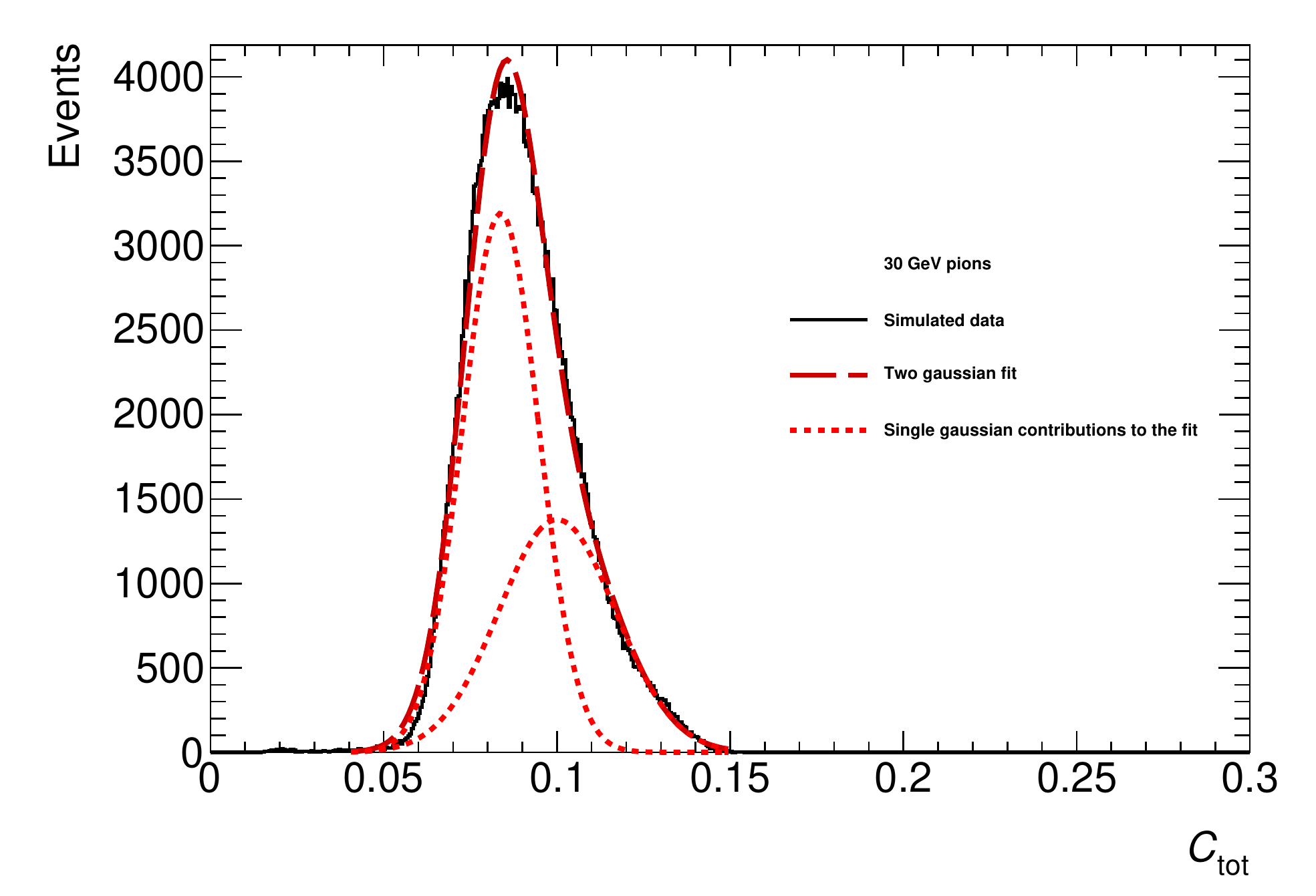}}
\caption{The black histograms show $C_\text{tot}$ distributions obtained using simulated pions with E$_\text{beam}$ equal to 18 GeV (a) and 30 GeV (b). The events were selected requiring C$_\text{long}$ $<$ 0.6. Two Gaussian functions fits, obtained using the method of the least squares, are superimposed on the data (red dashed curve). Red dotted curves show the individual Gaussian contributions. }
\label{fig:pi C_tot 18 30 MC C_long >= 0.6}       % Give a unique label
\end{figure}

\begin{figure}[ht]
\centering
\subfloat[]{\includegraphics[width=4.3cm,clip]{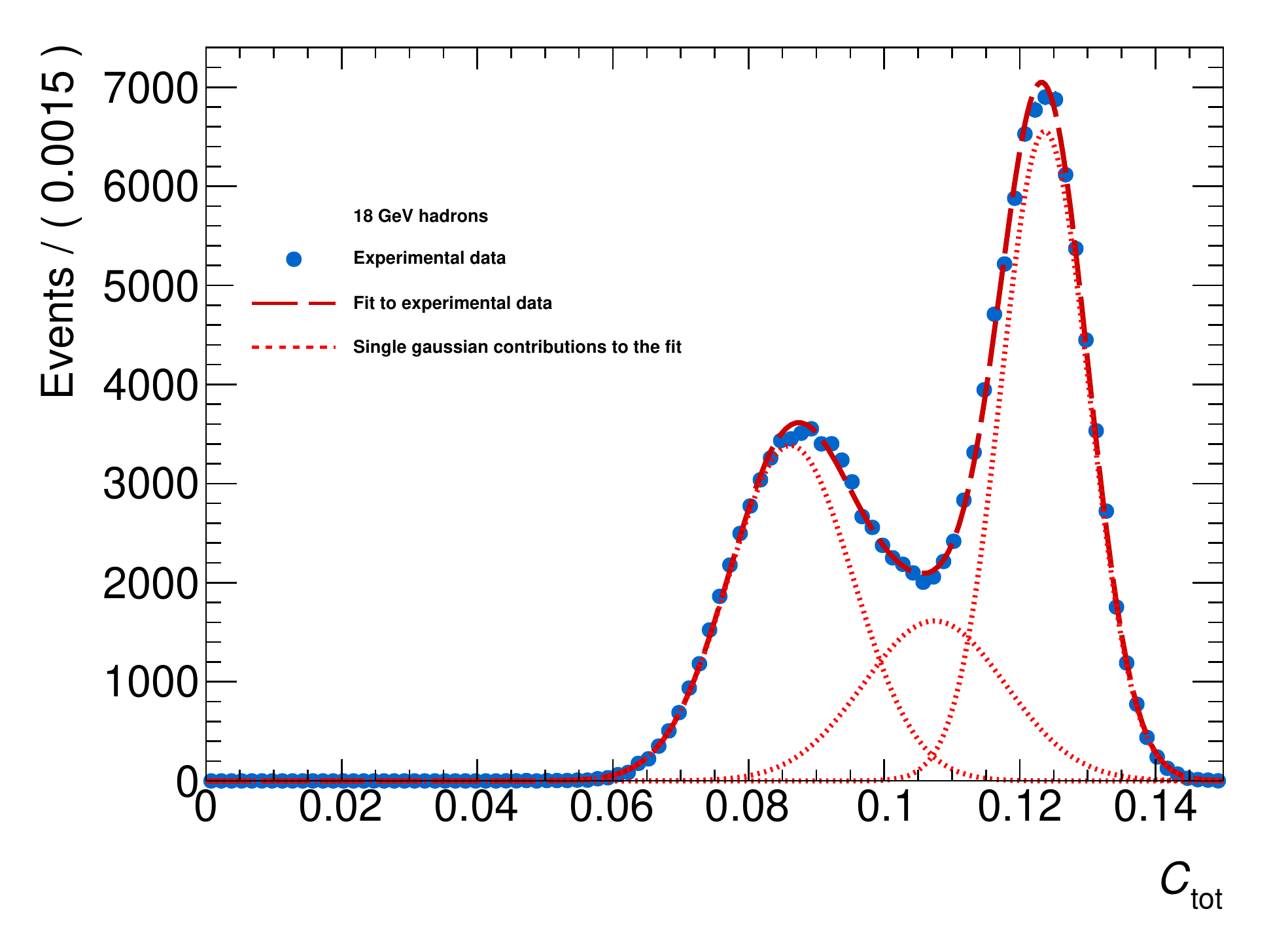}}
\subfloat[]{\includegraphics[width=4.3cm,clip]{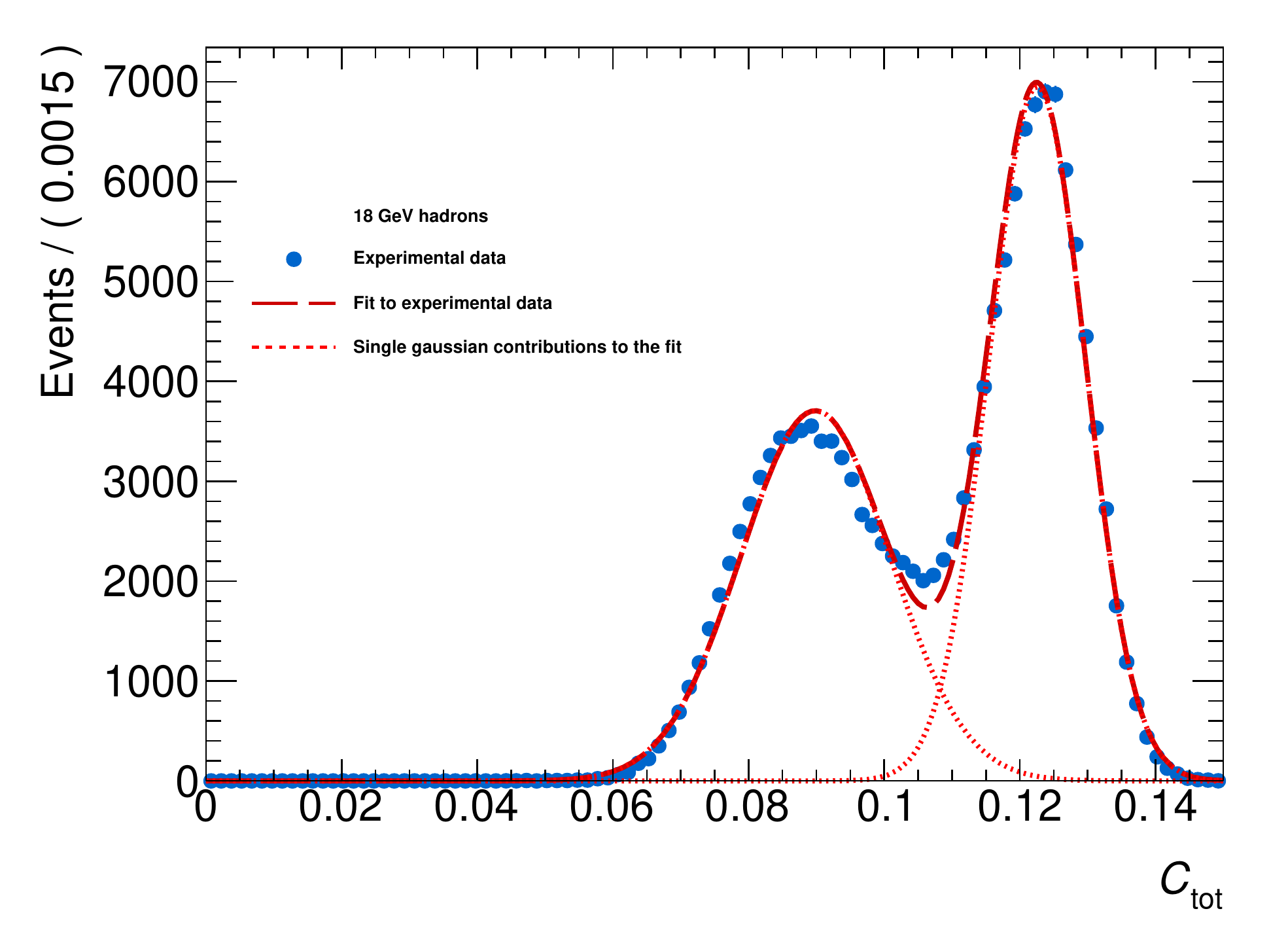}} \\
\subfloat[]{\includegraphics[width=4.3cm,clip]{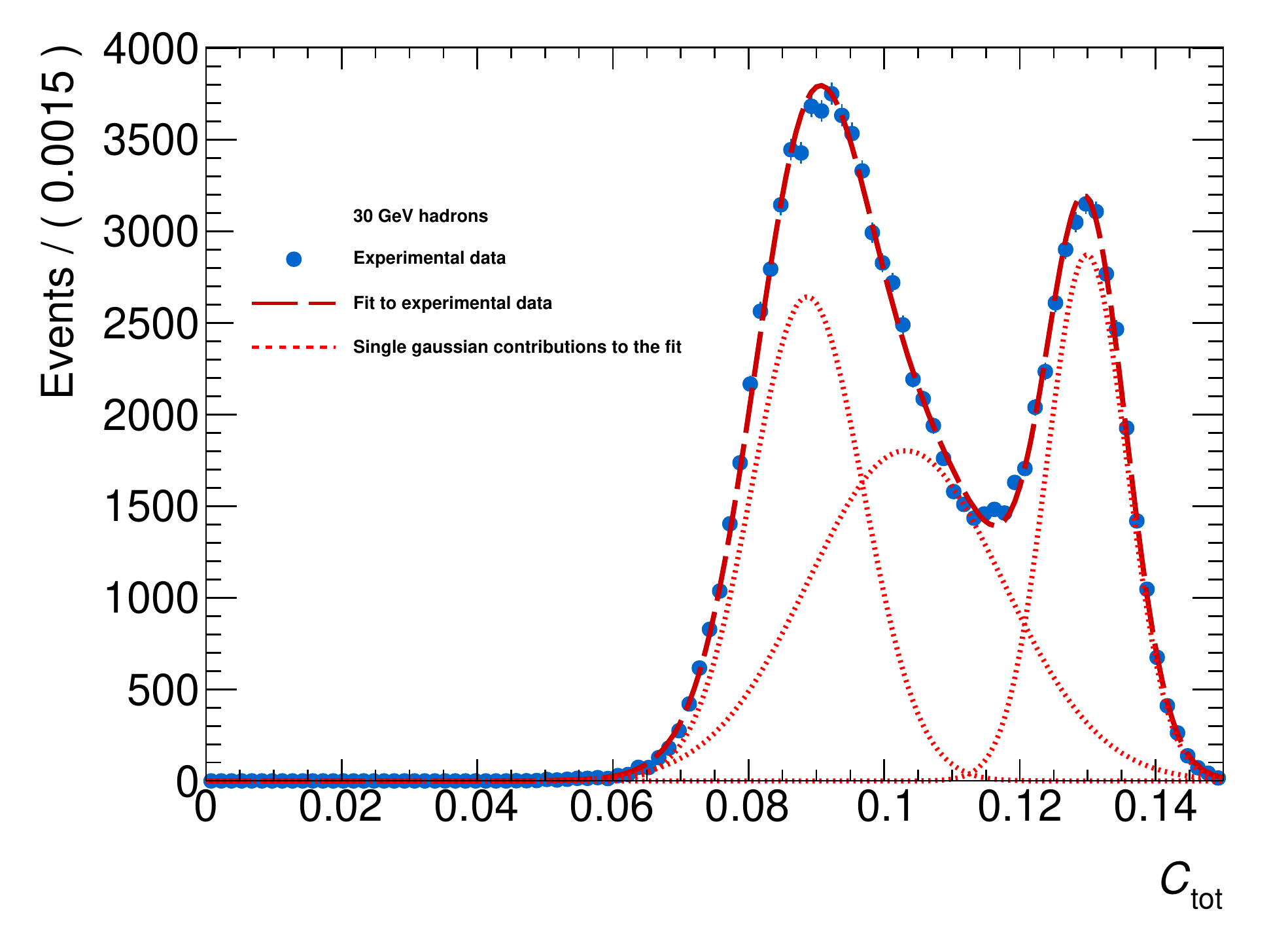}}
\subfloat[]{\includegraphics[width=4.3cm,clip]{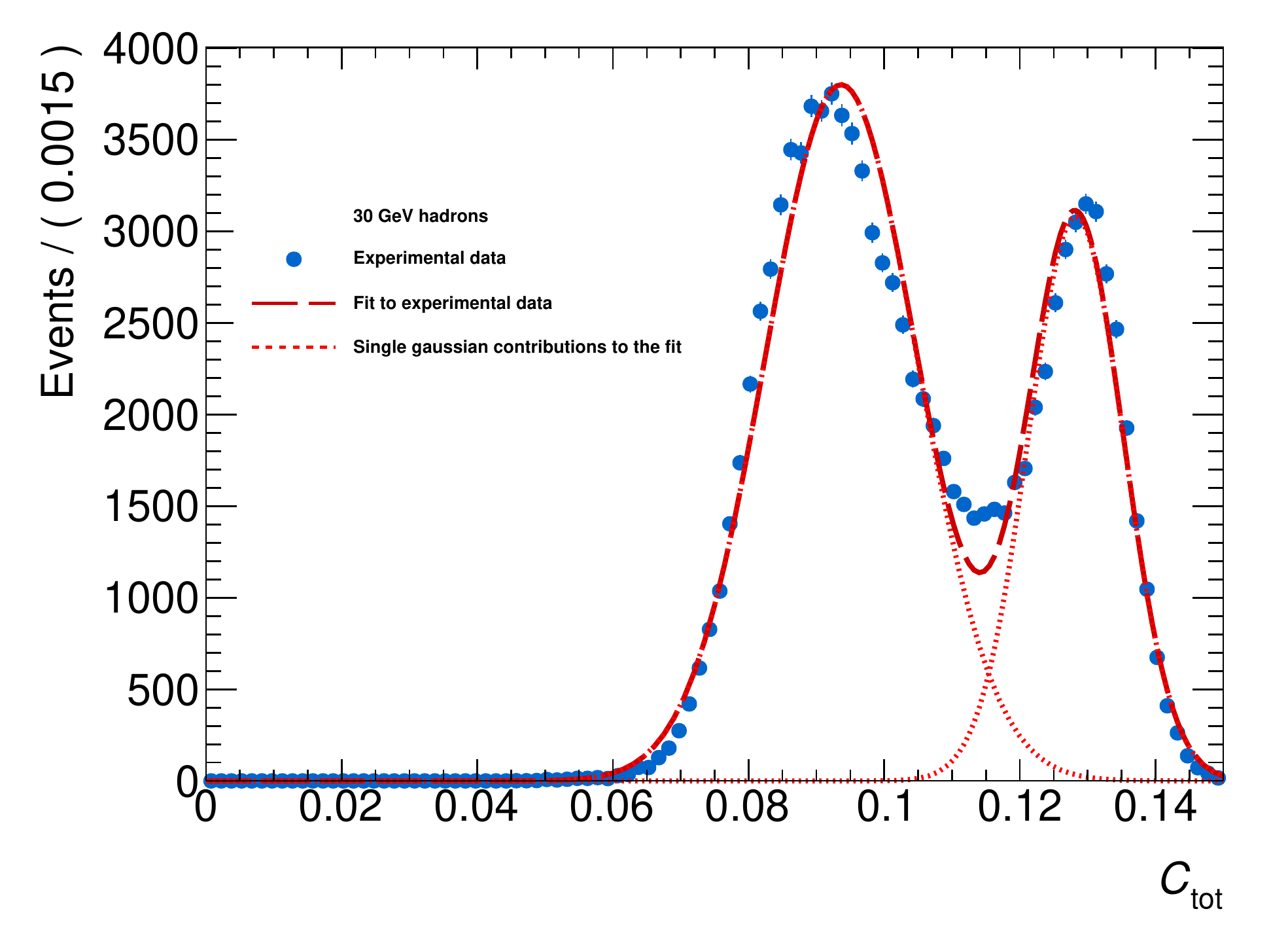}}
\caption{The dotted histograms (a), (b) and (c), (d) represent the $C_\text{tot}$ distributions of $e/\pi$ sample events with E$_\text{beam}$ equal to 18 GeV and 30 GeV respectively. The events were selected requiring C$_\text{long}$ $\ge$ 0.6. Three- and two-Gaussian fit functions (red dashed curves), are superimposed on histograms (a), (c) and (b), (d) respectively. The functions parameters were obtained using the method maximum likelihood. Red dotted curves show the individual Gaussian function contributions. In each histogram, the function with the largest value of the mean $\mu$ describes the electron contamination.}
\label{fig:C_tot 18 30 Exp}       % Give a unique label
\end{figure}

The analysis is based on the fact that electron (pion) C$_\text{tot}$ distributions are well described by one (two) Gaussian function. As an example, Figure~\ref{fig:e C_tot 18 30 MC 20 Exp}~(a) shows the experimental C$_\text{tot}$ distribution obtained using an enriched electron beam with  $E_\text{beam}$ = 20 GeV and \\ C$_\text{long}\ge
 $C$_\text{long}^\text{min} = 0.6$. 
%The events with C$_\text{long} \ge$~C$_\text{long}^\text{min} = 0.6$ were selected. 
The fit was performed in the region C$_\text{tot} \ge$ 1.125. Figures~\ref{fig:e C_tot 18 30 MC 20 Exp}~(b) and~\ref{fig:e C_tot 18 30 MC 20 Exp}~(c) demonstrate that also simulated electron C$_\text{tot}$ distributions at 18 and 30 GeV are well described by one Gaussian function. 
% ----------
%\FloatBarrier
% -----------
%\FloatBarrier
%\newpage
%\pagebreak
Pion $C_\text{tot}$ distributions are best described by two Gaussian functions. The distributions of $e/\pi$ data events with C$_\text{long}$ $<$ C$_\text{long}^\text{min}$ = 0.6 and $E_\text{beam}$ equal to 18 and 30 GeV, respectively, are shown in Figure~\ref{fig:C_tot 18 30 C_long < 0.6}. 
%Enriched pion event samples were selected requiring C$_\text{long}$ <~C$_\text{long}^\text{min}$.
Two Gaussian contributions fit is shown. Individual Gaussian contributions are also presented.
In Figure~\ref{fig:pi C_tot 18 30 MC C_long >= 0.6} $C_\text{tot}$ distributions of simulated pions with E$_\text{beam}$ equal to 18 GeV (a) and 30 GeV (b) are shown. They are also well described by the sum of two Gaussian functions. 
% ----------
% ----------
%\FloatBarrier
% ----------
%\FloatBarrier
% ----------
The number of electrons in the four $e/\pi$ samples were determined considering C$_\text{tot}$ distributions of the events with C$_\text{long} \ge$~C$_\text{long}^\text{min} = 0.6$. Examples of such distributions obtained in the case of events produced by beams of particle with energies equal to 18 GeV and 30 GeV are shown in Figure~\ref{fig:C_tot 18 30 Exp}. Three Gaussian functions were fitted to the experimental 
%C$_\text{tot}$ 
distributions
%of the $e/\pi$ events with C$_\text{long} \ge~C_\text{long}^\text{min}$ 
using the method maximum likelihood. The fit functions are superimposed on the histograms in Figures~\ref{fig:C_tot 18 30 Exp}~(a) and~\ref{fig:C_tot 18 30 Exp}~(c). 
%Fig.~\ref{fig:C_tot 18 30 Exp}. 
The individual Gaussian function contributions are also shown. The functions with the largest mean values $\mu$ describe the electron contributions. The numbers of the electrons reported in Table~\ref{tab:selection chain} are determined from the areas limited by such functions. The statistical uncertainties are equal to the corresponding diagonal terms of the fit error matrices. 
%\FloatBarrier
\subsection{Pion, Kaon and Proton identification}
\label{sec:hadrons_identification}
\begin{figure}[ht]
\centering
\subfloat[]{\includegraphics[width=4.3cm,clip]{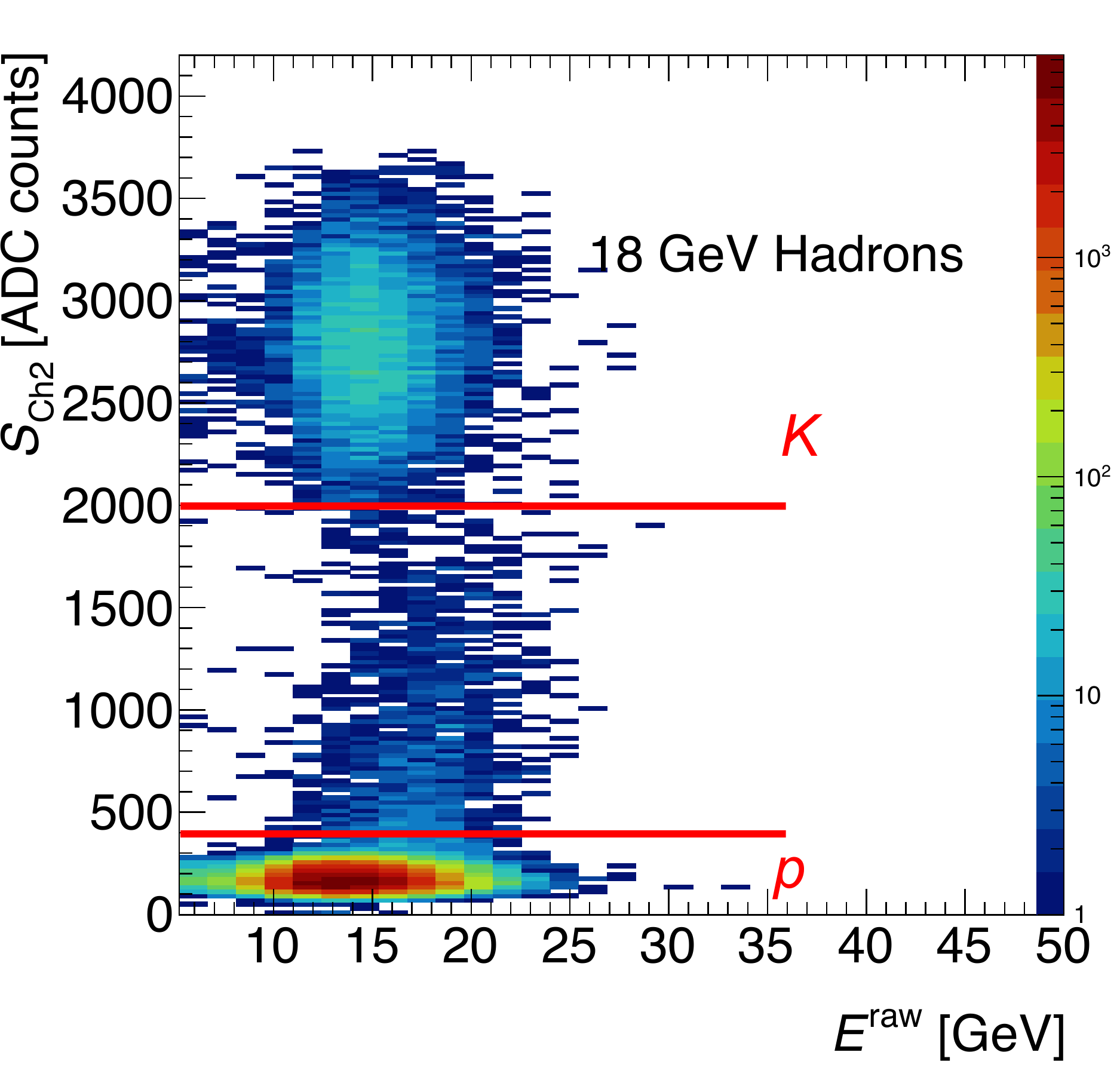}}
\subfloat[]{\includegraphics[width=4.3cm,clip]{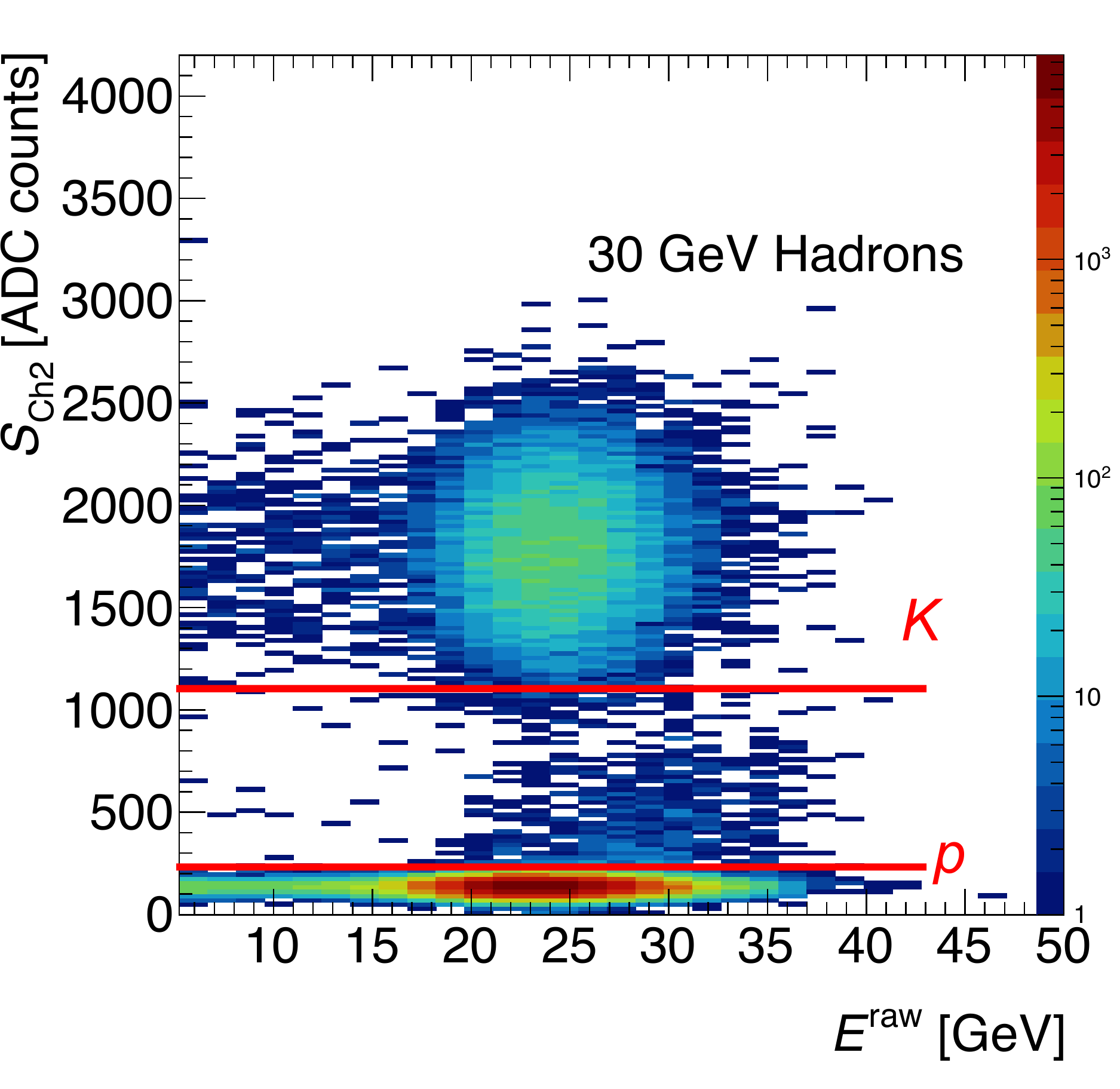}}
\caption{Scatter plot of the Ch2 signals, $S_\text{Ch2}$, in ADC counts units, vs the energy measured in the calorimeter, $E^\text{raw}$ obtained analyzing $K/p$ sample events produced by beams of particles with energy equal to 18 (a) and 30 (b) GeV. The cut values applied in the analysis to select kaon and proton events are shown. Colors are used in the plots to show the cell contents.}
\label{fig: Ch2 K p}       % Give a unique label
\end{figure}

The third set of selection criteria was specific to the type of hadronic particles being studied. For each E$_\text{beam}$ data set the number of pions reported in Table~\ref{tab:selection chain} was estimated by subtracting the number of electron events obtained using the method described in Section~\ref{subsubsec:electron_identification} from the number of events of the corresponding $e/\pi$ sample. 
%The numbers of pions are reported in Table~\ref{tab:selection chain}. 
The Ch2 signal measurements allow a separation of kaons and protons in the $K/p$ samples.
%The cases of samples produced by particle beam with $E_\text{beam}$ equal to 18 and 30 GeV are shown in Fig.~\ref{fig: Ch2 K p}.
The scatter plots of the Ch2 signals, $S_\text{Ch2}$, in ADC counts units vs the energy measured in the calorimeter, $E^\text{raw}$, obtained by analyzing data produced by beams of particles with energies equal to 18 and 30 GeV, are shown in Figure~\ref{fig: Ch2 K p}.
The $S_\text{Ch2}$ selection values in ADC count units are reported in Table~\ref{tab:Cs cuts}. The obtained numbers of kaons and protons are reported in Table~\ref{tab:selection chain}.
% ----------
%\FloatBarrier
% ----------
\subsection{Reconstruction of the energy deposited in the modules}
%\subsection{Shower energy reconstruction}
\label{subsec:shower_energy}

As already discussed in Section~\ref{subsec:detector}, the energy $E^\text{raw}$ deposited by incident particles in the detector was obtained as the sum of the energy measured in the calorimeter cells. In this study only cells with $\mid E_\text{c}^\text{raw}\mid$ $>$ 2$\sigma_\text{noise}$ were considered in the sum. For each run, the cell electronics noise $\sigma_\text{noise}$ was determined using random events collected between beam bursts (Pedestal Triggers). Typical noise values are of the order of 30 MeV. \\ 
No corrections for dead material, containment and non-compensation effects were applied.

\begin{figure}[ht]
\centering
\subfloat[]{\includegraphics[width=4.3cm,clip]{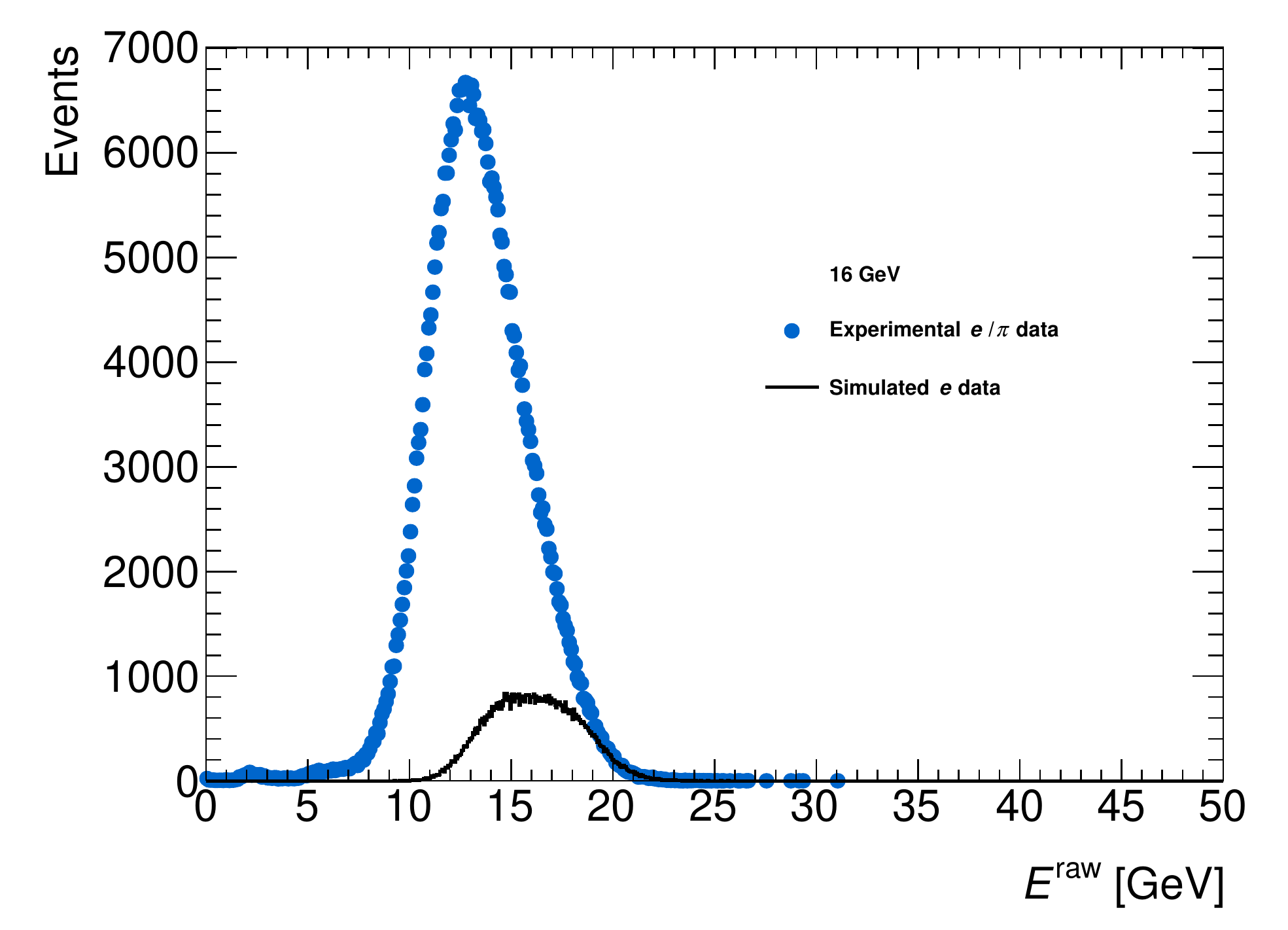}}
\subfloat[]{\includegraphics[width=4.3cm,clip]{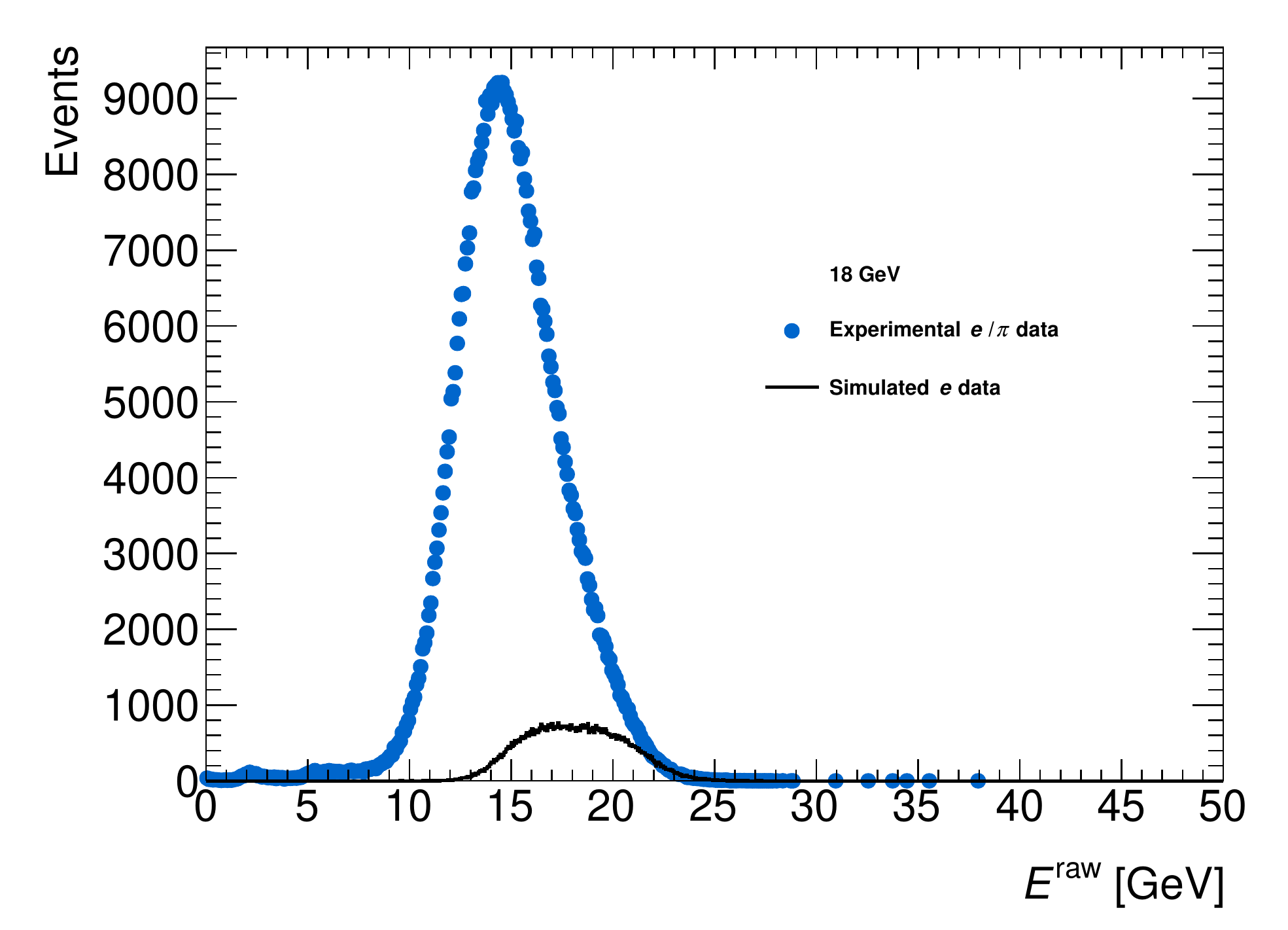}} \\
\subfloat[]{\includegraphics[width=4.3cm,clip]{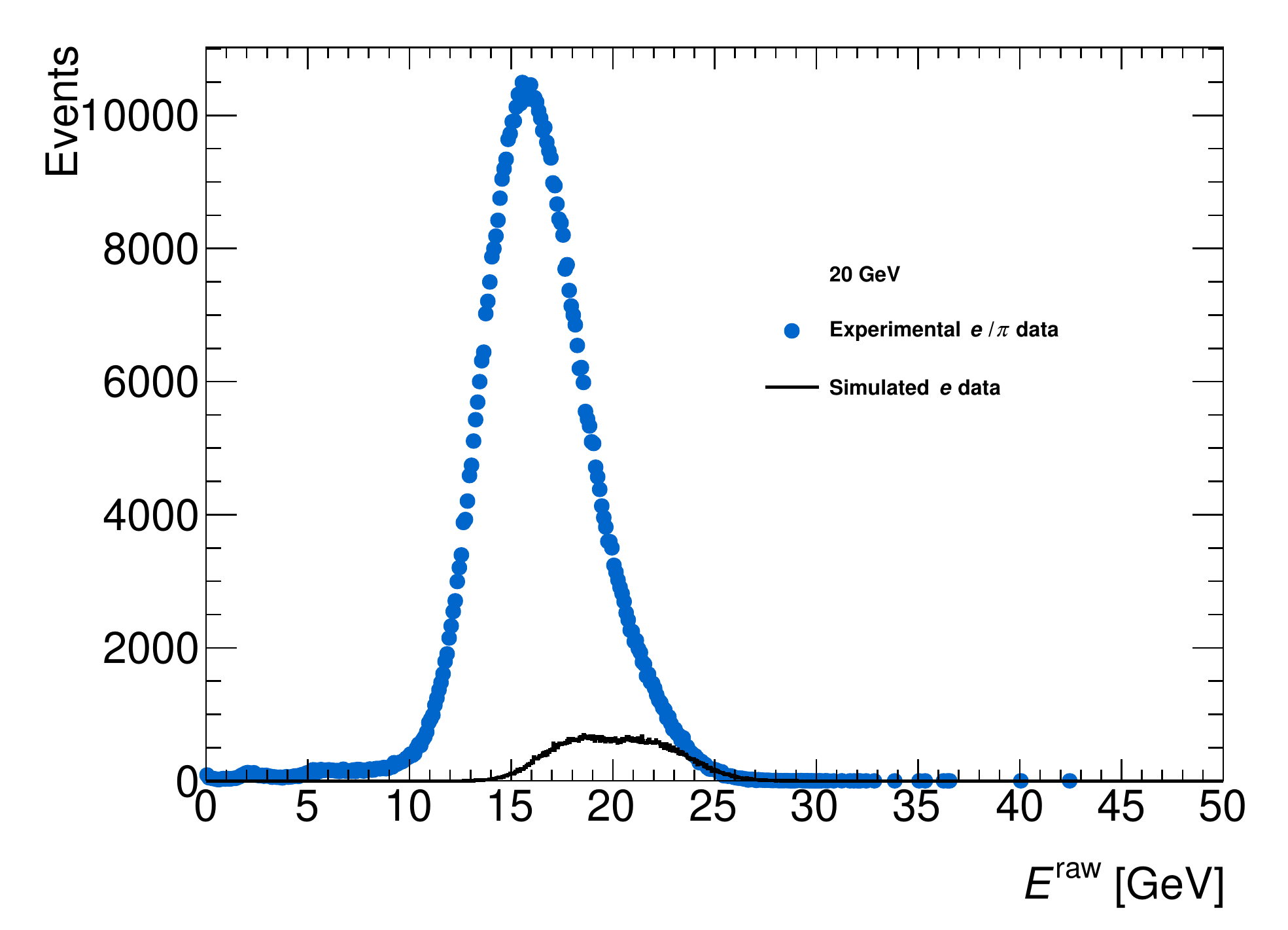}}
\subfloat[]{\includegraphics[width=4.3cm,clip]{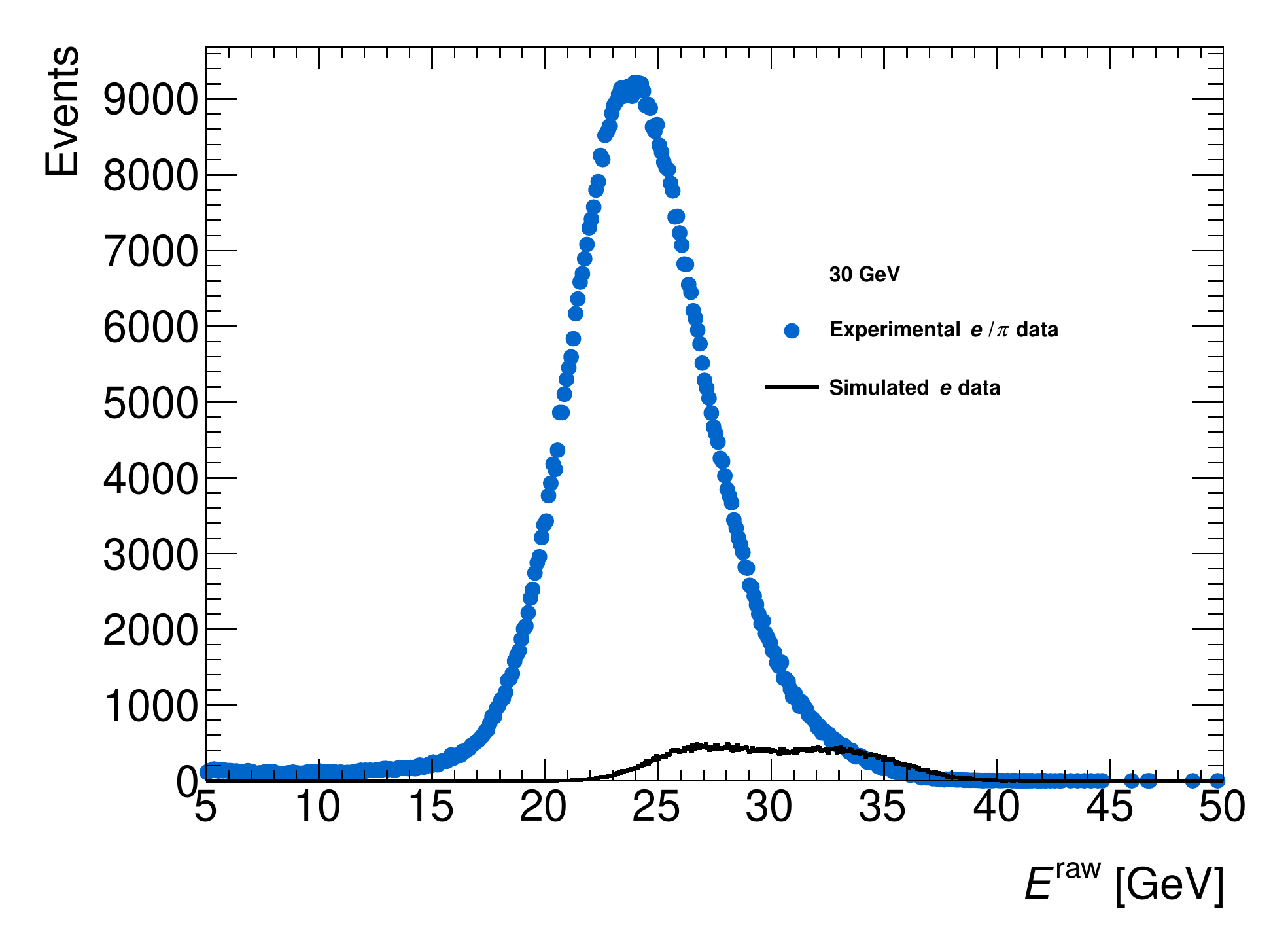}}
\caption{Distributions of the reconstructed energy $E^\text{raw}$ of the $e/\pi$ samples events with $E_\text{beam}$ equal to 16 GeV (a), 18 GeV (b) 20 GeV (c) and 30 GeV (d). The blue dotted histograms correspond to the experimental data. %Only statistical error are shown. 
The black histograms correspond to the expected distributions of the electrons contaminating the samples obtained using simulated events. The normalization procedure is described in the text.}
\label{fig: Electron subtraction}       % Give a unique label
\end{figure}
\begin{figure}[ht]
\centering
\subfloat[]{\includegraphics[width=4.3cm,clip]{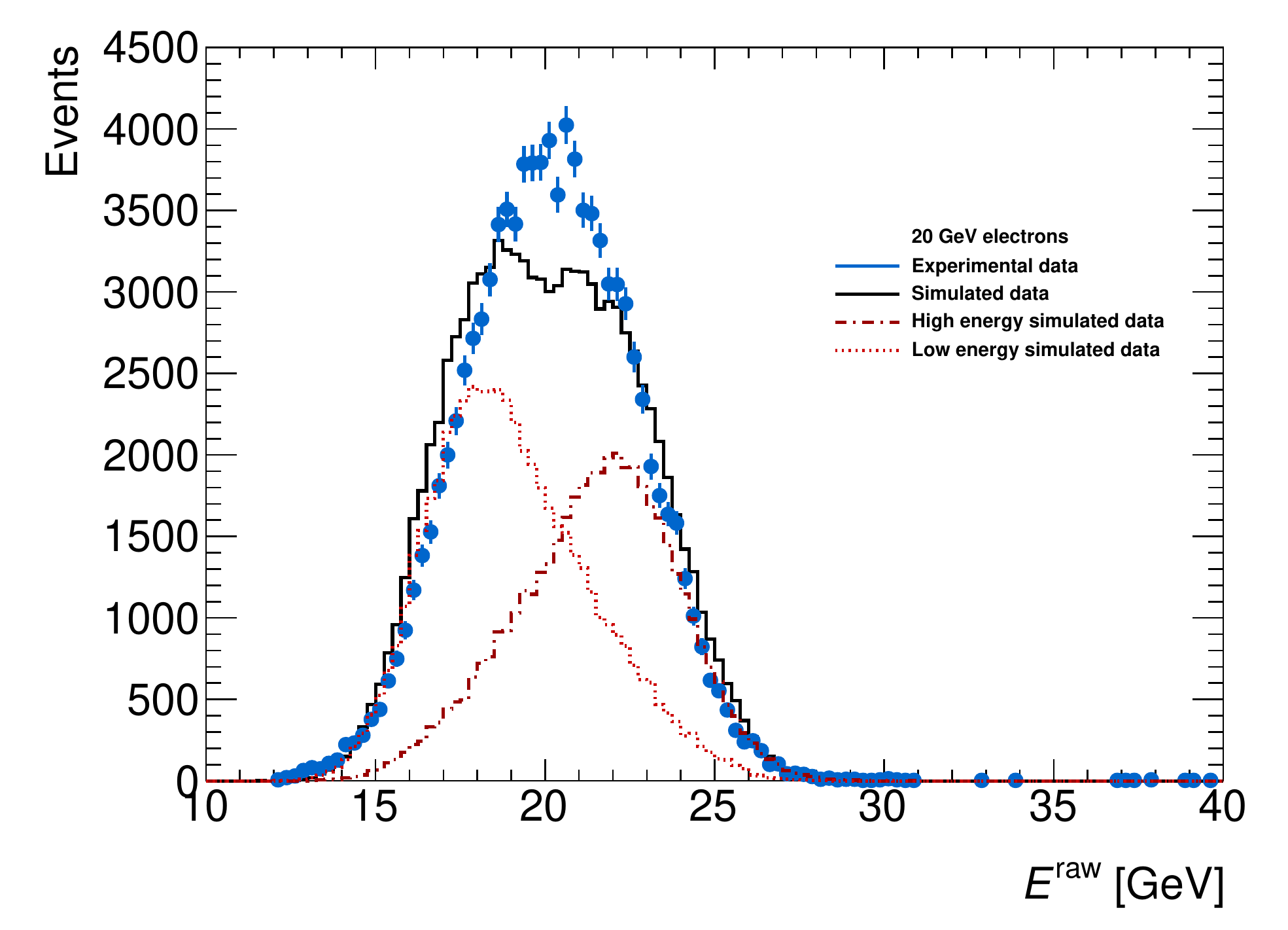}}
\subfloat[]{\includegraphics[width=4.3cm,clip]{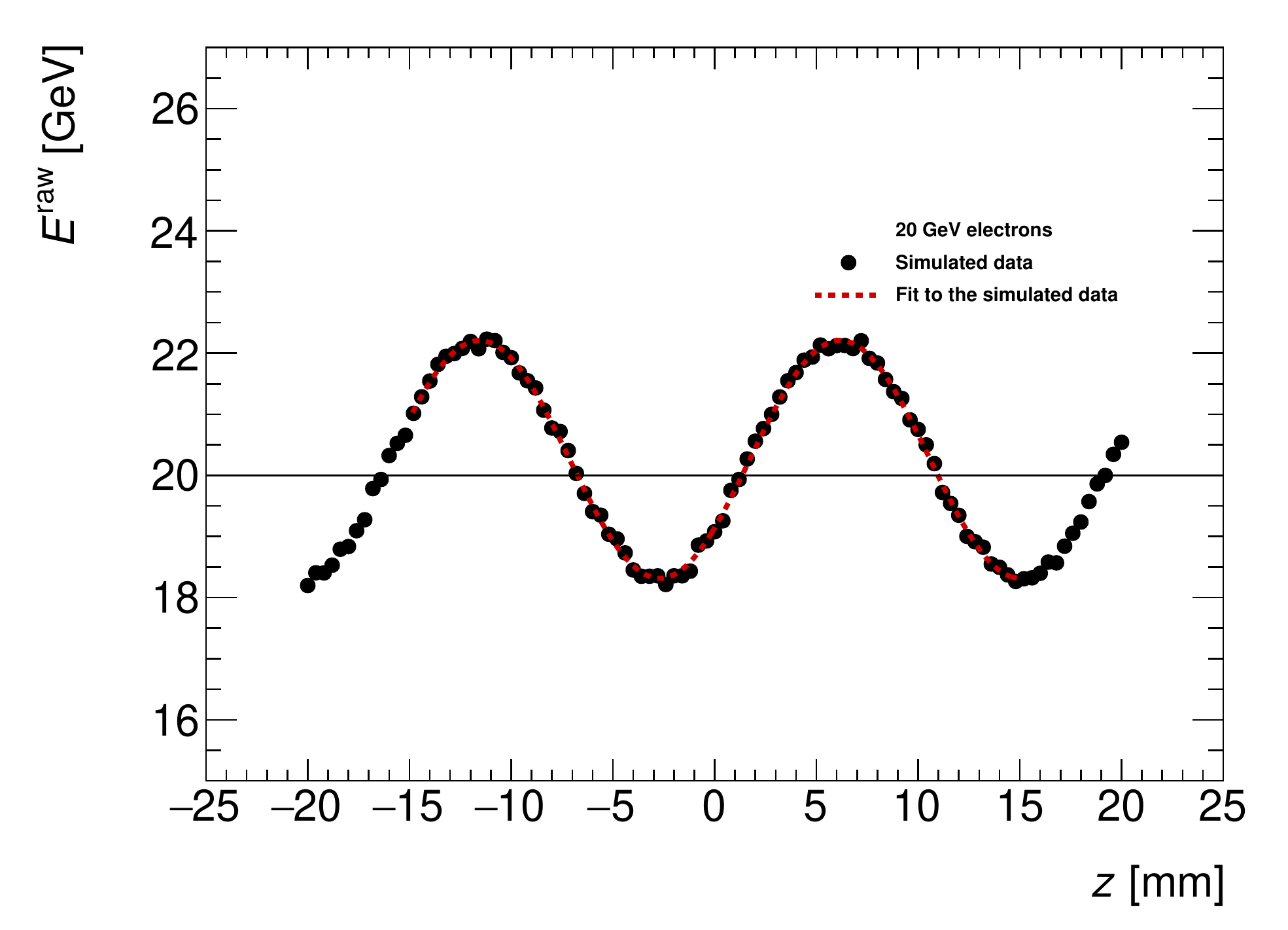}}\\
\subfloat[]{\includegraphics[width=4.3cm,clip]{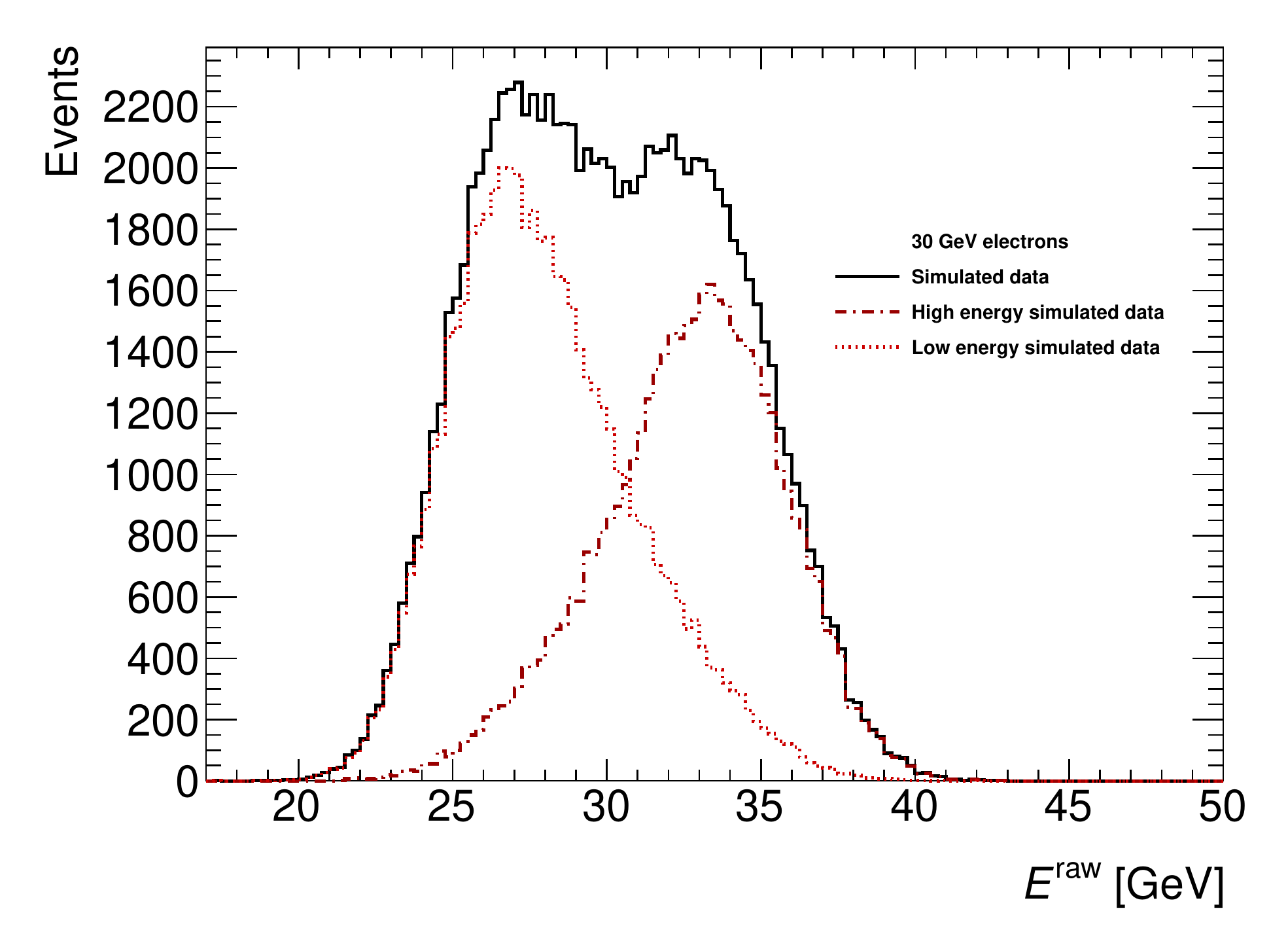}}
\subfloat[]{\includegraphics[width=4.3cm,clip]{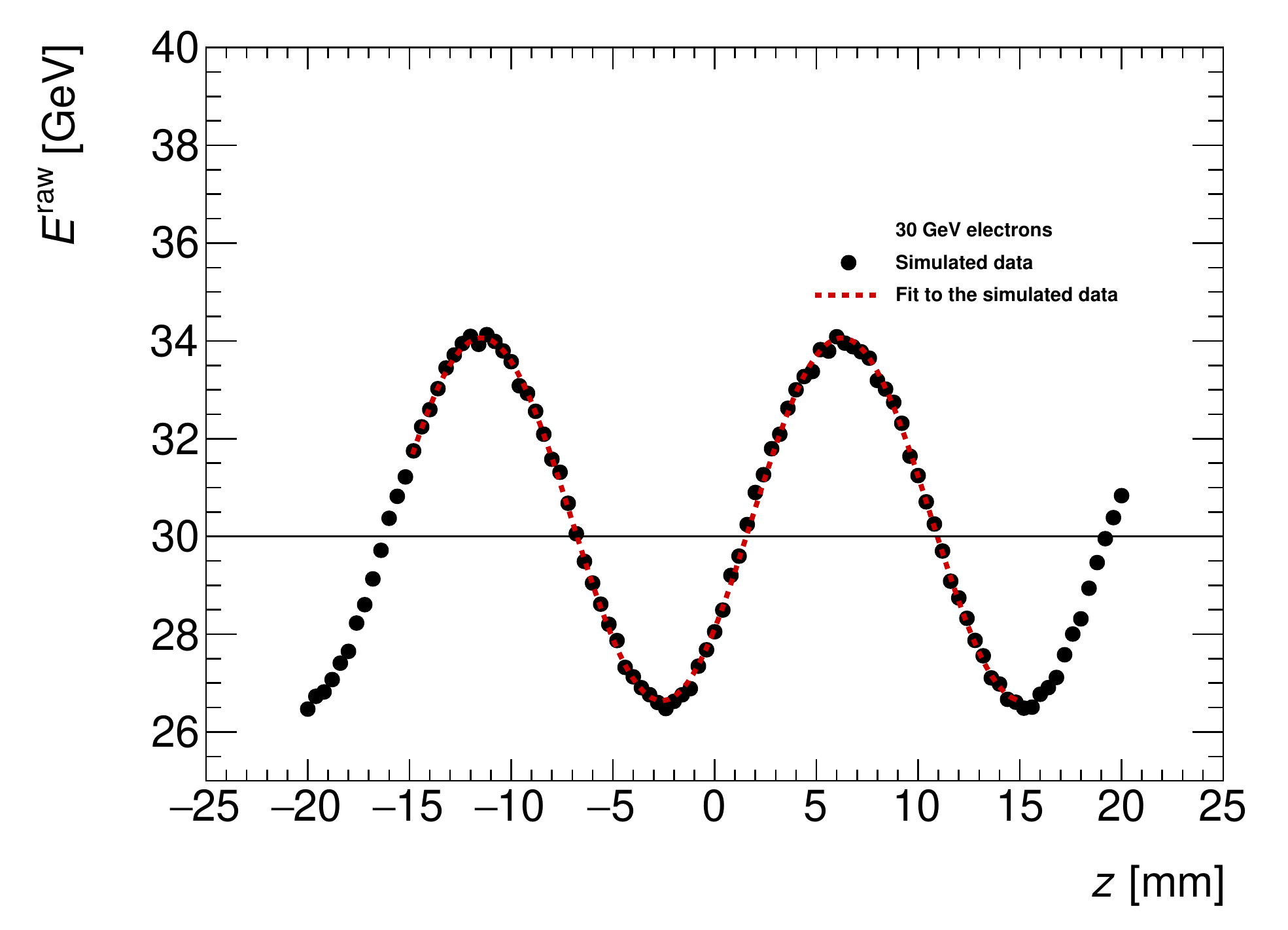}}
\caption{The black histograms in (a) and (c) show the distributions of the reconstructed energy $E^\text{raw}$ obtained analysing simulated data obtained using electron beams with $E_\text{beam}$ equal to 20 GeV and 30 GeV respectively. The blue dot distribution in (a) has been obtained using experimental data. In (a) and (c) are also shown the distributions obtained using “high energy events” (red dashed line) and “low energy events” (red dotted line) discussed in Section~\ref{subsec:hadron_response_ratio}. The histograms (b) and (d) show the oscillation of the electron response due to the sampling fraction variations as obtained using simulated electrons with $E_\text{beam}$ equal to 20 GeV and 30 GeV respectively. The dashed curves in red correspond to the fit of Eq.~\ref{eq:oscillation} to the data. The horizontal black line corresponds to the electron mean energy, $p_0$ (Eq.(\ref{eq:oscillation})).
}
\label{fig: 20 GeV Electron energy distribution}       % Give a unique label
\end{figure}
\begin{figure}[ht]
\centering
\subfloat[]{\includegraphics[width=4.3cm,clip]{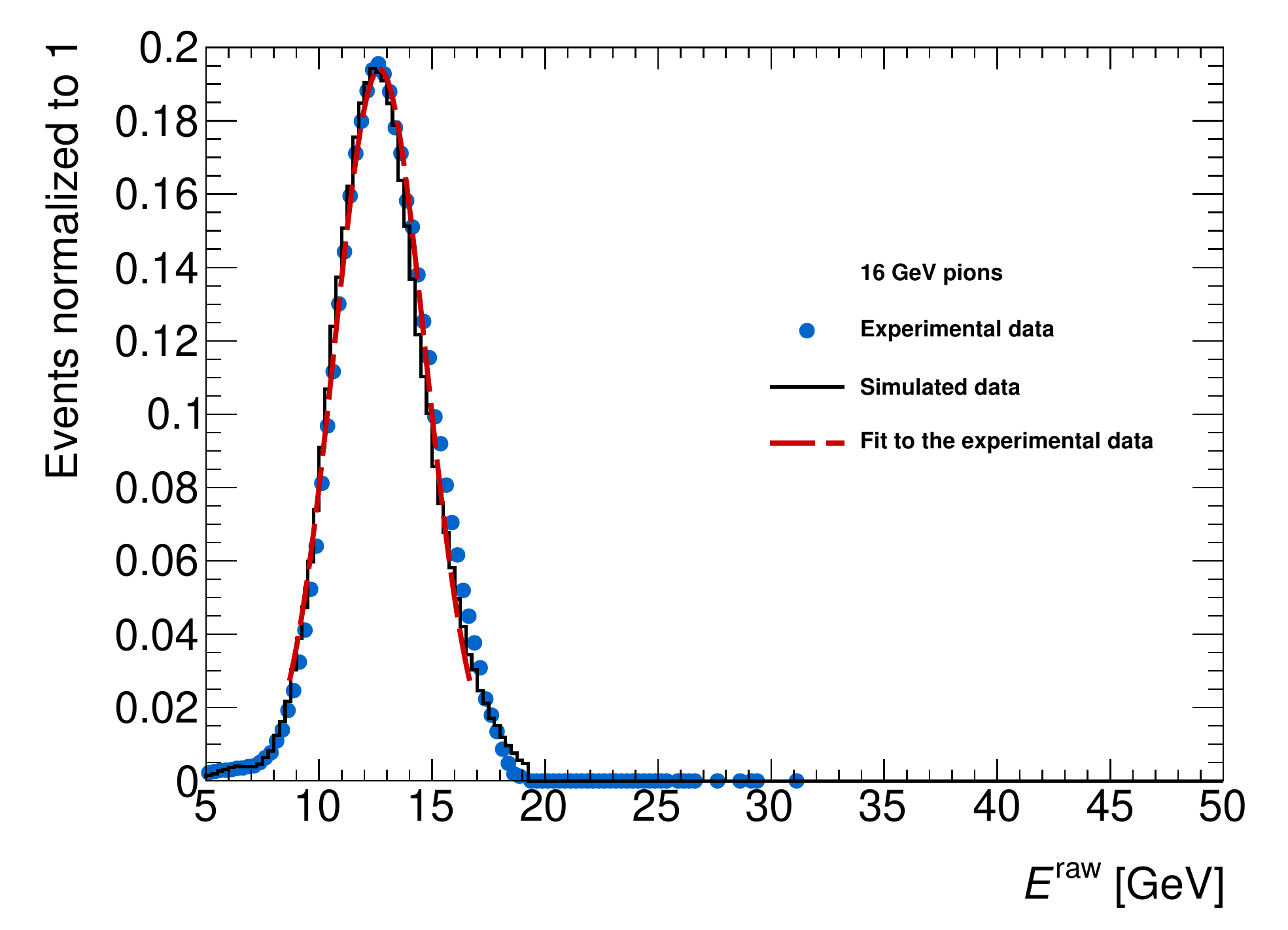}}
\subfloat[]{\includegraphics[width=4.3cm,clip]{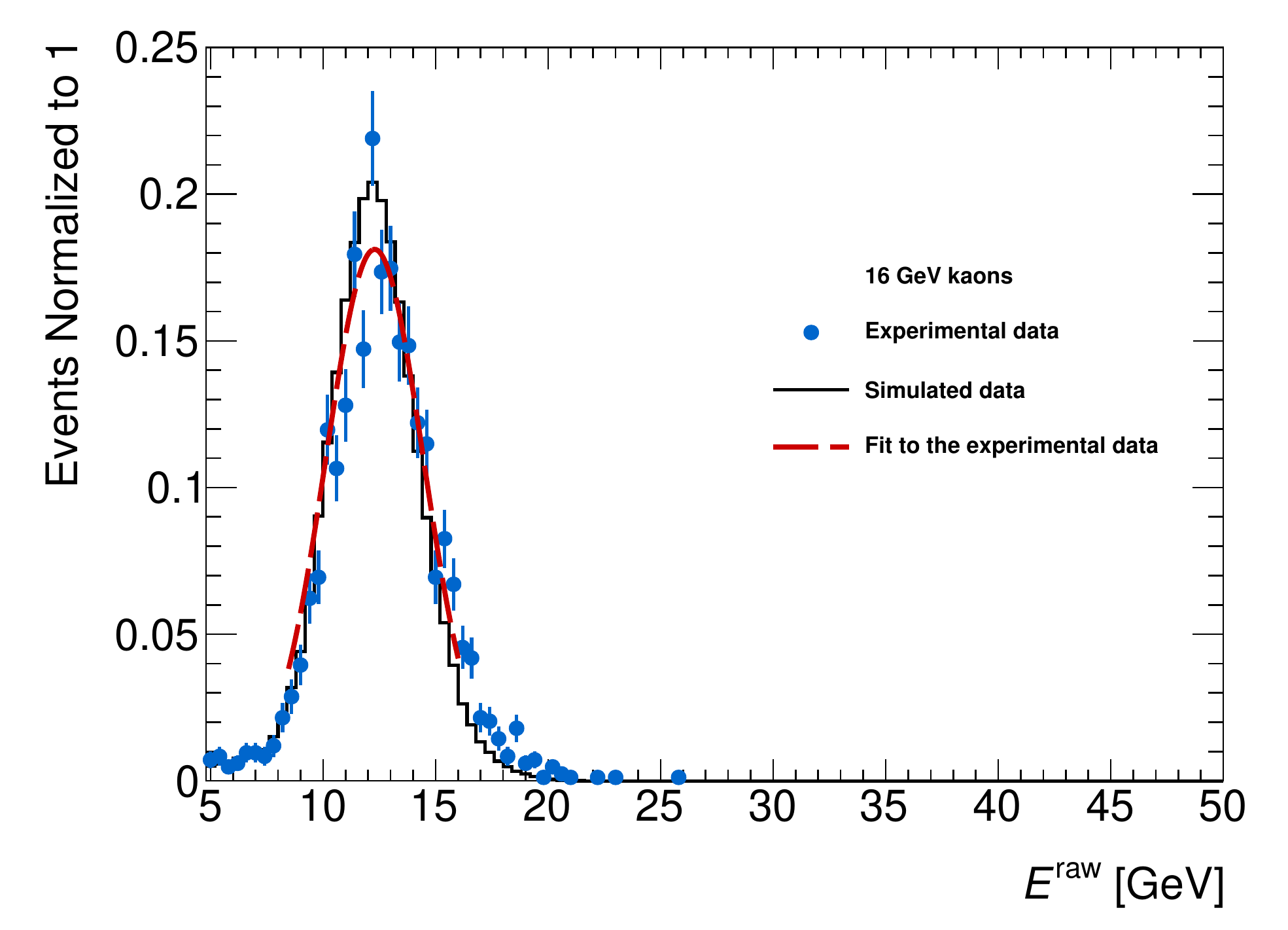}} \\
\subfloat[]{\includegraphics[width=4.3cm,clip]{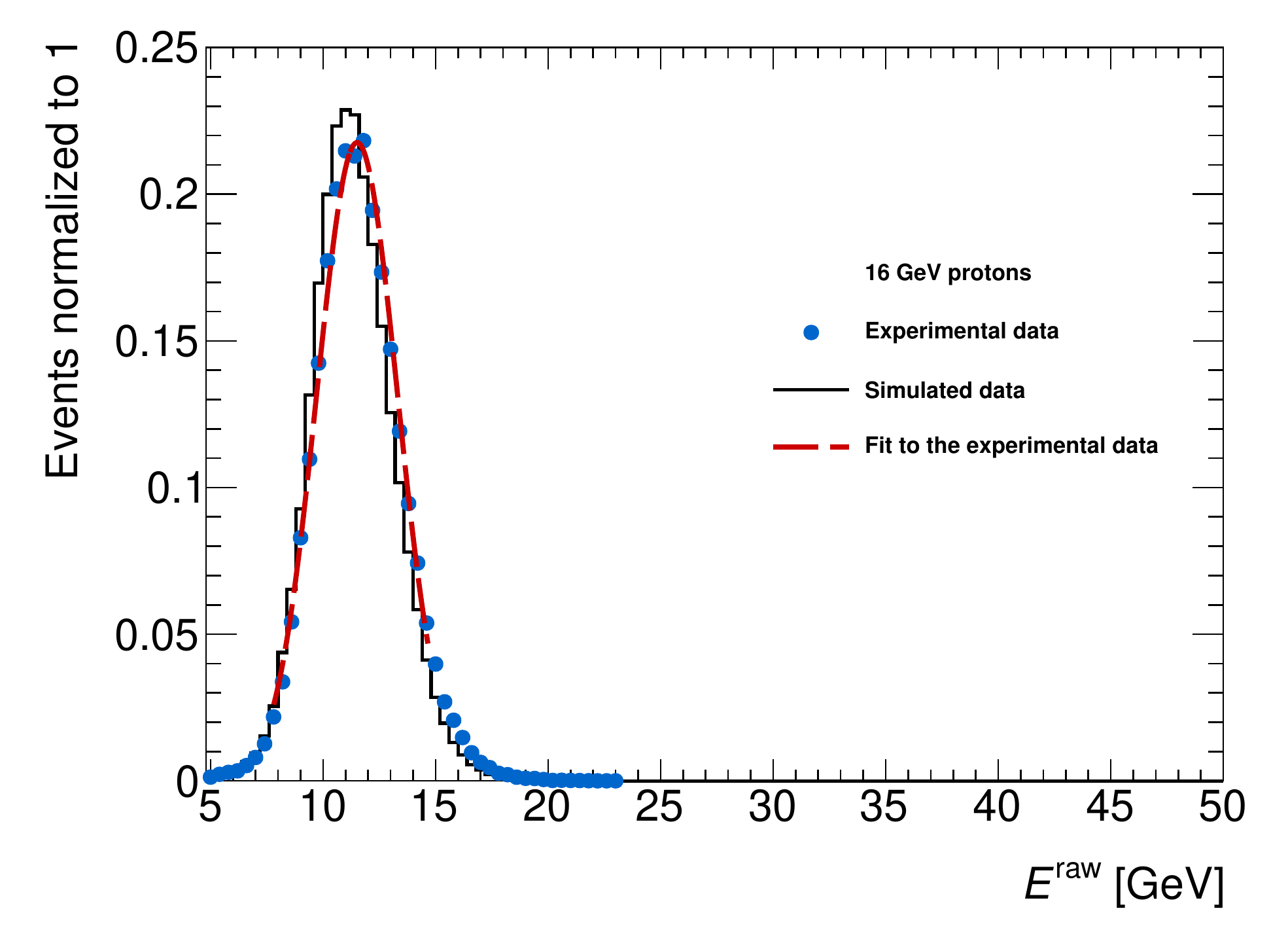}}
\caption{Distributions of the reconstructed energy $E^\text{raw}$ obtained analyzing pion (a), kaon (b) and proton (c) data with $E_\text{beam}$ = 16 GeV. The blue dotted histograms represent the experimental data. Only statistical uncertainties are shown.  The dashed curves in red correspond to the fit of a Gaussian function to the experimental data in a region ±2$\sigma$ around the peak value. The black histograms correspond to the predictions of the MC simulation. 
%\textcolor {red} {Take away the results}
}
\label{fig: Eraw_16_GeV}       % Give a unique label
\end{figure}
\begin{figure}[ht]
\centering
\subfloat[]{\includegraphics[width=4.3cm,clip]{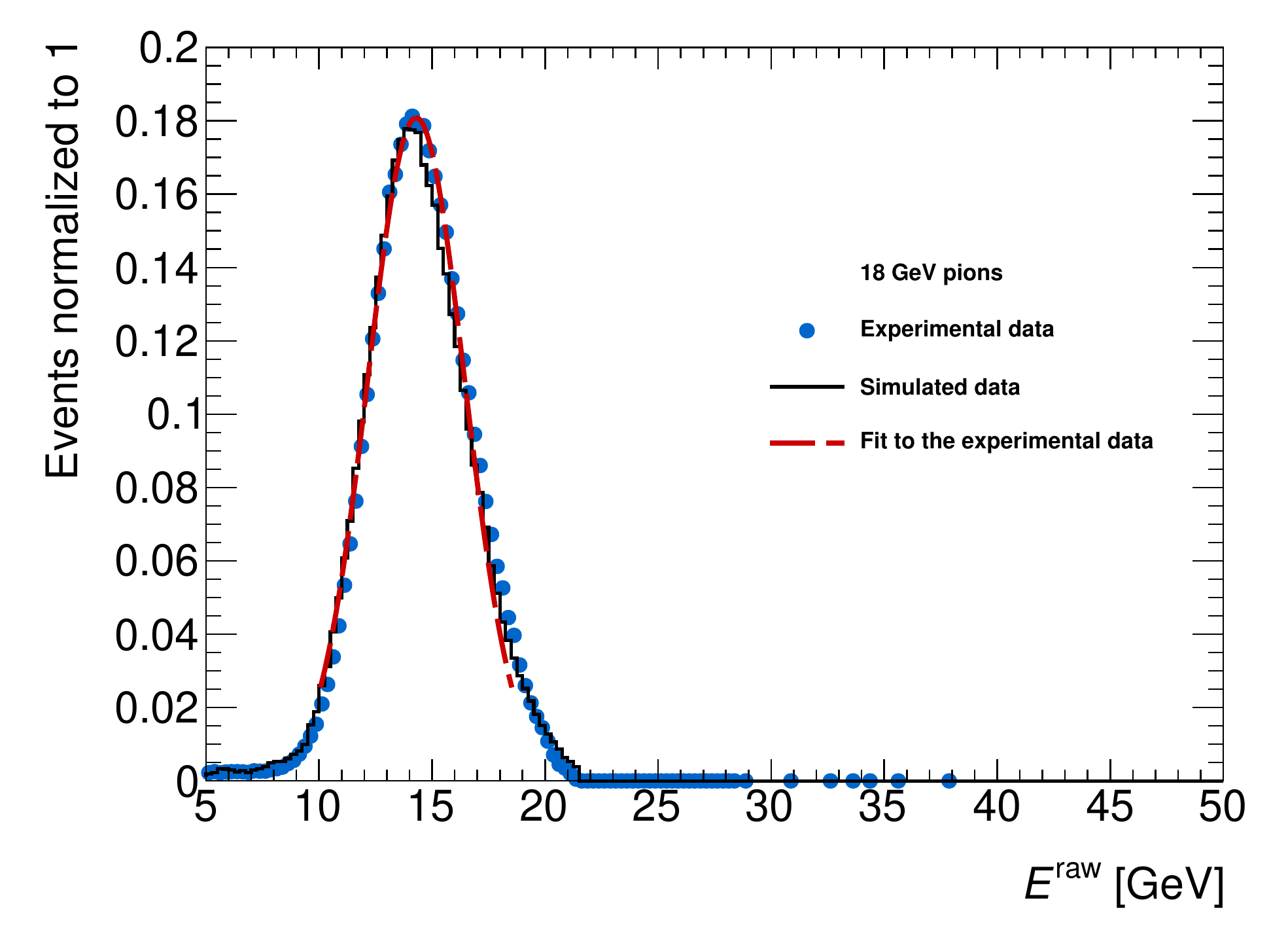}}
\subfloat[]{\includegraphics[width=4.3cm,clip]{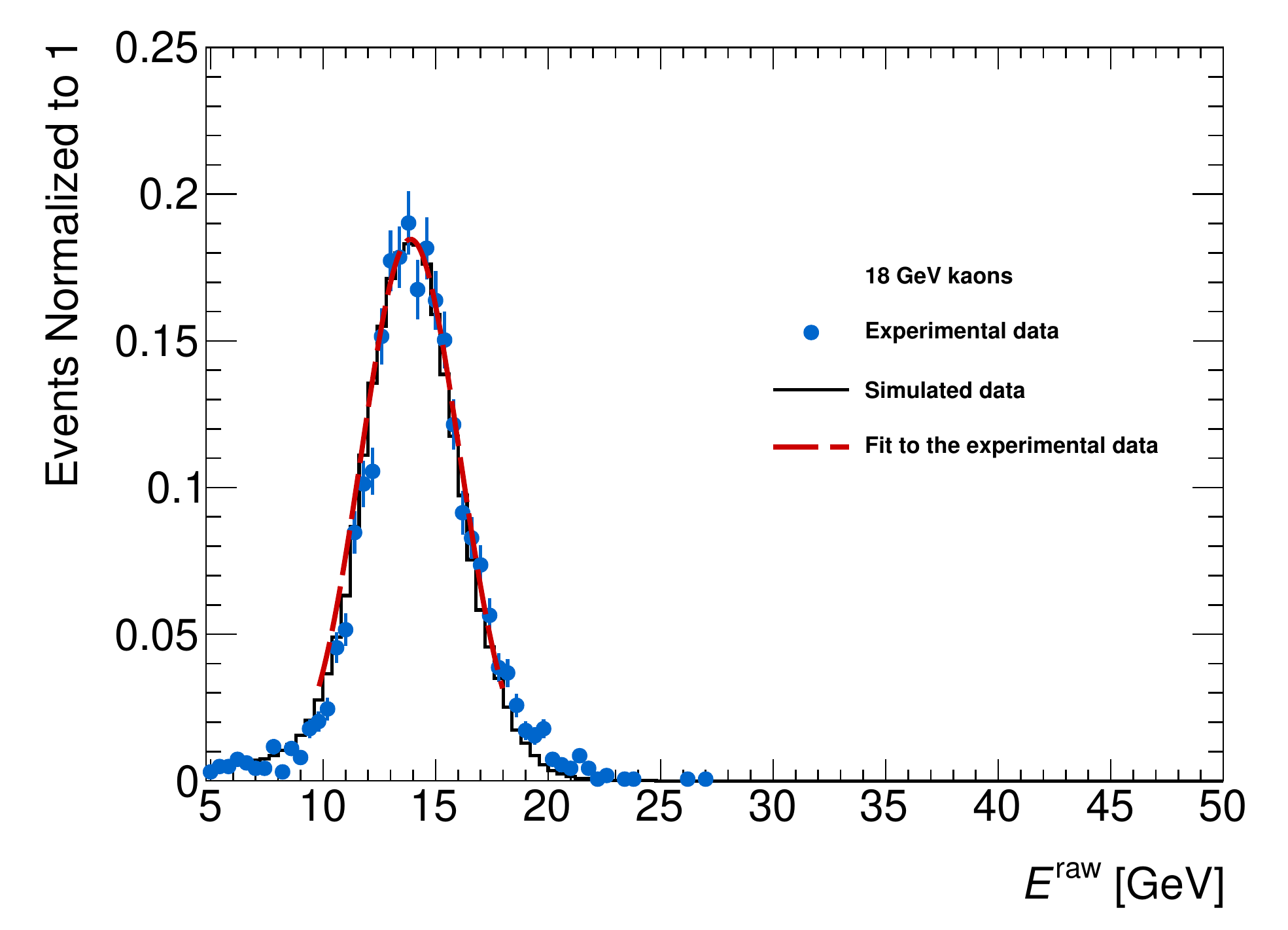}} \\
\subfloat[]{\includegraphics[width=4.3cm,clip]{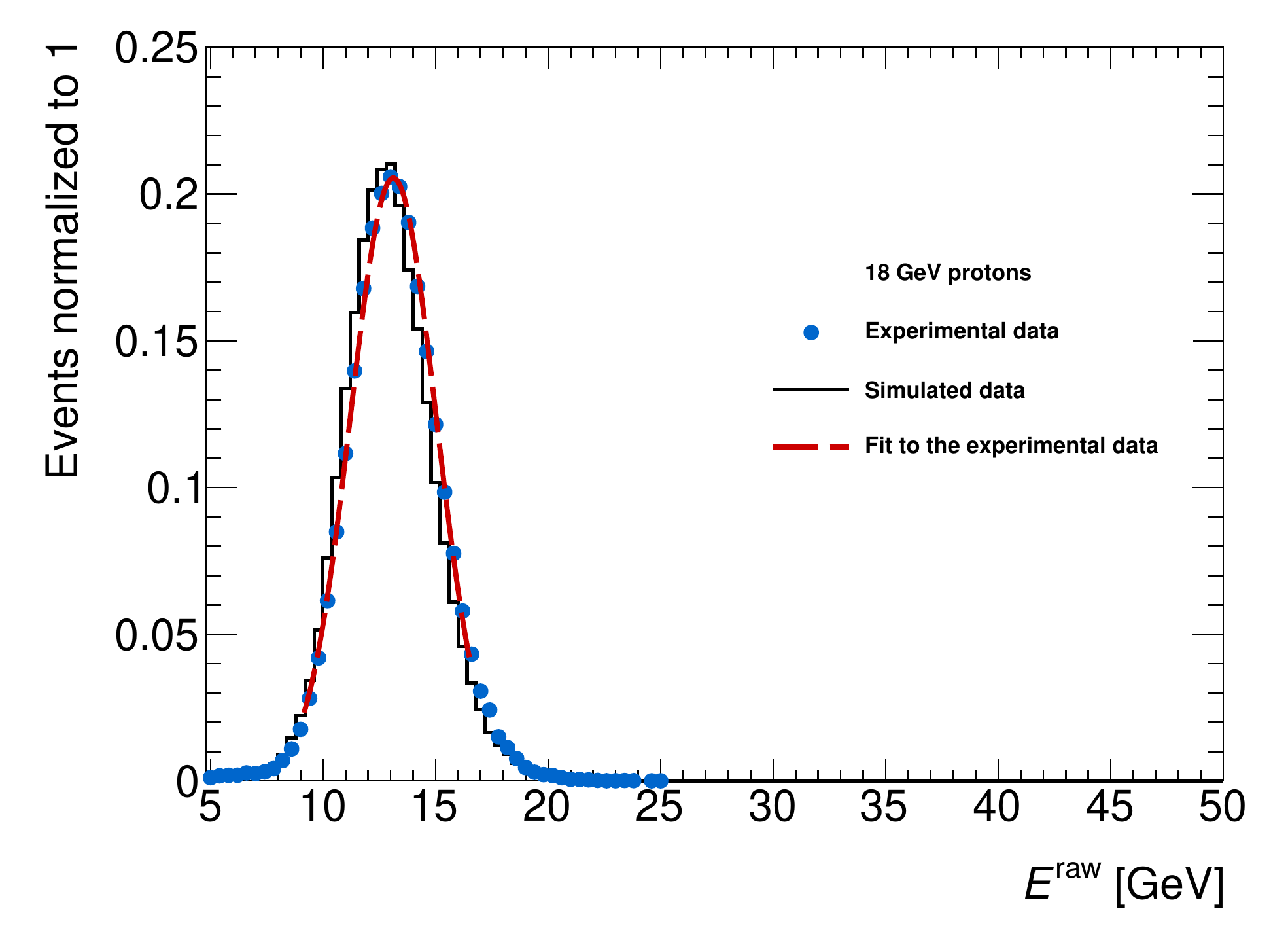}}
\caption{Distributions of the reconstructed energy $E^\text{raw}$ obtained analyzing pion (a), kaon (b) and proton (c) data with $E_\text{beam}$ = 18 GeV. The blue dotted histograms represent the experimental data. Only statistical uncertainties are shown.  The dashed curves in red correspond to the fit of a Gaussian function to the experimental data in a region ±2$\sigma$ around the peak value. The black histograms correspond to the predictions of the MC simulation.  
%\textcolor {red} {Take away the results}
}
\label{fig: Eraw_18_GeV}       % Give a unique label
\end{figure}
\begin{figure}[ht]
\centering
\subfloat[]{\includegraphics[width=4.3cm,clip]{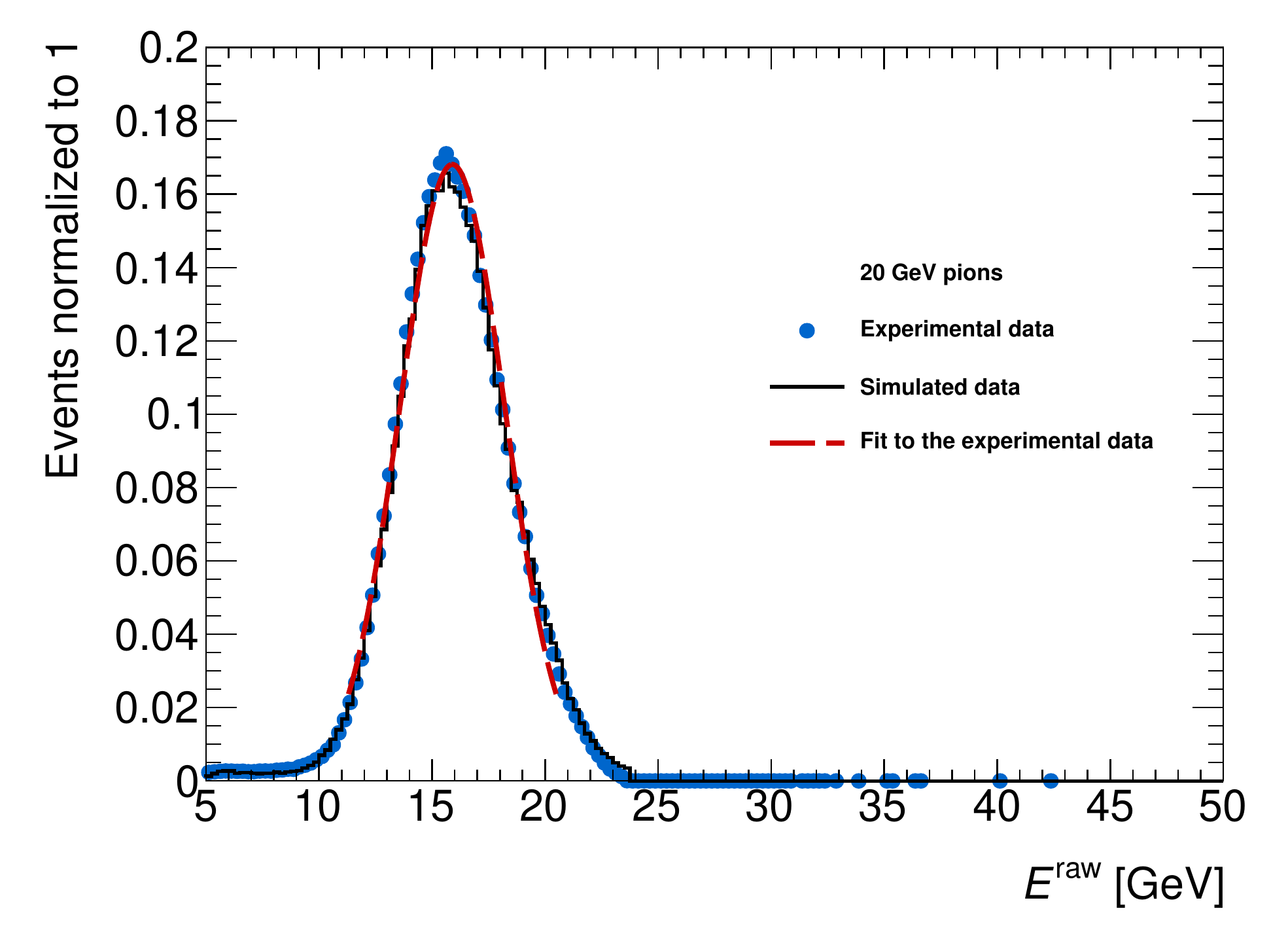}}
\subfloat[]{\includegraphics[width=4.3cm,clip]{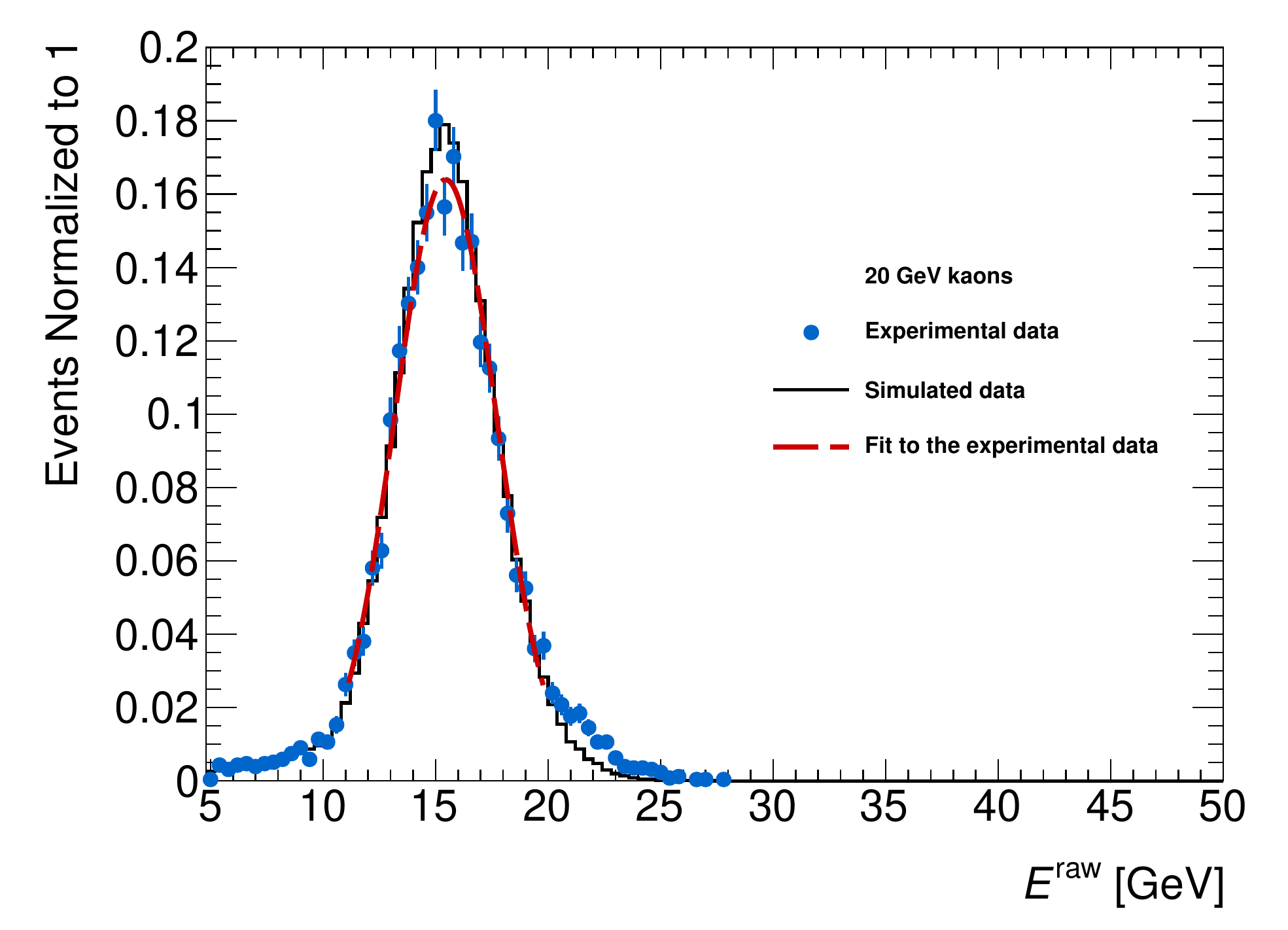}} \\
\subfloat[]{\includegraphics[width=4.3cm,clip]{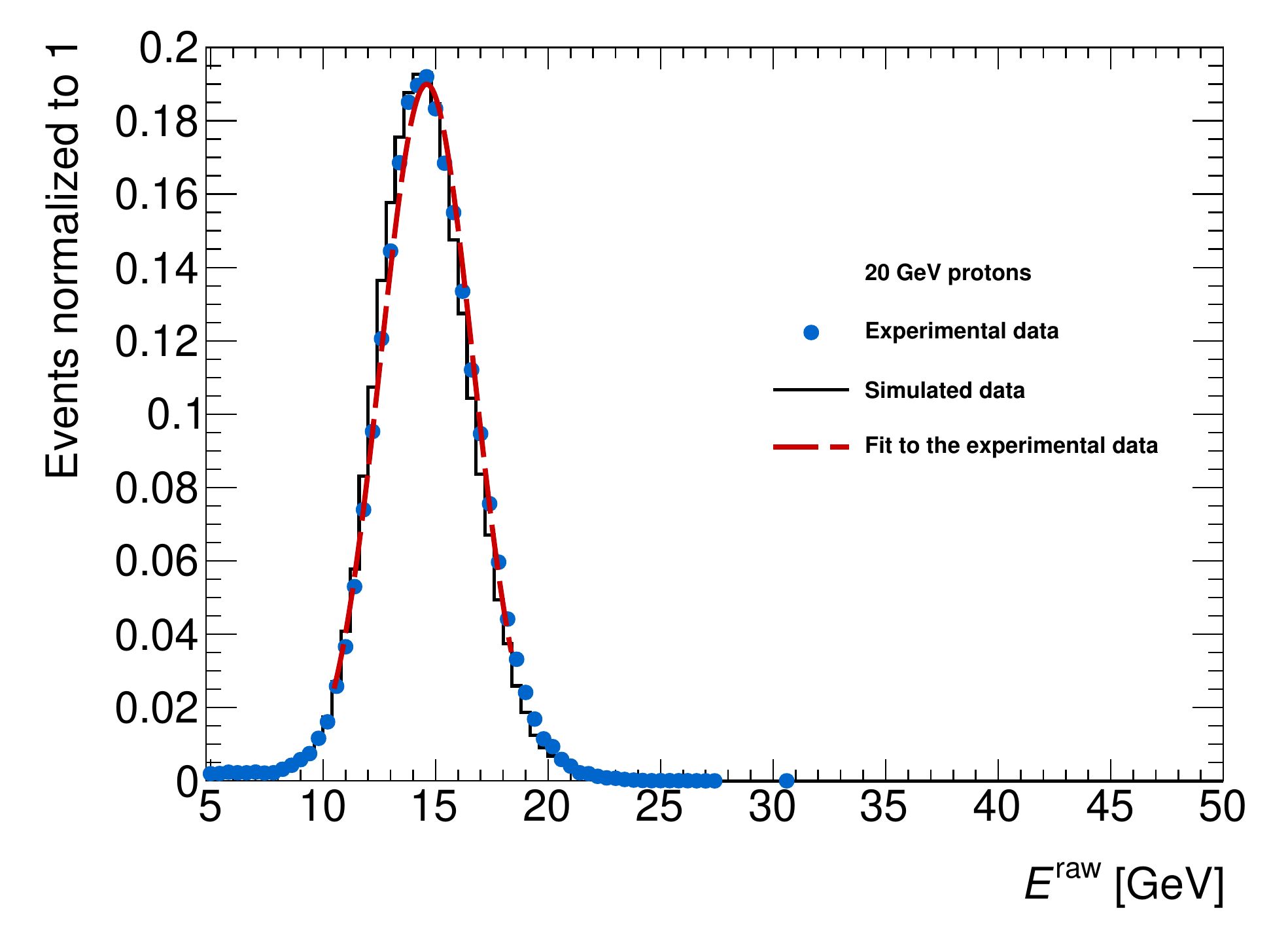}}
\caption{Distributions of the reconstructed energy $E^\text{raw}$ obtained analyzing pion (a), kaon (b) and proton (c) data with $E_\text{beam}$ = 20 GeV. The blue dotted histograms represent the experimental data. Only statistical uncertainties are shown.  The dashed curves in red correspond to the fit of a Gaussian function to the experimental data in a region ±2$\sigma$ around the peak value. The black histograms correspond to the predictions of the MC simulation. 
%\textcolor {red} {Take away the results}
}
\label{fig: Eraw_20_GeV}       % Give a unique label
\end{figure}
\begin{figure}[]
\centering
\subfloat[]{\includegraphics[width=4.3cm,clip]{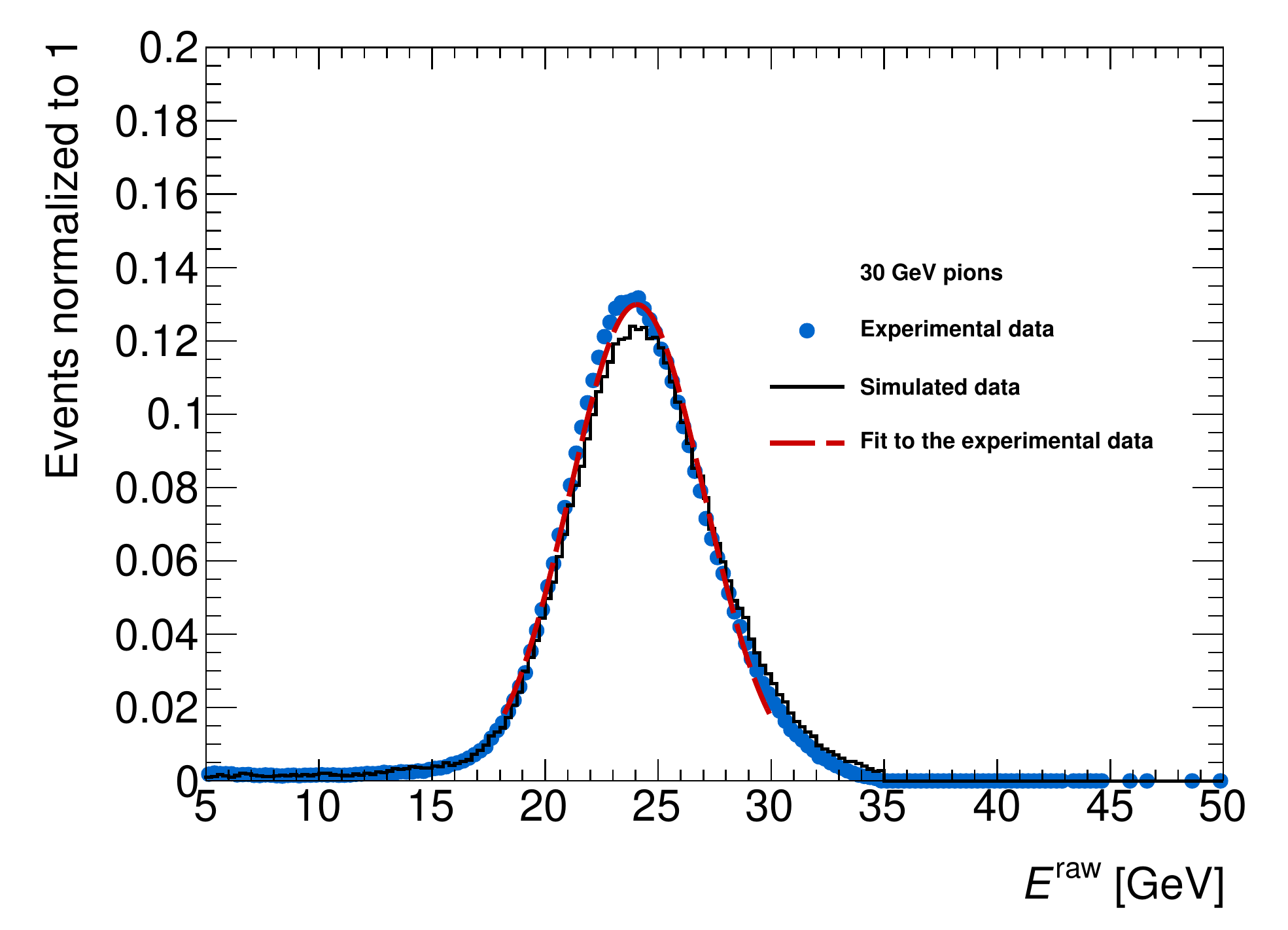}}
\subfloat[]{\includegraphics[width=4.3cm,clip]{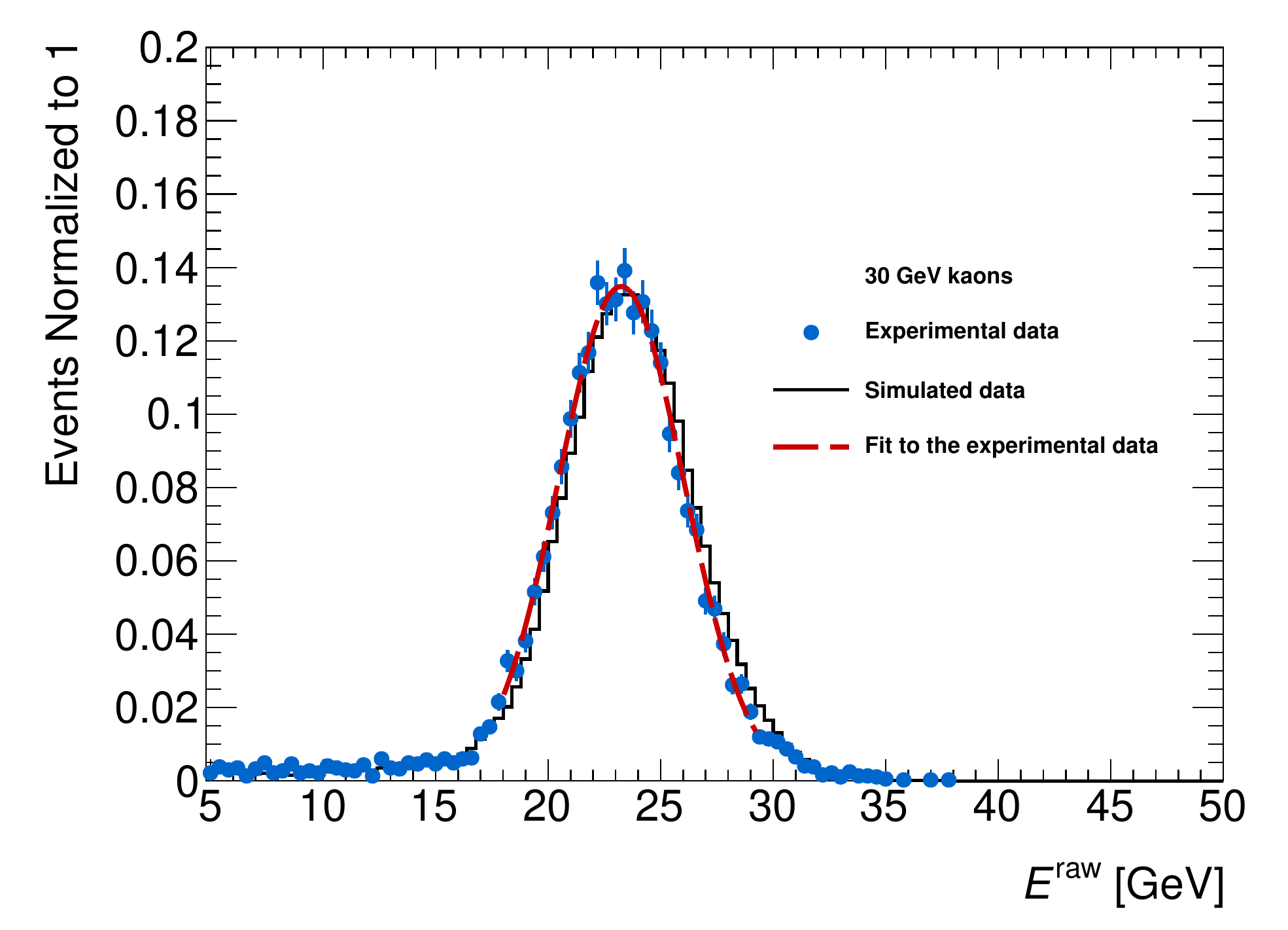}} \\
\subfloat[]{\includegraphics[width=4.3cm,clip]{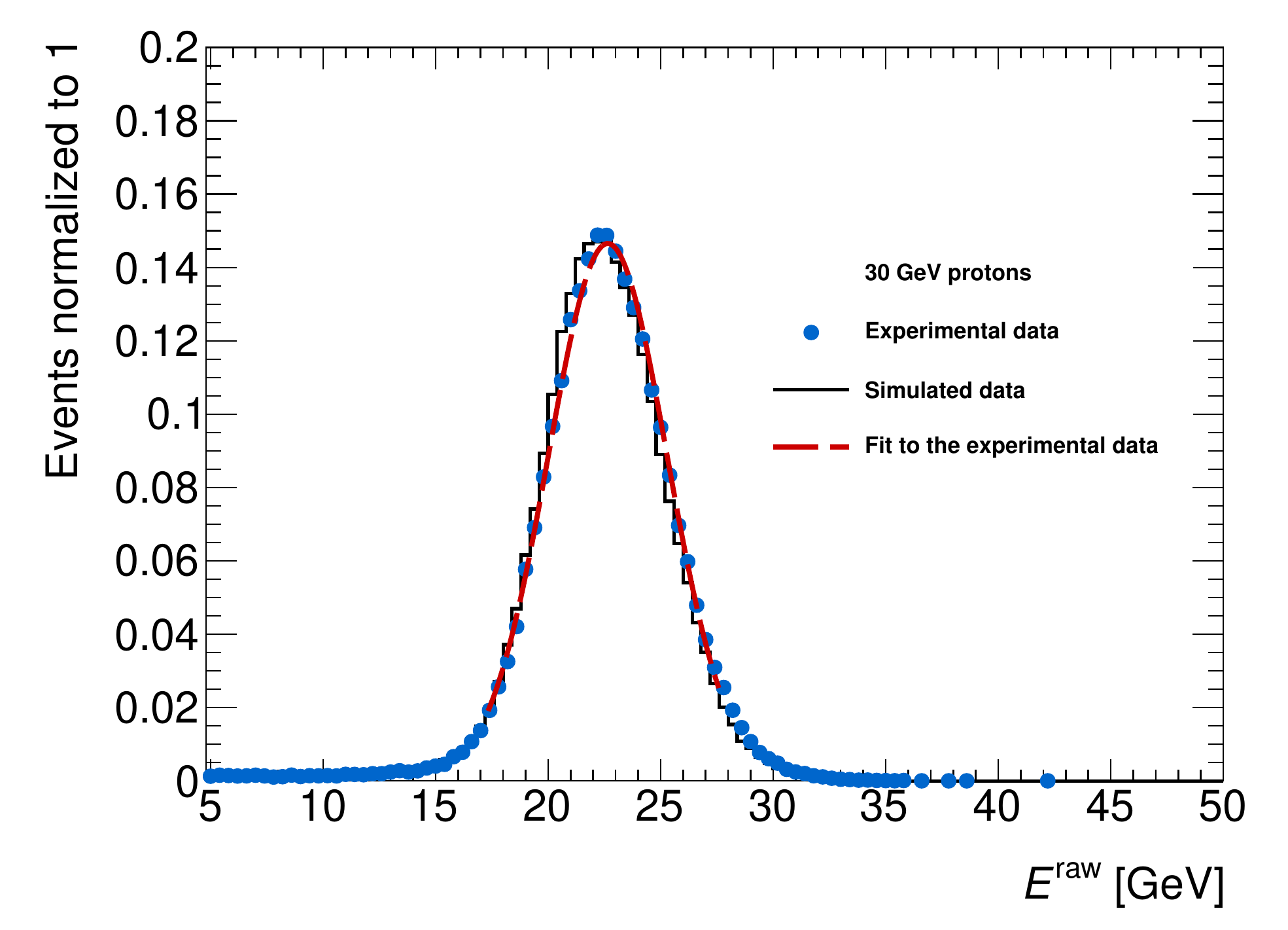}}
\caption{Distributions of the reconstructed energy $E^\text{raw}$ obtained analyzing pion (a), kaon (b) and proton (c) data with $E_\text{beam}$ = 30 GeV. The blue dotted histograms represent the experimental data. Only statistical uncertainties are shown.  The dashed curves in red correspond to the fit of a Gaussian function to the experimental data in a region ±2$\sigma$ around the peak value. The black histograms correspond to the predictions of the MC simulation. 
%\textcolor {red} {Take away the results}
}
\label{fig: Eraw_30_GeV}       % Give a unique label
\end{figure}

%\FloatBarrier
% ----------
Due to the electron contamination, as sketched in the Figure~\ref{fig: Electron subtraction}, the pion energy distributions $n_\pi(E^\text{raw})$ 
were obtained using, bin per bin, the formula   
\begin{equation}
n_\pi(E^\text{raw}) = n_{e/\pi}(E^\text{raw}) - N_e f_e(E^\text{raw}) .
\label{eq:subtraction}
\end{equation}
 where $n_{e/\pi}(E^\text{raw})$ is the number of $e/\pi$ events in the considered $E^\text{raw}$ bin, the electron distribution $f_e(E^\text{raw})$ is normalized to 1 and the number of electrons, $N_e$, was determined using the procedure described in Section~\ref{subsubsec:electron_identification}. Simulated electron distributions were used in the analysis because experimental data are available only for electron beam energy equal to 20 GeV. 
 %The histogram (a) in Fig.~\ref{fig: 20 GeV Electron energy distribution} allows 
 A comparison between the distributions obtained analyzing simulated and experimental electrons with the same beam energy, direction and impact point is shown in Figure~\ref{fig: 20 GeV Electron energy distribution} (a) . 
% ----------
%\FloatBarrier
% ----------
Figures~\ref{fig: Eraw_16_GeV} to~\ref{fig: Eraw_30_GeV} show the $E^\text{raw}$ distributions obtained in the case of beams of pions, kaons and protons with energies equal to 16, 18, 20 and 30 GeV, respectively.
% ----------
% ---------
%\FloatBarrier
% ----------
%\FloatBarrier
% -----------
%\FloatBarrier
% ----------
%\FloatBarrier
\begin{table}[]
\centering
\caption{Numbers of simulated and retained pion, kaon and proton events for each beam energy value. Selection criteria used in the analysis are discussed in the text.
}
\label{tab:selection chain MC}       % Give a unique label
% For LaTeX tables you can use 
\begin{tabular}{	|l|c|c|c|c|} 
\hline 
$E_\text{beam}$ [GeV] & 16 & 18 & 20 & 30 \\ \hline
%\hline
%\multicolumn{5}{|l|}{
%Experimental data & & & &  \\ \hline
%\multicolumn{5}{|l|}{Experimental data}\\ \hline 
%Physics Trigger & 694658 &  944460 &  1226756 &  1297099 \\ \hline
%Cut~1.: Beam line scintillators & 656262 & 895863 & 1155580 & 1230470 \\ 
%Cut 2: Beam line chambers impact point &  &  &  &  \\ \hline
%Cut 2.: Beam line chamber & 552179 & 771513 & 935131 & 1069709 \\ 
%Cut 3.: Muon and spurious  & 501013 & 700590 & 777386  & 983892 \\
%~~~~~~~~~~~~~events rejection & & & &%\multicolumn{4}{|c|}{}
%\\ 
%$e/\pi$ & 385718 & 556782 & 611687  & 723286 \\ 
%Electrons & 67647 $\pm$ 9198 & 70834$\pm$3665 & 62137$\pm$3548 & 28288 $\pm$2481 \\ 
%Pions & 318071 $\mp$ 9198 & 485948 $\mp$3665 & 549550$\mp$3548 & 694998$\mp$2481  \\ 
%$K/p$ & 86635 & 133071 & 154181  & 137119 \\ 
%Kaons & 2372 & 4674 & 6782 & 11296  \\ 
%Protons & 84263 & 128397 & 147399 & 125823 \\ %\hline
%\hline
%\multicolumn{5}{|l|}{Simulated data} \\\hline
%Simulated data & & & &  \\ \hline
%\multicolumn{5}{|l|}{Simulated data}\\ \hline 
Generated events & \multicolumn{4}{c|}{300000}\\ 
\hline 
%Pions & 300000 &  300000 & 300000 & 300000  \\
Pions   & 283222 & 285211 & 286574 & 291040 \\
Kaons   & 247559  & 253040 & 256514 & 269728 \\
Protons & 292412 & 293891 & 294596 & 296532 \\ \hline
\end{tabular}
%\end{center}
\end{table}
%\textcolor {red} {New references?}
\section{Analysis of simulated data}
\label{sec:MC_selection}
The experimental results obtained using positive pions and kaons and protons beams, with energies in the range 16--30 GeV, were compared to the predictions of the Geant4-based ATLAS simulation program \cite{Agostinelli:2002hh}, \cite{Allison:2006ve} and \cite{Costanzo:916030}. The FTFP\_BERT\_ATL hadronic showering model \cite{Bertini:1971xb} was used in the simulation. This is the model presently being used in the simulation of the ATLAS events collected during the LHC Run~1 and Run~2.
The number of generated events for each experimental data point is reported in Table~\ref{tab:selection chain MC}. The responses of the beam line detectors were not included in the simulation. The distributions of the transverse beam impact point coordinates in the detector were tuned to reproduce the ones measured using the BC1. The TB detector material and geometry were fully described (see Ref.~\cite{Allison:2006ve}).  The measured electronics noise in the different calorimeter cells and the effects of photo-statistics (70 photo electron per GeV) in the PM signals, are included in the MC simulation.
%To be consistent with the analysis of the experimental data, the scale of the cell energy measurements used was obtained making use of the response of simulated electrons. 
The simulated pion events were selected applying the $C_\text{long}$ and $C_\text{tot}$ cuts used in the analysis of experimental data. The numbers of the retained events for analyses are reported in Table~\ref{tab:selection chain MC}. The shower energy was reconstructed using the same procedure applied in the case of experimental data. The distributions of $E^\text{raw}$ obtained using simulated data are shown in Figures~\ref{fig: Eraw_16_GeV} to~\ref{fig: Eraw_30_GeV} for beam energies equal to 16, 18, 20 and 30 GeV respectively. 
\section{Determination of the energy response and resolution}
\label{sec:calorimeter_response}
%\subsection{Determination of the hadron responses}
%\label{subsec:hadron_response}
The experimental and simulated $E^\text{raw}$ distributions of pion, kaon and proton data are described reasonably well around the peak values by a Gaussian function.
As in Ref.~\cite{production_modules}, the $\mu$ and $\sigma$ parameters of Gaussian functions fitting the distributions in a region ±2$\sigma$ around the peak values were used to estimate the measurement responses $ \langle E^\text{raw} \rangle$ and resolutions $\sigma^\text{raw}$. An iterative procedure has been applied in order to get stable values of the parameters. The method of the least squares has been used. The fit functions obtained analysing experimental data are superimposed to the corresponding distributions in Figures~\ref{fig: Eraw_16_GeV} to~\ref{fig: Eraw_30_GeV}. The fit results obtained using experimental and simulated data are reported in Table~\ref{tab:response and resolution}. The statistical uncertainties correspond to the square root of the corresponding diagonal term of the fit error matrix. 
% ----------
\begin{table}[]
\centering
\caption{Energy response (resolution) obtained fitting Gaussian functions to the experimental and simulated $E^\text{raw}$ distributions obtained using pions ($ \langle E^\text{raw} \rangle$ ($\pi$) and ($\sigma^\text{raw}$($\pi$))), kaons ($ \langle E^\text{raw} \rangle$ ($K$) and ($\sigma^\text{raw}$($K$)) and protons ($ \langle E^\text{raw} \rangle$ ($p$) and ($\sigma^\text{raw}$($p$))) with different beam energy.
The statistical uncertainties correspond to the fit parameter uncertainties.}
%The statistical uncertainties correspond to the fit parameter uncertainties.}
\label{tab:response and resolution}       % Give a unique label
% For LaTeX tables you can use 
\begin{tabular}{	|c|c|c|} 
\hline
%\multicolumn{5}{|l|}{Pions} \\ \hline
\multicolumn{3}{|c|}{$\langle E^\text{raw} \rangle$ ($\pi$)} \\
\hline
$E_\text{beam}$ [GeV]  & Exp. Data  & Sim. Data \\ \hline
16 & 12.678$\pm$0.008 & 12.500$\pm$0.008 \\ \hline
18 & 14.294$\pm$0.007 & 14.134$\pm$0.009  \\ \hline
20 & 15.896$\pm$0.004 & 15.744$\pm$0.010  \\ \hline
30 & 24.058$\pm$0.004 & 24.110$\pm$0.013 \\ \hline 
\multicolumn{3}{|c|}{$\sigma^\text{raw}$ ($\pi$)} \\
\hline
$E_\text{beam}$ [GeV]  & Exp. Data  & Sim. Data \\ \hline
16 & 2.013$\pm$0.014 & 1.948$\pm$0.015 \\ \hline
18 & 2.139$\pm$0.012 & 2.122$\pm$0.017 \\ \hline
20 & 2.319$\pm$0.004 & 2.289$\pm$0.019 \\ \hline
30 & 2.962$\pm$0.004 & 2.966$\pm$0.026 \\ \hline \hline
\multicolumn{3}{|c|}{$ \langle E^\text{raw}$ $\rangle$ ($k$)} \\
\hline
$E_\text{beam}$ [GeV]  & Exp. Data  & Sim. Data \\ \hline
16 & 12.291$\pm$0.252 & 12.236$\pm$0.004 \\ \hline
18 & 13.886$\pm$0.114  & 13.899$\pm$0.005 \\ \hline
20 & 15.445$\pm$0.034  & 15.459$\pm$0.005 \\ \hline
30 & 23.244$\pm$0.035 & 23.636$\pm$0.006 \\ \hline 
\multicolumn{3}{|c|}{$\sigma^\text{raw}$($K$)} \\
\hline
$E_\text{beam}$ [GeV]  & Exp. Data  & Sim. Data \\ \hline
16 & 2.168$\pm$0.441 & 1.904$\pm$0.004 \\ \hline
18 & 2.175$\pm$0.226  & 2.059$\pm$0.005 \\ \hline
20 & 2.262$\pm$0.034  & 2.183$\pm$0.005 \\ \hline
30 & 2.790$\pm$0.035 & 2.831$\pm$0.006 \\ \hline
\hline
\multicolumn{3}{|c|}{$ \langle E^\text{raw}$ $\rangle$ ($p$)} \\
\hline
$E_\text{beam}$ [GeV]  & Exp. Data  & Sim. Data \\ \hline
16 & 11.511$\pm$0.008 & 11.234$\pm$0.003 \\ \hline
18 & 13.119$\pm$0.007 & 12.827$\pm$0.004 \\ \hline
20 & 14.606$\pm$0.006 & 14.429$\pm$0.004 \\ \hline
30 & 22.649$\pm$0.009 & 22.457$\pm$0.005 \\ \hline
 \multicolumn{3}{|c|}{$\sigma^\text{raw}$($p$)} \\
\hline
$E_\text{beam}$ [GeV]  & Exp. Data  & Sim. Data \\ \hline
16 & 1.795$\pm$0.007 & 1.729$\pm$0.004 \\ \hline
18 & 1.898$\pm$0.006 & 1.854$\pm$0.004 \\ \hline
20 & 2.047$\pm$0.006 & 1.989$\pm$0.004 \\ \hline
30 & 2.633$\pm$0.009 & 2.588$\pm$0.005 \\ \hline
\end{tabular}
%\end{center}
\end{table}
%\FloatBarrier
% ----------
\subsection{Energy responses and resolutions normalized to incident beam energy}
\label{subsec:hadron_response_ratio}
%The determinations of the energy response normalized to incident beam energy
Energy response normalized to incident beam energy
\begin{equation}
R^{\langle E^\text{raw}\rangle} =\frac{\langle E^{raw}\rangle}{E_\text{beam}}
\label{eq:E^raw}
\end{equation}
%and of the energy resolution normalized to incident beam energy 
and energy resolution normalized to incident beam energy

\begin{equation}
R^{\sigma^\text{raw}} = \frac{\sigma^\text{raw}}{E_\text{beam}}
\label{eq:sigma^raw}
\end{equation}
obtained for the different values of $E_\text{beam}$ are reported in Table~\ref{tab:response and resolution ratios}. In the case of experimental results, the first uncertainty value corresponds to the statistical uncertainty. The systematic uncertainty, second value,
was obtained combining in quadrature the contributions of the seven sources discussed in the following. In the case of simulated  data only statistical uncertainties are reported.
% ----------
\begin{table}[]
\centering
\caption{Measured energy response (resolution) normalized to incident beam energy obtained using pions ($R^{\langle E^\text{raw}\rangle}$($\pi)$ and ($R^{\sigma^\text{raw}}$($\pi$))), kaons ($R^{\langle E^\text{raw}\rangle}$($K$) and ($R^{\sigma^\text{raw}}$($K$))) and protons ($R^{\langle E^\text{raw}\rangle}$($p$) and ($R^{\sigma^\text{raw}}$($p$))) of different beam energy obtained analyzing experimental and simulated data. In the case of experimental data, statistical and systematic uncertainties are reported. The effects of the different sources of systematic sources discussed in the text were combined in quadrature. Only statistical uncertainties are reported in the case of simulated data.}
\label{tab:response and resolution ratios}       % Give a unique label
% For LaTeX tables you can use 
    % Give a unique label
% For LaTeX tables you can use 
\begin{tabular}{	|c|c|c|} 
\hline
%\multicolumn{5}{|l|}{Pions} \\ \hline
\multicolumn{3}{|c|}{$R^{\langle E^\text{raw}\rangle}$($\pi$)} \\
\hline
$E_\text{beam}$ [GeV]  & Exp. Data  & Sim. Data \\ \hline
16 & 0.7924$\pm$0.0005$\pm$0.0116 & 0.7812$\pm$0.0005  \\ \hline
18 & 0.7941$\pm$0.0004$\pm$0.0108 & 0.7852$\pm$0.0005  \\ \hline
20 & 0.7948$\pm$0.0002$\pm$0.0101 & 0.7872$\pm$0.0005  \\ \hline
30 & 0.8019$\pm$0.0001$\pm$0.0098 & 0.8036$\pm$0.0004 \\ \hline 
\multicolumn{3}{|c|}{$R^{\sigma^\text{raw}}$($\pi$)} \\
\hline
$E_\text{beam}$ [GeV]  & Exp. Data  & Sim. Data \\ \hline
16 & 0.1258$\pm$0.0009$\pm$0.0038 & 0.1217$\pm$0.0009 \\ \hline
18 & 0.1188$\pm$0.0007$\pm$0.0022 & 0.1179$\pm$0.0009 \\ \hline
20 & 0.1159$\pm$0.0002$\pm$0.0013 & 0.1144$\pm$0.0010 \\ \hline
30 & 0.0987$\pm$0.0001$\pm$0.0006 & 0.0988$\pm$0.0008 \\ \hline \hline
\multicolumn{3}{|c|}{$R^{\langle E^\text{raw}\rangle}$($K$)} \\
\hline
$E_\text{beam}$ [GeV]  & Exp. Data  & Sim. Data \\ \hline
16 & 0.7682$\pm$0.0158$\pm$0.0094 & 0.7647$\pm$0.0003 \\ \hline
18 & 0.7714$\pm$0.0064$\pm$0.0093 & 0.7721$\pm$0.0003 \\ \hline
20 & 0.7723$\pm$0.0017$\pm$0.0093 & 0.7729$\pm$0.0002 \\ \hline
30 & 0.7748$\pm$0.0012$\pm$0.0093 & 0.7878$\pm$0.0002 \\ \hline 
\multicolumn{3}{|c|}{$R^{\sigma^\text{raw}}$($K$)} \\
\hline
$E_\text{beam}$ [GeV]  & Exp. Data  & Sim. Data \\ \hline
16 & 0.1356$\pm$0.0276$\pm$0.000007 & 0.1190$\pm$0.0003 \\ \hline
18 & 0.1209$\pm$0.0126$\pm$0.0005 & 0.1144$\pm$0.0002 \\ \hline
20 & 0.1131$\pm$0.0017$\pm$0.0008 & 0.1091$\pm$0.0002 \\ \hline
30 & 0.0930$\pm$0.0012$\pm$0.0002 & 0.0943$\pm$0.0002 \\ \hline
\hline
\multicolumn{3}{|c|}{$R^{\langle E^\text{raw}\rangle}$($p$)} \\
\hline
$E_\text{beam}$ [GeV]  & Exp. Data  & Sim. Data \\ \hline
16 & 0.7195$\pm$0.0005$\pm$0.0086 & 0.7021$\pm$0.0002 \\ \hline
18 & 0.7288$\pm$0.0004$\pm$0.0087 & 0.7126$\pm$0.0002 \\ \hline
20 & 0.7303$\pm$0.0003$\pm$0.0088 & 0.7214$\pm$0.0002 \\ \hline
30 & 0.7549$\pm$0.0003$\pm$0.0091 & 0.7485$\pm$0.0001 \\ \hline
 \multicolumn{3}{|c|}{$R^{\sigma^\text{raw}}$($p$)}  \\
\hline
$E_\text{beam}$ [GeV]  & Exp. Data  & Sim. Data \\ \hline
16 & 0.1122$\pm$0.0004$\pm$0.000034 & 0.1081$\pm$0.0002 \\ \hline
18 & 0.1055$\pm$0.0003$\pm$0.0004 & 0.1030$\pm$0.0002 \\ \hline
20 & 0.1024$\pm$0.0003$\pm$0.0007 & 0.0994$\pm$0.0002 \\ \hline
30 & 0.0877$\pm$0.0003$\pm$0.0002 & 0.0862$\pm$0.0001 \\ \hline
\end{tabular}
\end{table}
%\FloatBarrier
% ----------
Seven sources of systematic uncertainties were considered in the study:
\begin{enumerate}
\item
Systematic Uncertainty 1. affects only pion determinations. It corresponds to the statistical uncertainty on the determination of the number of electrons contaminating the $e/\pi$ samples discussed in Section~\ref{subsubsec:electron_identification}.  
\item
As discussed in the same section 
%Subsection~\ref{subsubsec:electron_identification} 
the electron contamination was determined studying the C$_\text{tot}$ distributions of the $e/\pi$ sample events with C$_\text{long}\ge$~C$_\text{long}^\text{min}$

= 0.6. Results obtained with different values of C$_\text{long}$ %C$_\text{long}$ > 0.5 and C$_\text{long}$ > 0.7 
were used for uncertainty estimations. Systematic Uncertainty 2. values reported in Table~\ref{tab:systematic errors} correspond to half of the differences of the determinations of $R^{\langle E^\text{raw} \rangle}$ and $R^{\sigma^\text{raw}}$ obtained using  C$_\text{long}^\text{min} = 0.5$ and C$_\text{long}^\text{min} = 0.7$ respectively.
\item
Effects due to the missmodeling of the $C_\text{tot}$ distributions used to determine the number of electrons contaminating the $e/\pi$ samples was estimated comparing the results obtained using three Gaussian functions fits (see Section~\ref{subsubsec:electron_identification}) with the ones obtained using two Gaussian functions fits. The estimated percentage of electrons increases from a value of 11\% at 16 GeV up to 28\% at 30 GeV. Systematic Uncertainty 3. values,  affecting only pion determinations, are reported in Table~\ref{tab:systematic errors} for each of the four beam energy samples. It is equal to the differences of the values of $R^{\langle E^\text{raw} \rangle}$ and $R^{\sigma^\text{raw}}$ obtained using the two fitting functions.
\item
As discussed in Section~\ref{subsec:shower_energy} the experimental $E^\text{raw}$ distributions of pions were obtained using Eq.~(\ref{eq:subtraction}). In Figure~\ref{fig: 20 GeV Electron energy distribution} are shown electron distributions obtained in the case of simulated data with beam energies equal to 20 and 30 GeV. Due to the regularly spaced scintillating tiles (see Figure~\ref{fig:module}) and the compactness of electromagnetic showers, the electron response varies with the periodicity of sampling fraction and thus depends on the coordinate of the impact point of the beam particles along the front face of the calorimeter module ($z$). 
%This is shown in Figs.~\ref{fig: 20 GeV Electron energy distribution} (b) and (d). The 
In Figures~\ref{fig: 20 GeV Electron energy distribution} (b) and (d)
is shown that the variation is reasonably well described by a simple periodic function~\cite{production_modules}

\begin{equation}
E^\text{raw}(z) = p_0[1+p_1\sin(2\pi z/p_2)+p_3]~.
\label{eq:oscillation}
\end{equation}

The parameter $p_0$ corresponds to the mean reconstructed energy.
%shown in the histograms (b) and (d) of Fig.~\ref{fig: 20 GeV Electron energy distribution}. 
The relative amplitude of the oscillation is described by $p_1$.
The parameter $p_2$ corresponds to the periodic thickness as seen by the beam at a given $z$ value and $p_3$ is a phase. The behavior is responsible of the two peak structure of the $E^\text{raw}$ distributions evident, in particular, in the case of $E_\text{beam}$ = 30 GeV simulated data in Figure~\ref{fig: 20 GeV Electron energy distribution}~(c). The effects of the uncertainty on the distribution of the $z$ coordinates of the electron impact point on the determinations of $R^{\langle E^\text{raw} \rangle}$ and $R^{\sigma^\text{raw}}$ was estimated using the $E^\text{raw}$ distributions of the events with a $z$ value corresponding to $E^\text{raw}$ $>$ $p_0$, “high energy events”, and $E^\text{raw}$ $<$ $p_0$, “low energy events”,  respectively. The distributions are shown in Figures~\ref{fig: 20 GeV Electron energy distribution} (a) and~\ref{fig: 20 GeV Electron energy distribution} (c). Systematic Uncertainty 4. values, reported in Table~\ref{tab:systematic errors},  correspond to half of the differences of the values obtained using the two distributions. This uncertainty affects only pion determinations.
\item
The 30 GeV scatter plot $S_\text{Ch1}$ vs. $S_\text{Ch3}$ in Figure~\ref{fig:Ch3 vs Ch1 18 30 GeV} shows two spots in the $K/p$ region. Their origin is not clear. Systematic Uncertainty 5. values reported in Table~\ref{tab:systematic errors} correspond to the differences of the values of $R^{\langle E^\text{raw} \rangle}$ and $R^{\sigma^\text{raw}}$ obtained using the events with $S_\text{Ch1}$ $\le$ 400 [ADC counts] and $S_\text{Ch1}$ $\le$ 250 [ADC counts], respectively. Although the other three energy data points do not show the two spot structure, a systematic uncertainty was determined also for them using the described procedure with the same selection criterion values. 
\item
As it appears in Figure~\ref{fig: Ch2 K p}, proton $S_\text{Ch2}$ distributions show large tails. Their origin is not understood. Systematic Uncertainty 6. values in Table~\ref{tab:systematic errors}, correspond to the differences of the values of $R^{\langle E^\text{raw} \rangle}$ and $R^{\sigma^\text{raw}}$ obtained using for each of the four proton beam energies, the upper values of the $S_\text{Ch2}$ signals of Table~\ref{tab:Cs cuts}, and the ones obtained selecting the events with $S_\text{Ch2}$ $\le$~2000 ADC counts at 16 GeV, 18 GeV and 20 GeV and 1000 ADC counts at 30 GeV respectively. The same effect could also be present in the case of kaons. Since they produce a signal in Ch2, the effect may not be visible. For this reason the systematic uncertainty obtained for protons is also applied in kaon determinations. %\textcolor {red} {To verify}
\item
The effect of the uncertainty of the scale of the reconstructed cell energy $\Delta C_\text{c}^\text{EM}$ on the measurements was also investigated.
An estimation of the uncertainty on the energy response can be obtained using the formula:
%The Systematic Uncertainty 7 in Table~\ref{tab:systematic errors} is defined by
\begin{equation}
\Delta \langle E^\text{raw}\rangle^\text{EM}= \Delta C_\text{c}^\text{EM}\sqrt{\sum_{i}\langle E_{c}^\text{raw}\rangle_i^2}
\label{eq:syst_error}
\end{equation}
 where $\Delta C_\text{c}^\text{EM}$ is equal to 2.4\% (see Section~\ref{subsec:detector}) and $\langle E_c^\text{raw} \rangle_i$ is the average energy deposited in the cell $i$. $E_\text{beam}$ is known at few per mile and one obtains the values of $\Delta$ $R^{\langle E^\text{raw} \rangle}$ reported in Table~\ref{tab:systematic errors} for the twelve data points (Systematic Uncertainty 7.). No significant dependence of the values on the beam energies was found. The uncertainty on $C_\text{c}^\text{EM}$ affects in a negligible way the determinations of $R^{\sigma ^\text{raw}}$. 
\end{enumerate}
% ----------
\begin{table}[]
\centering
\caption{Systematic uncertainties on the estimations of $R^{\langle E^\text{raw} \rangle}$ and $R^{\sigma^\text{raw}}$ in percent. The pion measurements are affected by the uncertainty on the number of electrons contaminating the $e/\pi$ samples (Systematic uncertainties 1., 2. and 3.), on the $E^\text{raw}$ shape of the contaminating electrons (Systematic uncertainty 4.). The kaon and proton measurements are affected by the uncertainty on the Ch1 (Systematic uncertainty 5.) and Ch2 (Systematic uncertainty 6.) selection criteria. The uncertainty on the determination of the cell energy response non-uniformity, Systematic uncertainty 7., affects the determinations obtained for the three particle beams. 
%The hyphens in the cases indicate negligible systematic uncertainties. 
}
\label{tab:systematic errors}       % Give a unique label
% For LaTeX tables you can use 
\begin{tabular}{	|c|c|c|c|c|c|} 
\hline
\multicolumn{2}{|c|}{$E_\text{beam}$ [GeV]} & \multicolumn{2}{c|}{16} & \multicolumn{2}{c|}{18} 
%\multicolumn{2}{c|}{20} & \multicolumn{2}{c|}{30} 
\\ 
\hline
Syst. & Beam & $R^{\langle E^\text{raw} \rangle}$  & $R^{\sigma^\text{raw}}$ & $R^{\langle E^\text{raw} \rangle}$  & $R^{\sigma^\text{raw}}$ 
%& $R^{\langle E^\text{raw} \rangle}$  & $R^{\sigma^\text{raw}}$ & $R^{\langle E^\text{raw} \rangle}$  & $R^{\sigma^\text{raw}}$ 
\\ 
Uncer. & Part. & [\%] & [\%] & [\%] & [\%] \\ \hline
1. & $\pi$ & 0.227 & 2.854 & 0.067  & 1.665 
%& 0.152 & 0.614 & 0.066 & 0.317 
\\ \hline
2. & $\pi$ & 0.088 & 0.315 & 0.012  & 0.043 
%& 0.039 & 0.158 & 0.021 & 0.101 
\\ \hline
3. & $\pi$ & 0.235 & 0.854 & 0.213 & 0.786 
%& 0.215 & 0.875 & 0.098 & 0.470
\\ \hline
4. & $\pi$ & 0.753 & 0.431 & 0.603 & 0.266 
%& 0.345 & 0.041 & 0.184 & 0.077 
\\ \hline
5. & $K$ & 0.221 & - & 0.029 & 0.145 
%& 0.005 & 0.011 & 0.047 & 0.049
\\ 
& $p$ & - & 0.031 & 0.011 & 0.044 
%& 0.001 & 0.007 & - & 0.010 
\\ \hline
6. & $K$ & 0.002 & 0.006 & 0.067 & 0.377 
%& 0.152 & 0.682 & 0.038 & 0.207 
\\ 
& $p$ & 0.002 & 0.006 & 0.067 & 0.377 
%& 0.152 & 0.682 & 0.038 & 0.207 
\\ \hline
7. & $\pi$ & 1.138 & - & 1.138 & - 
%& 1.138 &  - & 1.138 & -
\\ 
& $K$ & 1.174 & - & 1.174 & - 
%& 1.174 &  - & 1.174 & -
\\
& $p$ & 1.234 & - & 1.234 & - 
%& 1.234 &  - & 1.234 & -
\\ 
\hline
\hline
\multicolumn{2}{|c|}{$E_\text{beam}$ [GeV]} & %\multicolumn{2}{c|}{16} & \multicolumn{2}{c|}{18} 
\multicolumn{2}{c|}{20} & \multicolumn{2}{c|}{30} 
\\ 
\hline
Syst. & Beam & $R^{\langle E^\text{raw} \rangle}$  & $R^{\sigma^\text{raw}}$ & $R^{\langle E^\text{raw} \rangle}$  & $R^{\sigma^\text{raw}}$ 
%& $R^{\langle E^\text{raw} \rangle}$  & $R^{\sigma^\text{raw}}$ & $R^{\langle E^\text{raw} \rangle}$  & $R^{\sigma^\text{raw}}$ 
\\ 
Uncer. & Part. & [\%] & [\%] & [\%] & [\%] \\ \hline
1. 
& $\pi$ 
%& 0.227 & 2.854 & 0.067  & 1.665 
& 0.152 & 0.614 & 0.066 & 0.317 
\\ \hline
2. 
& $\pi$ 
%& 0.088 & 0.315 & 0.012  & 0.043 
& 0.039 & 0.158 & 0.021 & 0.101 
\\ \hline
3. 
& $\pi$ 
%& 0.235 & 0.854 & 0.213 & 0.786 
& 0.215 & 0.875 & 0.098 & 0.470
\\ \hline
4. 
& $\pi$ 
%& 0.753 & 0.431 & 0.603 & 0.266 
& 0.345 & 0.041 & 0.184 & 0.077 
\\ \hline
5. & $K$ 
%& 0.221 & - & 0.029 & 0.145 
& 0.005 & 0.011 & 0.047 & 0.049
\\ 
& $p$ 
%& - & 0.031 & 0.011 & 0.044 
& 0.001 & 0.007 & - & 0.010 
\\ \hline
6. & $K$ 
%& 0.002 & 0.006 & 0.067 & 0.377 
& 0.152 & 0.682 & 0.038 & 0.207 
\\ 
& $p$ 
%& 0.002 & 0.006 & 0.067 & 0.377 
& 0.152 & 0.682 & 0.038 & 0.207 
\\ \hline
7. & $\pi$
%& 1.138 & - & 1.138 & - 
& 1.138 &  - & 1.138 & -
\\ 
& $K$ 
%& 1.174 & - & 1.174 & - 
& 1.174 &  - & 1.174 & -
\\
& $p$ 
%& 1.234 & - & 1.234 & - 
& 1.234 &  - & 1.234 & -
\\ 
\hline
\end{tabular}
%\end{center}
\end{table}
%\FloatBarrier
% ------------

The effects of each of the seven considered sources of systematic uncertainties on the four energy  determinations are correlated. The uncertainty in the energy response normalized to incident beam energy is dominated by the systematic effects due to cell response non uniformity (Systematic Uncertainty 7.).

The systematic uncertainties in Table~\ref{tab:response and resolution ratios} were obtained by combining in quadrature the effects of the seven sources reported in Table~\ref{tab:systematic errors}. Eleven values of the twelve energy response normalized to incident beam energy determinations have a total uncertainty smaller than 1.4\%. It is mainly defined by the uncertainty in the calibration of the energy response of the relatively small part of the calorimeter involved in the study. In the case of kaons with $E_\text{beam}$ = 16 GeV, due to the large statistical error, the uncertainty on the determination of $R^{\langle E^\text{raw} \rangle}$, is equal to 2.4\%. Nine of the twelve determinations of the energy resolution normalized to incident beam energy, $R^{\sigma^\text{raw}}$, have a total uncertainty smaller than 1.9\%. The uncertainty values of the determinations of $R^{\sigma^\text{raw}}$ obtained in the case of 16 GeV pion and kaon and 18 GeV kaon beams are equal to 3.1\%, 20.3\% and 10.4\% respectively.

\begin{figure}[ht]
\centering
\subfloat[]{\includegraphics[width=4.3cm,clip]{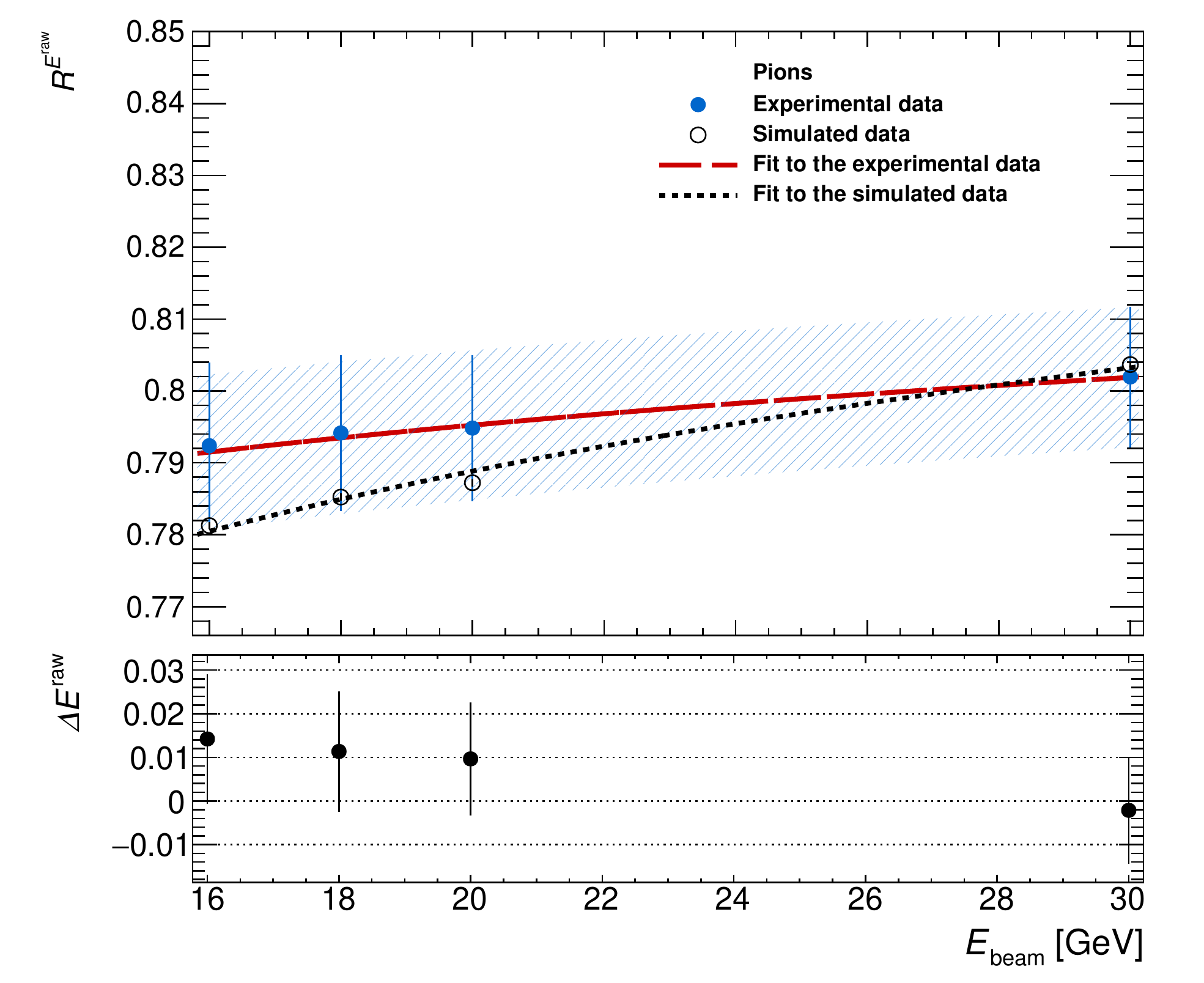}}
\subfloat[]{\includegraphics[width=4.3cm,clip]{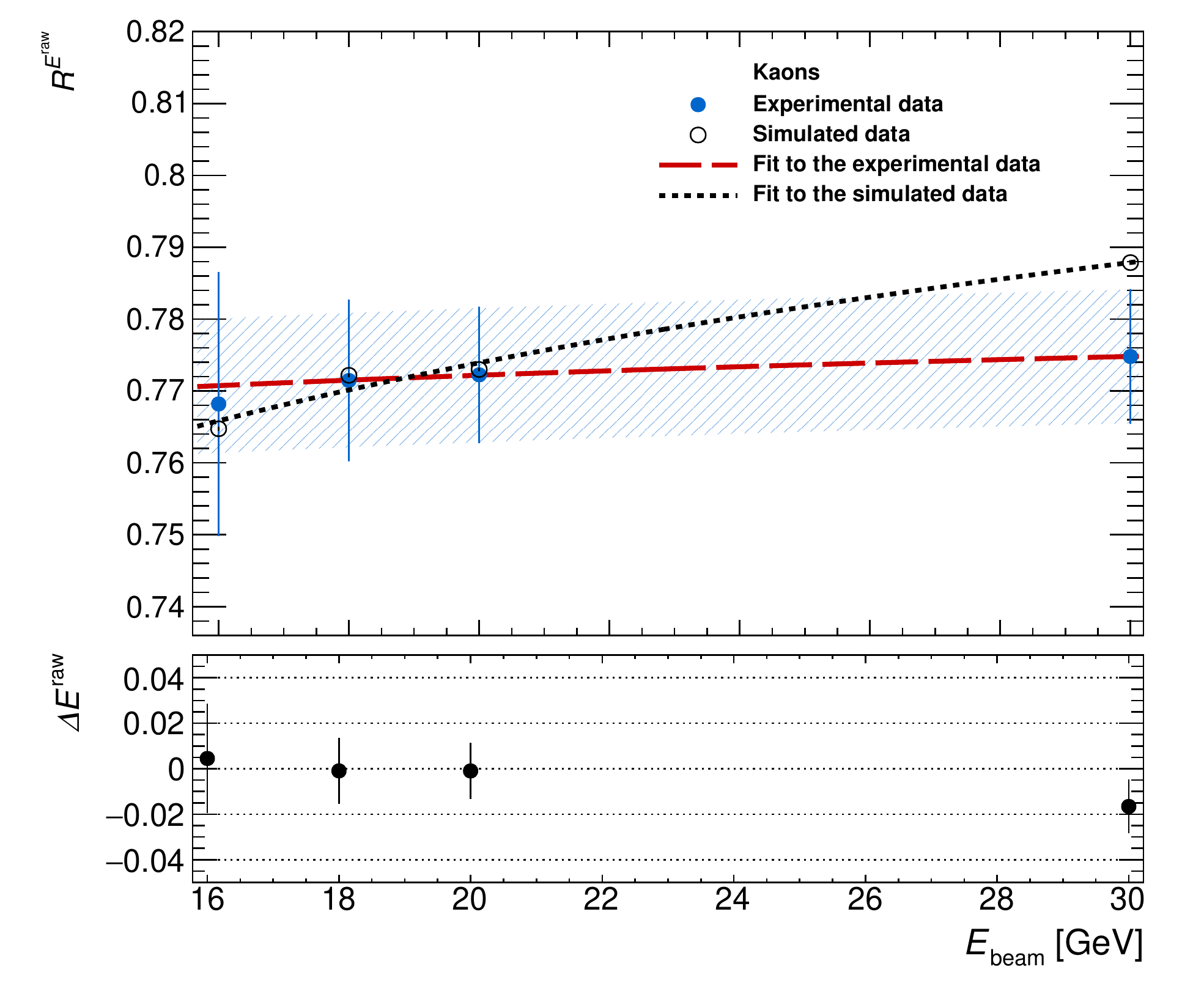}} \\
\subfloat[]{\includegraphics[width=4.3cm,clip]{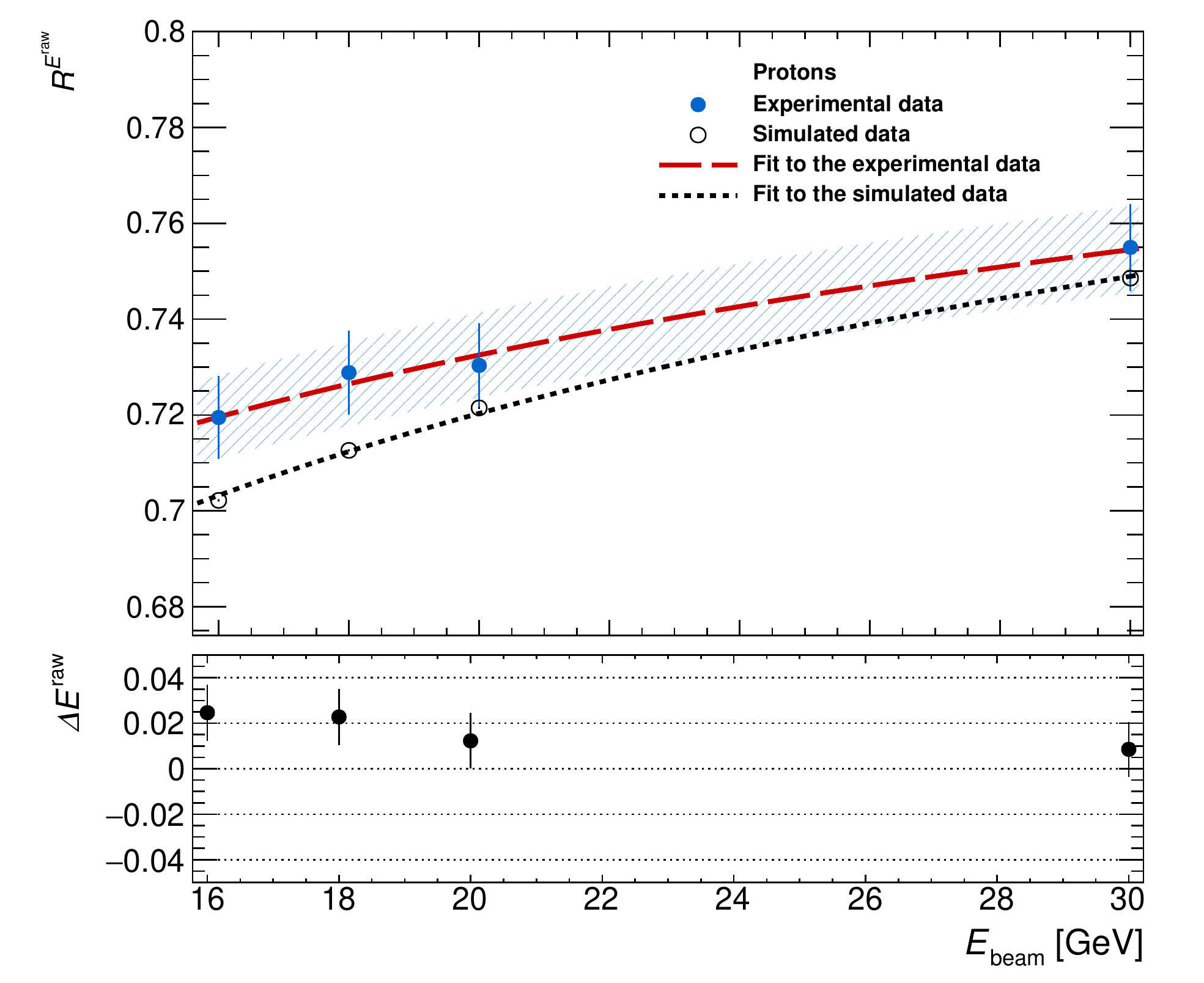}}
\caption{Energy response normalized to incident beam energy, $R^{\langle E^\text{raw} \rangle}$, measured (blue dots) and predicted by MC simulation (black circles) as a function of beam energy obtained in the case of  pion (a),  kaon (b) and  proton (c) beams. The experimental uncertainties include statistical and systematic effects combined in quadrature. Only statistical uncertainties affect simulated results. The red dashed (black dotted) curves are fits of the Eq.~(\ref{eq:parametrization_results}) to the experimental (simulated) data points. In case of experimental determinations the dashed blue strips display the correlated systematic uncertainties. In the bottom of the histograms are shown the fractional differences $\Delta E^{\langle \text{raw} \rangle}$ defined in Eq.~(\ref{eq:Delta_E^raw}). 
%The dashed horizontal lines indicate the $\pm$ 5\% region. 
The uncertainties include statistical and systematic effects combined in quadrature.
}
\label{fig:Response means}       % Give a unique label
\end{figure}

\begin{figure}[ht]
\centering
\subfloat[]{\includegraphics[width=4.3cm,clip]{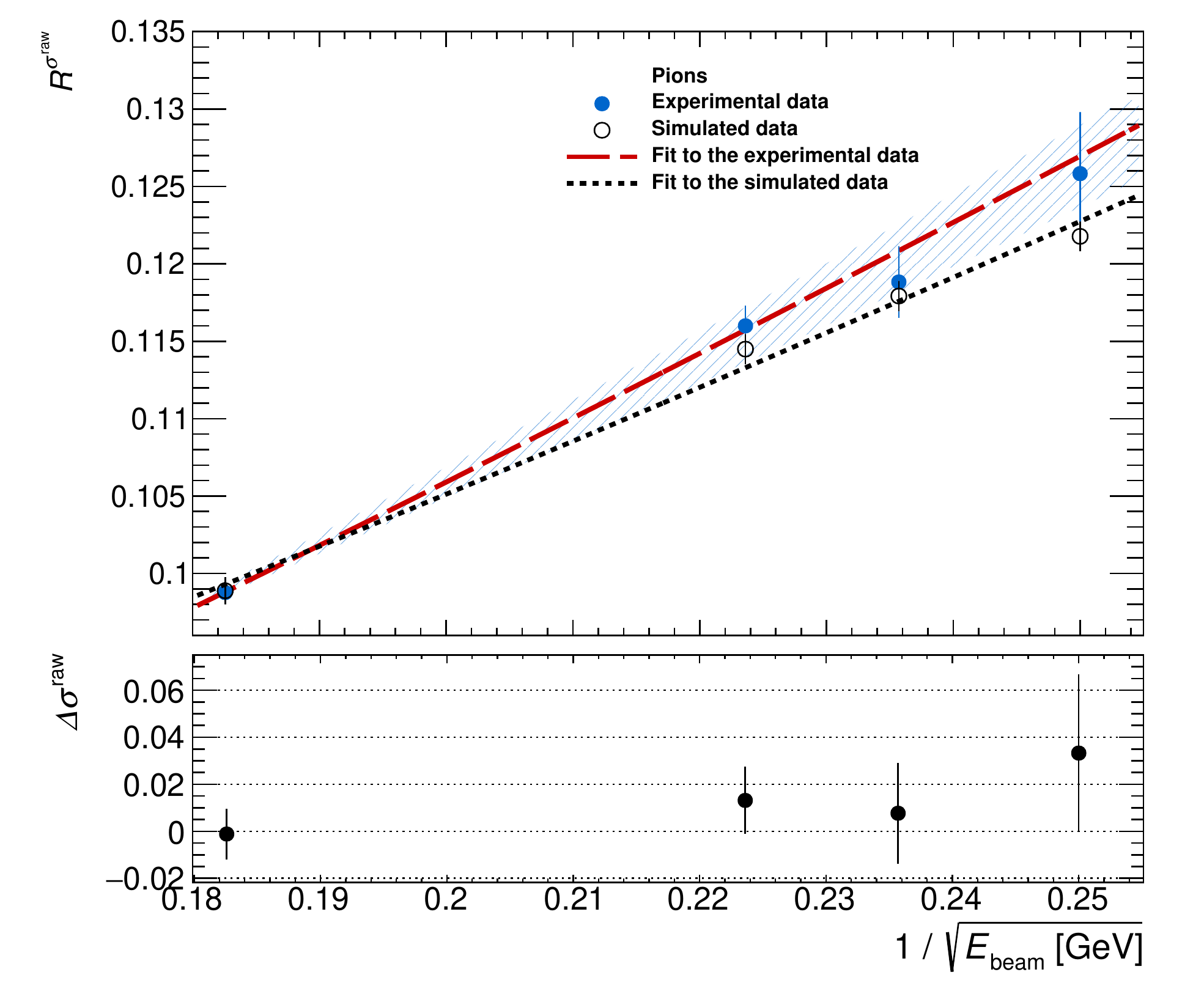}}
\subfloat[]{\includegraphics[width=4.3cm,clip]{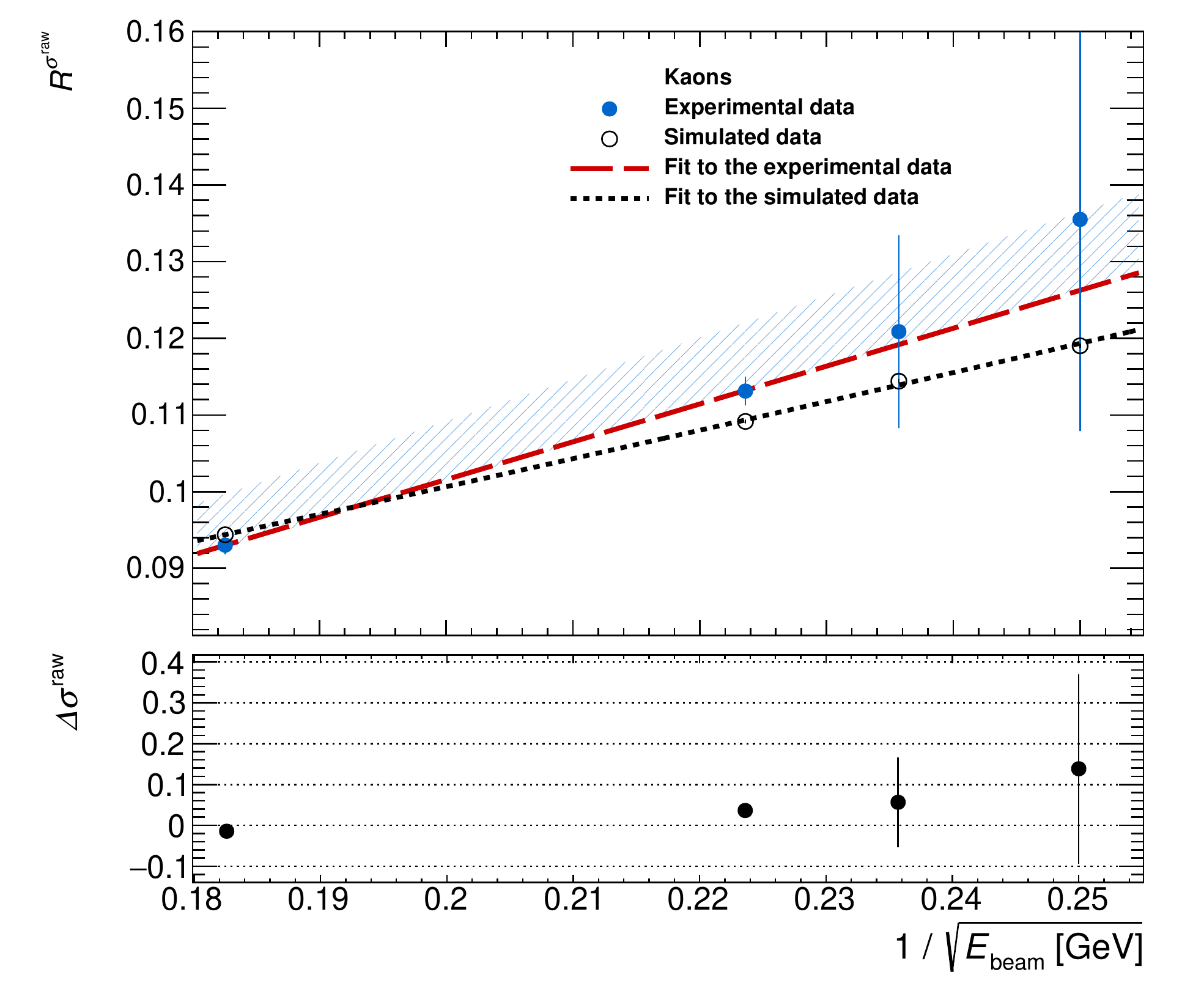}} \\
\subfloat[]{\includegraphics[width=4.3cm,clip]{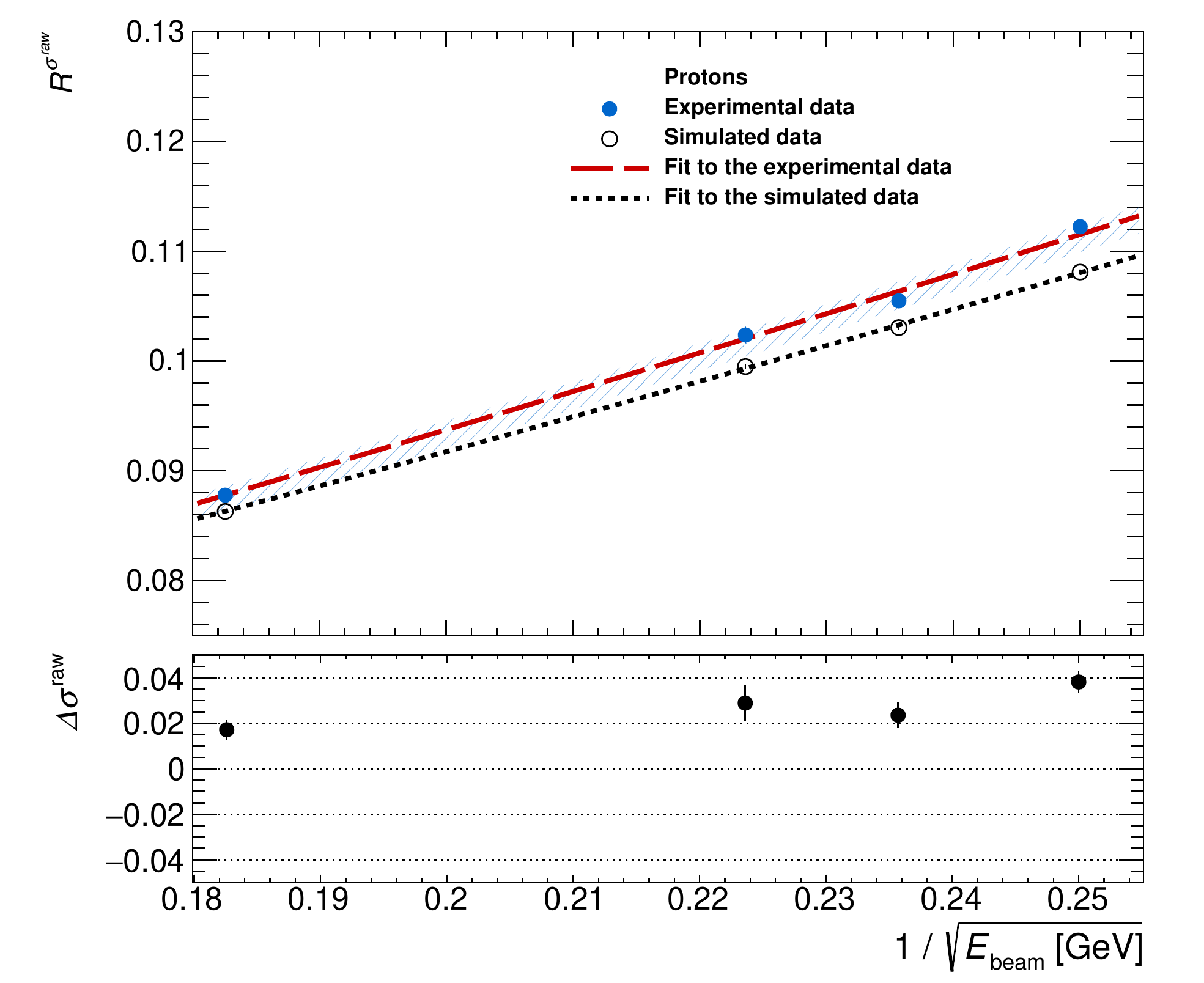}}
\caption{Energy resolution normalized to incident beam energy, $R^{\sigma^\text{raw}}$, measured (blue dots) and predicted by MC simulation (black circles) as a function of 1⁄$\sqrt{E_\text{beam}}$ obtained in the case of pion (a), kaon (b) and  proton (c) beams.  
%The values of $E_\text{beam}$ in the square roots are in GeV. 
The experimental uncertainties include statistical and systematic effects combined in quadrature. Only statistical uncertainties affect simulated results. The red dashed (black dotted) curves are fits of the Eq.~(\ref{eq:fractional_resolution}) to the experimental (simulated) data points. In case of experimental determinations the dashed blue strips display the correlated systematic uncertainties. In the bottom of the histograms are shown the fractional differences $\Delta\sigma^\text{raw}$ defined in Eq.~(\ref{eq:Delta_sigma^raw}). 
%The dashed horizontal lines indicate the $\pm$5\% region. 
The uncertainty includes statistical and systematic effects combined in quadrature. %\textcolor {red} {}
%The band must be defined only by the systematic errors
}
\label{fig:Response sigmas}       % Give a unique label
\end{figure}

The determinations of $R^{\langle E^\text{raw} \rangle}$ ($R^{\sigma^\text{raw}}$) as a function of $E_\text{beam}$ (1⁄$\sqrt{E_\text{beam}[\text{GeV}]}$) are reported in the histograms of Figure~\ref{fig:Response means} (Figure~\ref{fig:Response sigmas}) . In the case of experimental results, statistical and systematic uncertainties are combined in quadrature. In the case of simulated results only statistical uncertainty are shown.
%\FloatBarrier
% -----------
%\FloatBarrier
\subsection{Comparison between experimental and simulated results}
\label{subsec:comparison}
A quantitative comparison between experimental and simulated results can be obtained using the quantities

\begin{equation}
\Delta \langle E^\text{raw}\rangle = \frac {\langle E^\text{raw}\rangle}{\langle E_\text{MC}^\text{raw} \rangle} - 1 
\label{eq:Delta_E^raw}
\end{equation}
and
\begin{equation}
\Delta \sigma^\text{raw} = \frac {\sigma^\text{raw}}{\sigma_\text{MC}^\text{raw}} - 1 .
\label{eq:Delta_sigma^raw}
\end{equation}

The results are reported in Table~\ref{tab:Comparison with MC} where statistical and systematic uncertainties are shown separately.
The statistical uncertainties include the experimental and simulated uncertainties combined in quadrature. The results are also shown in Figures~\ref{fig:Response means} and \ref{fig:Response sigmas} where statistical and systematic uncertainties are combined in quadrature. 

%The determinations obtained using experimental and simulated data agree within the uncertainties. The average of the absolute values of the difference of all the energy response measurements obtained using experimental and simulated data was found to be 1.0\% with an average total uncertainty of 1.4\%. The average of the absolute values of the difference of all the resolution measurements was found to be 3.4\%. The average total uncertainty of the resolution measurements is 6\% in the case of pions and kaons and 0.6\% in the case of protons.

The average of the absolute values of the difference of all the energy response (resolution) measurements obtained using experimental and simulated data was found to be 1.1\% (3.4\%). In the case of the response determinations and the resolution determinations of pions and kaons, the differences are consistent within the uncertainties. The uncertainties of the proton resolution determinations are about one order of magnitude smaller.  

\begin{table}[]
\centering
\caption{Comparison of the energy response (top) and resolution (bottom) obtained analyzing experimental and simulated data in the case of pion, kaon and proton beams with different beam energies. Statistical uncertainties (first value) and systematic uncertainties (second value) are reported.
%\textcolor {red} {Check the number of digits}
}
\label{tab:Comparison with MC}       % Give a unique label
% For LaTeX tables you can use 
\begin{tabular}{	|c|c|c|} 
\hline
\multicolumn{3}{|c|}{Pions} \\ \hline
$E_\text{beam}$ & $\Delta E^\text{raw} =  \frac{\langle E^\text{raw}\rangle}{\langle E_\text{MC}^\text{raw}\rangle} - 1$ & $\Delta \sigma^\text{raw} = \frac{\sigma^\text{raw}}{\sigma_\text{MC}^\text{raw}} - 1$ \\ 
~[GeV] &  &  \\ \hline
16 & 0.0142$\pm$0.0009$\pm$0.0146 & 0.0333$\pm$0.0110$\pm$0.0317 \\ \hline
18 & 0.0113$\pm$0.0008$\pm$0.0137 & 0.0077$\pm$0.0102$\pm$0.0187  \\ \hline
20 & 0.0097$\pm$0.0007$\pm$0.0128 & 0.0131$\pm$0.0088$\pm$0.0113  \\ \hline
30 & -0.0021$\pm$0.0006$\pm$0.0121 & -0.0012$\pm$0.0090$\pm$0.0061 \\ \hline
\multicolumn{3}{|c|}{Kaons} \\ \hline
$E_\text{beam}$ & $\Delta E^\text{raw} =  \frac{\langle E^\text{raw}\rangle}{\langle E_\text{MC}^\text{raw}\rangle} - 1$ & $\Delta \sigma^\text{raw} = \frac{\sigma^\text{raw}}{\sigma_\text{MC}^\text{raw}} - 1$ \\  
~[GeV] &  &  \\ \hline
16 & 0.0045$\pm$0.0207$\pm$0.0122 & 0.1385$\pm$0.2319$\pm$0.0001  \\ \hline
18 & -0.0009$\pm$0.0082$\pm$0.0120 & 0.0565$\pm$0.1099$\pm$0.0042  \\ \hline
20 & -0.0010$\pm$0.0022$\pm$0.0121 & 0.0363$\pm$0.0157$\pm$0.0071  \\ \hline
30 & -0.0166$\pm$0.0015$\pm$0.0118 & -0.0144$\pm$0.0125$\pm$0.0021  \\ \hline
\multicolumn{3}{|c|}{Protons} \\ \hline
$E_\text{beam}$ & $\Delta E^\text{raw} =  \frac{\langle E^\text{raw}\rangle}{\langle E_\text{MC}^\text{raw}\rangle} - 1$ & $\Delta \sigma^\text{raw} = \frac{\sigma^\text{raw}}{\sigma_\text{MC}^\text{raw}} - 1$ \\ 
~[GeV] &  &  \\ \hline
16 & 0.0246$\pm$0.0008$\pm$0.0122 & 0.0382$\pm$0.0048$\pm$0.0003  \\ \hline
18 & 0.0228$\pm$0.0006$\pm$0.0123 & 0.0235$\pm$0.0041$\pm$0.0039  \\ \hline
20 & 0.0123$\pm$0.0005$\pm$0.0122 & 0.0289$\pm$0.0037$\pm$0.0070  \\ \hline
30 & 0.0085$\pm$0.0005$\pm$0.0121 & 0.0171$\pm$0.0041$\pm$0.0021  \\ \hline
\end{tabular}
%\end{center}
\end{table}
%\FloatBarrier
\subsection{Comparison between pion, kaon and proton energy responses and resolutions}
\label{subsec:ratios K/pi and p/pi}
\begin{figure}[ht]
\centering
\subfloat[]{\includegraphics[width=4.3cm,clip]{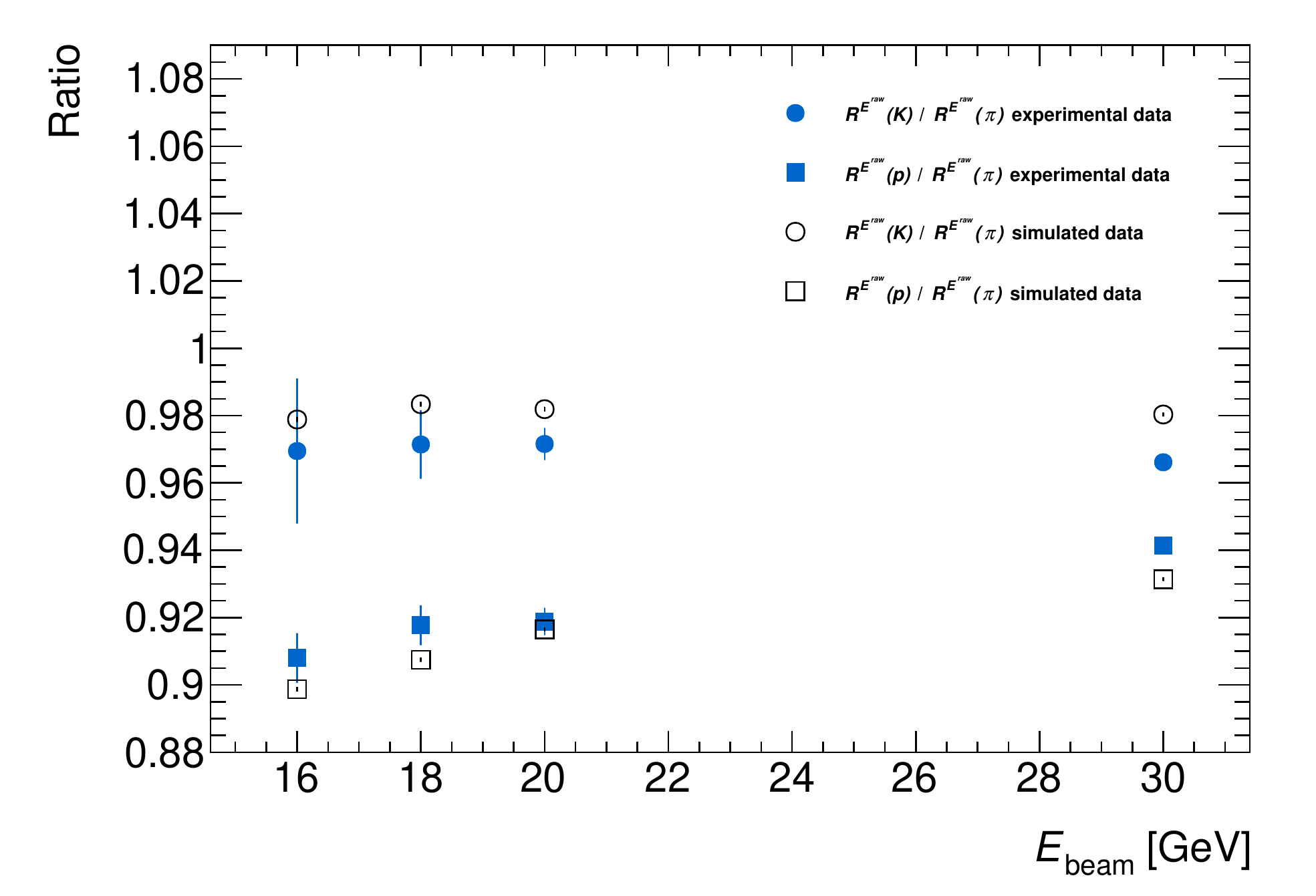}}
\subfloat[]{\includegraphics[width=4.3cm,clip]{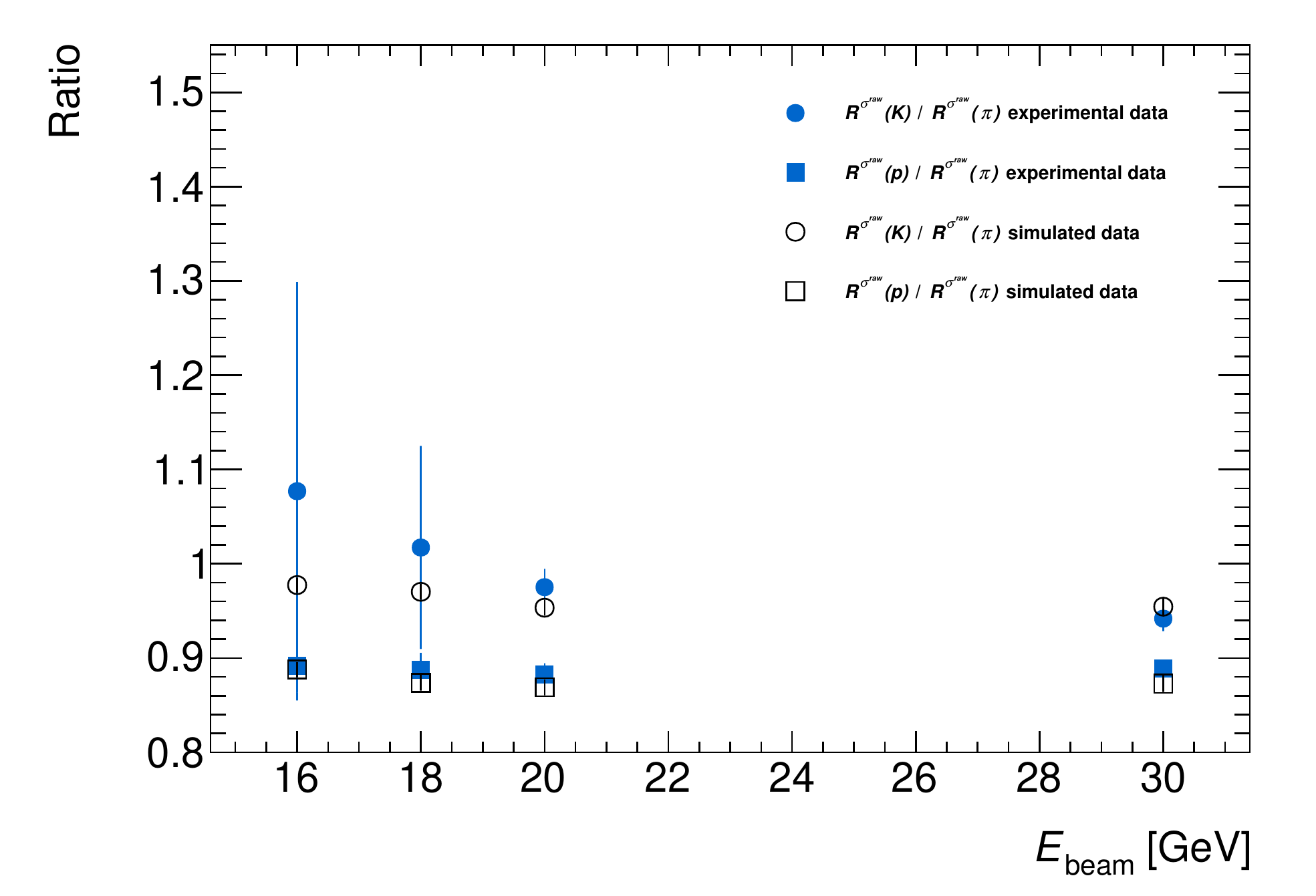}}
\caption{(a) Ratios of the kaon and proton responses over the pion ones as a function of $E_\text{beam}$. The blue dots (black empty circles) show the ratios $R^{\langle E^\text{raw}\rangle}$($K$)⁄$R^{\langle E^\text{raw}\rangle}$($\pi$) obtained using experimental (simulated) data. The blue full (black empty) squares show the ratios  $R^{\langle E^\text{raw}\rangle}$($p$)⁄$R^{\langle E^\text{raw}\rangle}$($\pi$) obtained using experimental (simulated) data. 
%$R^{E^{raw}}$($p$)⁄$R^{E^{raw}}$($K$) (green) 
(b) Ratios of the kaon and proton resolutions over the pion ones as a function of $E_\text{beam}$. The blue dots (black empty circles) show the ratios $R^{\sigma^\text{raw}}$($K$)/$R^{\sigma^\text{raw}}$($\pi$) obtained using experimental (simulated) data. The blue full (black empty) squares show the ratios  $R^{\sigma^\text{raw}}$($p$)⁄$R^{\sigma^\text{raw}}$($\pi$) obtained using experimental (simulated) data.  In the case of experimental results uncertainties include statistical and systematic effects combined in quadrature. In the case of simulated results only statistical uncertainty are reported.
}
\label{fig:Response ratios}       % Give a unique label
\end{figure}

The values of the ratios
\begin{equation}
\frac{R^{\langle E^\text{raw}\rangle(K)}}{R^{\langle E^\text{raw}\rangle(\pi)}}
\label{eq:Eraw K/pi}
\end{equation}
\begin{equation}
\frac{R^{\sigma^\text{raw}(K)}}{R^{\sigma^\text{raw}(\pi)}}
\label{eq:sigma K/pi}
\end{equation}
\begin{equation}
\frac{R^{\langle E^\text{raw}\rangle(p)}}{R^{\langle E^\text{raw}\rangle(\pi)}}
\label{eq:Eraw p/pi}
\end{equation}
\begin{equation}
\frac{R^{\sigma^\text{raw}(p)}}{R^{\sigma^\text{raw}(\pi)}}
\label{eq:sigma p/pi}
\end{equation}
obtained using experimental and simulated data are reported in Table~\ref{tab:K/pion and p/K}. The statistical (first value) and the systematic (second value) uncertainties are shown separately in the case of experimental results. The systematic uncertainty was obtained combining in quadrature the contribution of the seven sources of systematic uncertainties discussed in Section~\ref{subsec:hadron_response_ratio} . The uncertainty on the scale of the reconstructed cell energy, $C_\text{c}^\text{EM}$,  affects in a correlated way the reconstruction of the energy deposited in the modules by pions, kaons and protons. It follow that its effects on the energy response ratio determinations are negligible. In the case of simulated data only statistical uncertainties are reported. The determinations are also shown as a function of $E_\text{beam}$ in Figure~\ref{fig:Response ratios}. In the case of results obtained analyzing experimental data, the error bars were obtained combining in quadrature statistical and systematic uncertainties. In the case of results obtained analyzing simulated data only statistical uncertainty are shown. 
% ------------
\begin{table}[ht]
\centering
\caption{Values of the ratios (\ref{eq:Eraw K/pi})--(\ref{eq:sigma p/pi}) obtained using experimental and simulated data produced by particles with $E_\text{beam}$ equal to 16, 18, 20 and 30 GeV. In the case of experimental determinations statistical (first value) and correlated systematic uncertainties (second value) are reported separately. Only statistical errors affect the MC determinations. 
%\textcolor {red} {Put statistic and systematic errors separately}
}
\label{tab:K/pion and p/K}       % Give a unique label

\begin{tabular}{	|c|c|c|} 
\hline
\multicolumn{3}{|c|} {$R^{\langle E^\text{raw}\rangle}$($K$)⁄$R^{\langle E^\text{raw}\rangle}$($\pi$)}   \\ 
\hline
$E_\text{beam}$[GeV] & Experimental data  & Simulation data  \\ \hline
16 & 0.9694$\pm$0.0199$\pm$0.0081 & 0.9788$\pm$0.0007 \\ \hline
18 & 0.9714$\pm$0.0080$\pm$0.0062 & 0.9833$\pm$0.0007  \\ \hline
20 & 0.9715$\pm$0.0021$\pm$0.0042 & 0.9819$\pm$0.0007 \\ \hline
30 & 0.9661$\pm$0.0014$\pm$0.0021 & 0.9803$\pm$0.0006 \\ \hline 
\hline
\multicolumn{3}{|c|} {$R^{\langle E^\text{raw}\rangle}$($p$)⁄$R^{\langle E^\text{raw}\rangle}$($\pi$)}   \\ 
\hline
$E_\text{beam}$[GeV] & Experimental data  & Simulation data   \\ \hline
16 & 0.9079$\pm$0.0008$\pm$0.0073 & 0.8987$\pm$0.0007  \\ \hline
18 & 0.9177$\pm$0.0006$\pm$0.0059 & 0.9075$\pm$0.0006  \\ \hline
20 & 0.9188$\pm$0.0004$\pm$0.0040 & 0.9164$\pm$0.0006 \\ \hline
30 & 0.9414$\pm$0.0004$\pm$0.0020 & 0.9314$\pm$0.0005 \\ \hline 
\hline
\multicolumn{3}{|c|} {$R^{\sigma^\text{raw}}$  ($K$)⁄$R^{\sigma^\text{raw}}$($\pi$)}   \\ 
\hline
$E_\text{beam}$[GeV] & Experimental data  & Simulation data   \\ \hline
16 & 1.0769$\pm$0.2195$\pm$0.0331 & 0.9774$\pm$0.0080 \\ \hline
18 & 1.0172$\pm$0.1059$\pm$0.0193 & 0.9702$\pm$0.0084 \\ \hline
20 & 0.9752$\pm$0.0147$\pm$0.0127 & 0.9533$\pm$0.0084\\ \hline
30 & 0.9418$\pm$0.0118$\pm$0.0061 & 0.9544$\pm$0.0088\\ \hline 
\hline
\multicolumn{3}{|c|} {$R^{\sigma^\text{raw}}$  ($p$)⁄$R^{\sigma^\text{raw}}$($\pi$)}   \\ 
\hline
$E_\text{beam}$[GeV] & Experimental data  & Simulation data   \\ \hline
16 & 0.8918$\pm$0.0074$\pm$0.0274 & 0.8876$\pm$0.0072  \\ \hline
18 & 0.8876$\pm$0.0059$\pm$0.0168 & 0.8738$\pm$0.0075  \\ \hline
20 & 0.8824$\pm$0.0029$\pm$0.0115 & 0.8689$\pm$0.0076 \\ \hline
30 & 0.8887$\pm$0.0033$\pm$0.0057 & 0.8727$\pm$0.0079\\ \hline 
\end{tabular}
%\begin{tabular}{	|c|c|c|c|c|} 
%\hline
%$E_\text{beam}$ & \multicolumn{2}{c|} {$R^{\langle E^\text{raw}\rangle}$($K$)⁄$R^{\langle E^\text{raw}\rangle}$($\pi$)}  & \multicolumn{2}{c|}{$R^{\sigma^\text{raw}}$  ($K$)⁄$R^{\sigma^\text{raw}}$($\pi$)} \\ \cline {2-5}
%[GeV] & Exp. Data  & Sim. Data & Exp. Data & Sim. Data  \\ \hline
%16 & 0.9694$\pm$0.0199$\pm$0.0081 & 0.9788$\pm$0.0007 & 1.0769$\pm$0.2195$\pm$0.0331 & 0.9774$\pm$0.0080 \\ \hline
%18 & 0.9714$\pm$0.0080$\pm$0.0062 & 0.9833$\pm$0.0007 & 1.0172$\pm$0.1059$\pm$0.0193 & 0.9702$\pm$0.0084 \\ \hline
%20 & 0.9715$\pm$0.0021$\pm$0.0042 & 0.9819$\pm$0.0007 & 0.9752$\pm$0.0147$\pm$0.0127 & 0.9533$\pm$0.0084\\ \hline
%30 & 0.9661$\pm$0.0014$\pm$0.0021 & 0.9803$\pm$0.0006 & 0.9418$\pm$0.0118$\pm$0.0061 & 0.9544$\pm$0.0088\\ \hline \hline
%$E_\text{beam}$ & \multicolumn{2}{c|} {$R^{\langle E^\text{raw}\rangle}$($p$)⁄$R^{\langle E^\text{raw}\rangle}$($\pi$)}  & \multicolumn{2}{c|}{$R^{\sigma^\text{raw}}$  ($p$)⁄$R^{\sigma^\text{raw}}$($\pi$)} \\ \cline {2-5}
%[GeV] & Exp. Data  & Sim. Data & Exp. Data & Sim. Data  \\ \hline
%16 & 0.9079$\pm$0.0008$\pm$0.0073 & 0.8987$\pm$0.0007 & 0.8918$\pm$0.0074$\pm$0.0274 & 0.8876$\pm$0.0072  \\ \hline
%18 & 0.9177$\pm$0.0006$\pm$0.0059 & 0.9075$\pm$0.0006 & 0.8876$\pm$0.0059$\pm$0.0168 & 0.8738$\pm$0.0075  \\ \hline
%20 & 0.9188$\pm$0.0004$\pm$0.0040 & 0.9164$\pm$0.0006 & 0.8824$\pm$0.0029$\pm$0.0115 & 0.8689$\pm$0.0076 \\ \hline
%30 & 0.9414$\pm$0.0004$\pm$0.0020 & 0.9314$\pm$0.0005 & 0.8887$\pm$0.0033$\pm$0.0057 & 0.8727$\pm$0.0079\\ \hline 
%\end{tabular}
%\end{center}
\end{table}

%In the considered beam energy range, the ratio of the energy responses of kaons and pions is constant and equal to about 0.97. The ratios of the energy responses of protons and pions range from 0.91 at $E_\text{beam}$~=16~GeV to 0.94 at $E_\text{beam}$~=~30 GeV. The differences between the values obtained using experimental and simulated data are smaller than 1\%. The ratios of the energy resolutions of kaons and pions range from 1.01 at $E_\text{beam}$~=~16 GeV to 0.94 at $E_\text{beam}$ =~30~GeV. The ratio of the energy response of protons and pions is constant and equal to about 0.89. The differences between the values obtained using experimental and simulated data are smaller than 1.5\%.

In the considered $E_\text{beam}$ range, the measured ratios of the kaon over pion energy responses is constant with a weighted average equal to 0.967 $\pm$ 0.002 (-0.014). In parenthesis are reported the differences with the determinations obtained using simulated data. The rations of the energy responses of protons and pions range between 0.908 $\pm$ 0.008 (+0.009) at $E_\text{beam}$ = 16~GeV to 0.941 $\pm$ 0.001 (+0.010) at $E_\text{beam}$ = 30~GeV. The values of the ratios of the energy resolution determinations are constants. The weighted averages values are $R^{\sigma^\text{raw}}$($K$)⁄$R^{\sigma^\text{raw}}$($\pi$) = 0.95 $\pm$ 0.01 (-0.011) and \\ $R^{\sigma^\text{raw}}$($p$)⁄$R^{\sigma^\text{raw}}$($\pi$)  = 0.888 $\pm$ 0.005 (+0.011).

The results allow an extension down to 16 GeV of previous determinations of the ratios of the energy responses and of the resolutions of protons and pions obtained by ATLAS Collaboration using beams with energy above 50 GeV~\cite{production_modules}. The response to protons was also reported systematically lower than that of negative or positive pions in Ref.~\cite{Abdullin:2009zz}. The measurements were performed in the momentum range from 3 to 300 GeV/c. In the same paper results concerning the response of charged kaons and anti-protons in the momentum range below 9 GeV/c are reported. 
%\FloatBarrier
%------------
%\FloatBarrier
%------------
\section {Comaparison with hadronic cascade model predictions}
\label{sec:cascade_models}  

\subsection{Parametrization of the energy response normalized to incident beam energy as a function of the beam energy}
\label{subsec:response_parametrization}

The calorimeter response for pions, kaons and protons can be described in terms of the calorimeter

non-compensation and leading particle effects~\cite{Wigmans:2000vf}.~The hadron energy response normalized to incident beam energy as a function of the beam energy can be parametrized according to

\begin{equation}
R^{\langle E^\text{raw} \rangle} = (1 - F_\text{h}) + F_\text{h} \times (\frac{e} {h})^{-1}
\label{eq:fractional_response}
\end{equation}

where $F_\text{h}$ represents the non-electromagnetic energy component of showers induced by incident hadrons of energy $E_\text{beam}$ and $e/h$ is the ratio between the responses to the purely EM and hadronic components of showers. 
The measurements allow a determination of the ratios of the non-electromagnetic energy component of showers induced by incident pions ($F_\text{h}(\pi)$), kaons ($F_\text{h}(K)$) and protons ($F_\text{h}(p)$) for the same value of $E_\text{beam}$. Using Eq.~(\ref{eq:fractional_response}) one obtains

\begin{equation}
\frac{F_\text{h} (K)} {F_\text{h} (\pi)} = \frac {1-R^{\langle E^\text{raw} \rangle}(K)} {1-R^{\langle E^\text{raw} \rangle}(\pi)}
\label{eq:ratio_Fh(K)_over_Fh(pi)}
\end{equation}

and

\begin{equation}
\frac{F_\text{h} (p)} {F_\text{h} (\pi)} = \frac {1-R^{\langle E^\text{raw} \rangle}(p)} {1-R^{\langle E^\text{raw} \rangle}(\pi)}
\label{eq:ratio_Fh(p)_over_Fh(pi)}
\end{equation}

The determinations obtained using experimental and simulated data are reported in Table~\ref{tab:FK/Fpi and Fp/Fpi}. The statistical (first value) and the systematic (second value) uncertainties are shown separately in the case of experimental results. The systematic uncertainties were obtained combining in quadrature the effects of the seven sources discussed in Section~\ref{subsec:hadron_response_ratio}. In the case of simulated  data, only statistical uncertainties are reported. Data show constant ratios $F_\text{h} (K)/F_\text{h}(\pi)$. The weighted average numerical value is 1.13 $\pm$ 0.01 (1.072 $\pm$ 0.001). The values in parenthesis were obtained analysing simulated data. The ratio $F_\text{h} (p)/F_\text{h}(\pi)$ decreases from 1.351 $\pm$ 0.04  (1.361 $\pm$ 0.003) at $E_\text{beam}$~16~GeV to 1.24 $\pm$ 0.01 (1.281 $\pm$ 0.003) at $E_\text{beam}$~30~GeV.    
%A constant value of the ratio $F_\text{h} (K)/F_\text{h}(\pi)$ around 1.12 (1.07) was obtained. The ratio $F_\text{h} (p)/F_\text{h}(\pi)$ decreases from 1.35 (1.36) at $E_\text{beam}$~16~GeV to 1.24 (1.28) at $E_\text{beam}$~30~GeV. The values in parenthesis were obtained using simulated data. 
%------------
\begin{table}[ht]
\centering
\caption{Values of the ratios $F_\text{h}(K)/F_\text{h} (\pi)$ and $F_\text{h}(p)/F_\text{h} (\pi)$  obtained using experimental and simulated data for the four values of the beam energies $E_\text{beam}$. In the case of experimental determinations statistical (first value) and correlated systematic (second value) uncertainties are reported separately. Only statistical uncertainties affects the MC determinations.
%\textcolor {red} {Put statistic and systematic errors separately}
}
\label{tab:FK/Fpi and Fp/Fpi}       % Give a unique label

\begin{tabular}{|c|c|c|} 
\hline
 \multicolumn{3}{|c|} {$F_\text{h}(K)/F_\text{h} (\pi)$} \\ 
 \hline
$E_\text{beam}$~[GeV] & Experimental Data  & Simulated Data  \\
\hline
16 & 1.1165$\pm$0.0761$\pm$0.0354 & 1.0756$\pm$0.0028 \\ \hline
18 & 1.1102$\pm$0.0309$\pm$0.0276 & 1.0608$\pm$0.0028  \\ \hline
20 & 1.1101$\pm$0.0083$\pm$0.0186 & 1.0669$\pm$0.0028 \\ \hline
30 & 1.1371$\pm$0.0059$\pm$0.0098 & 1.0805$\pm$0.0027 \\ \hline
\hline
\multicolumn{3}{|c|}{$F_\text{h}(p)/F_\text{h} (\pi)$} \\
\hline
$E_\text{beam}$~[GeV] & Experimental Data  & Simulated Data  \\
\hline
16 & 1.3512$\pm$0.0039$\pm$0.0418 & 1.3617$\pm$0.0034\\ \hline
18 & 1.3173$\pm$0.0030$\pm$0.0326 & 1.3382$\pm$0.0033 \\ \hline
20 & 1.3144$\pm$0.0019$\pm$0.0217 & 1.3091$\pm$0.0033\\ \hline
30 & 1.2373$\pm$0.0018$\pm$0.0107 & 1.2807$\pm$0.0031\\ \hline
\end{tabular}
%\begin{tabular}{|c|c|c|c|c|} 
%\hline
%$E_\text{beam}$ & \multicolumn{2}{c|} {$F_\text{h}(K)/F_\text{h} (\pi)$}  & \multicolumn{2}{c|}{$F_\text{h}(p)/F_\text{h} (\pi)$} \\ \cline {2-5}
%[GeV] & Exp. Data  & Sim. Data & Exp. Data & Sim. Data  \\ \hline
%16 & 1.1165$\pm$0.0761$\pm$0.0354 & 1.0756$\pm$0.0028 & 1.3512$\pm$0.0039$\pm$0.0418 & 1.3617$\pm$0.0034\\ \hline
%18 & 1.1102$\pm$0.0309$\pm$0.0276 & 1.0608$\pm$0.0028 & 1.3173$\pm$0.0030$\pm$0.0326 & 1.3382$\pm$0.0033 \\ \hline
%20 & 1.1101$\pm$0.0083$\pm$0.0186 & 1.0669$\pm$0.0028 & 1.3144$\pm$0.0019$\pm$0.0217 & 1.3091$\pm$0.0033\\ \hline
%30 & 1.1371$\pm$0.0059$\pm$0.0098 & 1.0805$\pm$0.0027 & 1.2373$\pm$0.0018$\pm$0.0107 & 1.2807$\pm$0.0031\\ \hline
%\end{tabular}
%\end{center}
\end{table}

%The results of the analysis of pion and kaon  experimental (simulated) data shows a constant value of the ratio $F_\text{h} (K)/F_\text{h}(\pi)$ around 1.12 (1.07) in the range 16 GeV < $E_\text{beam}$ < 30 GeV. The analysis of pion and proton experimental (simulated) data shows a decrease of the ratio $F_\text{h} (p)/F_\text{h} (\pi)$ from 1.35 to 1.24 (1.36 to 1.28). 
The ratio $F_\text{h} (p)/F_\text{h} (\pi)$, as obtained in Refs.~\cite{GROOM2007633} and \cite{GROOM2008638} from the copper/quartz-fiber calorimeter data~\cite{AKCHURIN1998380}, varies from 1.22 at 200 GeV to 1.15 at 370 GeV. In Ref.~\cite{GABRIEL1994336}, a constant value of $F_\text{h} (p)/F_\text{h}(\pi)$ in the range between 1.15 and 1.20 is predicted.

The determinations of $F_\text{h}(K)/F_\text{h} (\pi)$ and

\begin{figure}[ht]
\centering
\subfloat[]{\includegraphics[width=4.3cm,clip]{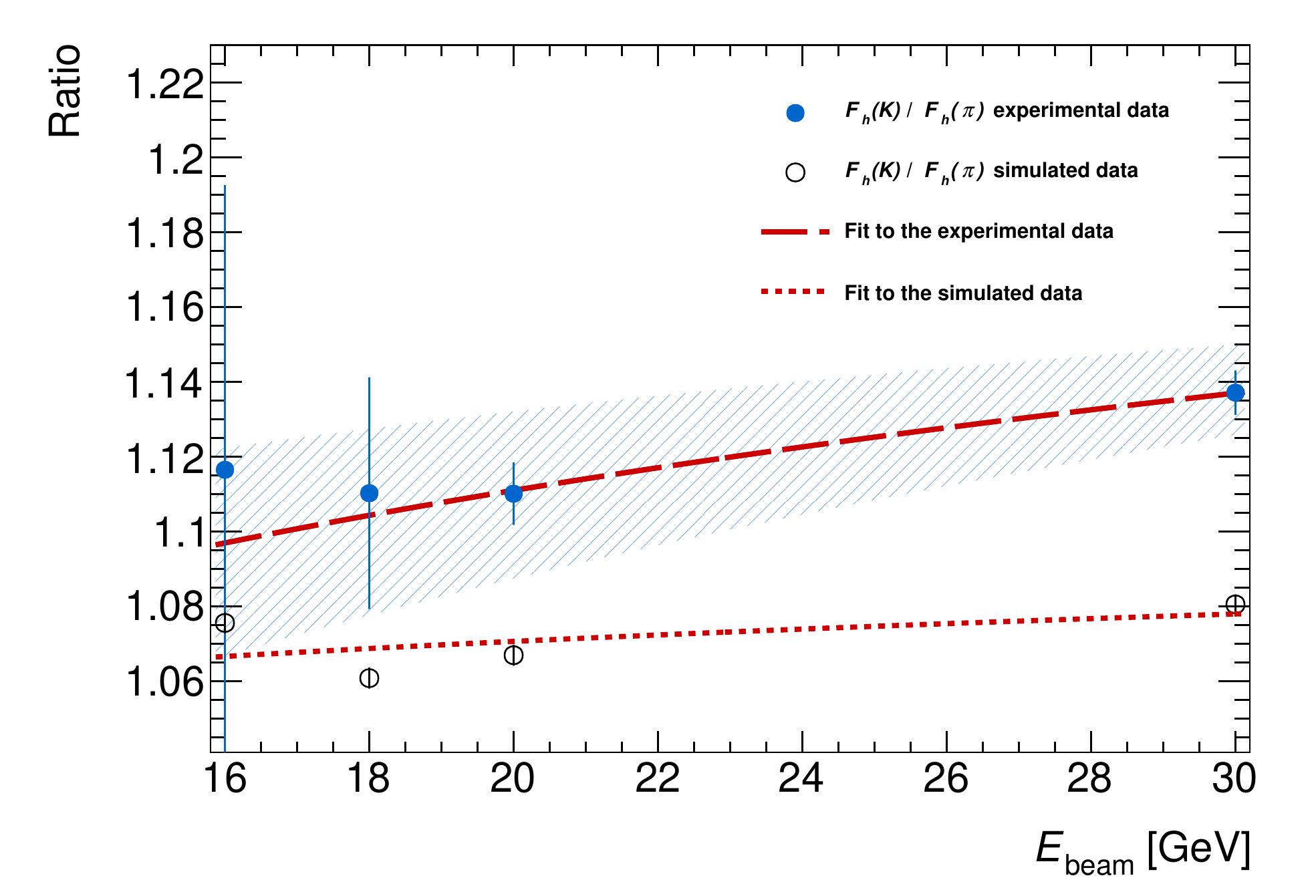}}
\subfloat[]{\includegraphics[width=4.3cm,clip]{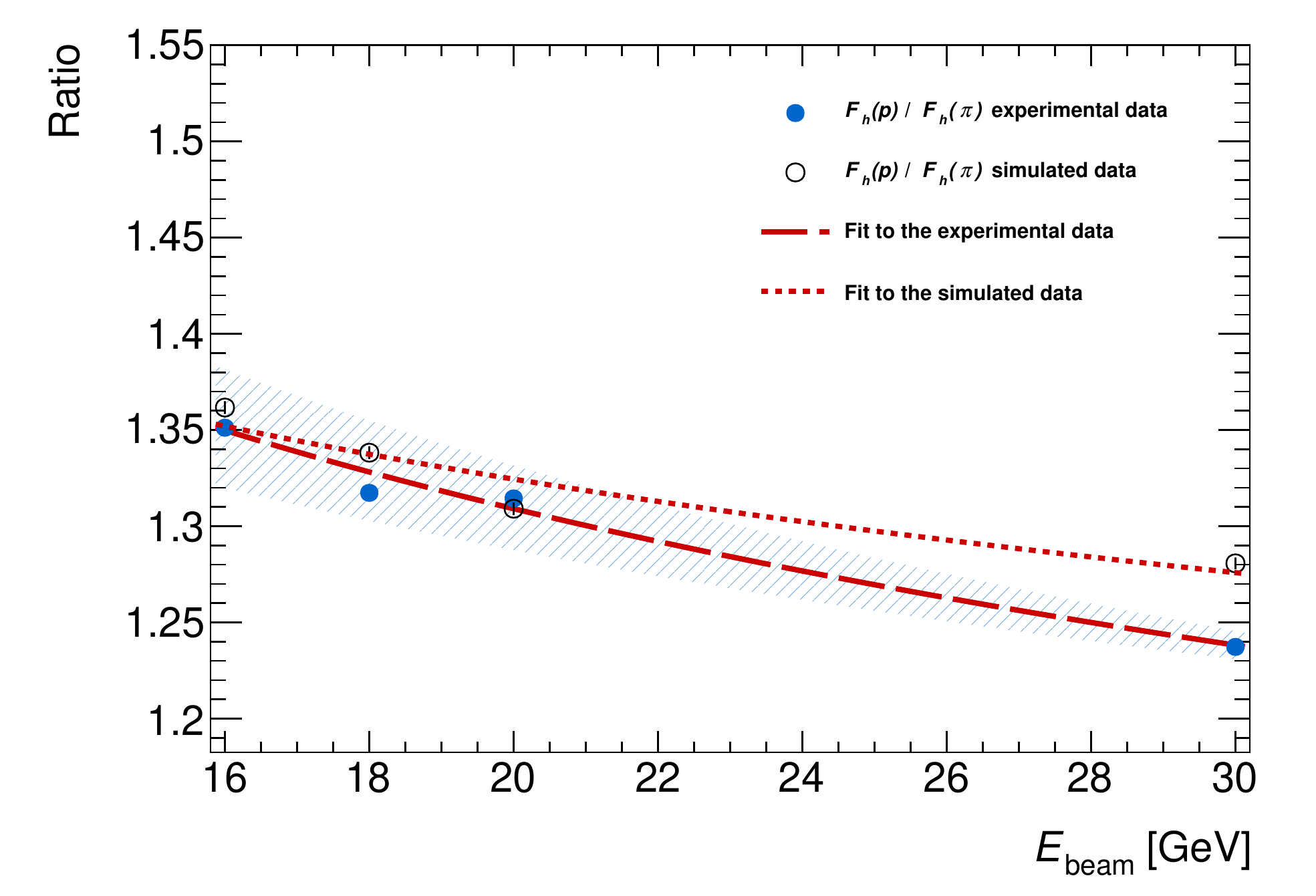}}
\caption{(a) $F_\text{h} (K)/F_\text{h} (\pi)$ as a function of $E_\text{beam}$
obtained using experimental, blue dots, and simulated data, black empty circles. 
(b) $F_\text{h} (p)/F_\text{h} (\pi)$ as a function of $E_\text{beam}$ 
obtained using experimental, blue dots, and simulated data, black empty circles. In the case of experimental results uncertainties include statistical and systematic effects combined in quadrature. In the case of simulated results only statistical uncertainty are shown. The dashed (dotted) red curves are fits of the functions~(\ref{eq:parameterization ratio_Fh(K)_over_Fh(pi)}) and (\ref{eq:parameterization ratio_Fh(p)_over_Fh(pi)}) to the experimental (simulated) data points. In case of experimental determinations the dashed red strips display the correlated systematic uncertainties.}
\label{fig:FK/Fpi and Fp/Fpi}       % Give a unique label
\end{figure}

$F_\text{h}(p)/F_\text{h} (\pi)$ as a function of $E_\text{beam}$ are also reported in the histograms of Figure~\ref{fig:FK/Fpi and Fp/Fpi} (a) and (b) respectively. In the case of experimental results, statistical and systematic uncertainties are combined in quadrature. In the case of simulated results only statistical uncertainty are shown.

In Groom's parametrization,~\cite{GABRIEL1994336}, \cite{GROOM2007633} and \cite{GROOM2008638}, one has

\begin{equation}
F_\text{h} = (\frac{E_\text{beam}}{E_0})^{m-1}
\label{eq:Groom}
\end{equation}

where the quantity $E_0$ is the energy at which multiple pion production becomes significant and the parameter $m$ describes the relation between the average multiplicity of secondary particles produced in the collision and the fraction of energy going into $\pi^0$'s in one collision.
%must be determined empirically for a given calorimeter. 
One obtains

\begin{equation}
R^{\langle E^\text{raw} \rangle} = 1 + \frac{1} {(E_0)^{m-1}}[(\frac{e} {h})^{-1}-1] (E_\text{beam})^{m-1} .
\label{eq:parametrization_results}
\end{equation}

and

%Using Eq.~(\ref{eq:Groom}) one obtains
\begin{equation}
\frac{F_\text{h} (K)} {F_\text{h} (\pi)} = \frac {E_0(\pi)^{m(\pi)-1}} {E_0(K)^{m(K)-1}} \times (E_\text{beam})^{m(K)-m(\pi)}
\label{eq:parameterization ratio_Fh(K)_over_Fh(pi)}
\end{equation}

%and
\begin{equation}
\frac{F_\text{h} (p)} {F_\text{h} (\pi)} = \frac {E_0(\pi)^{m(\pi)-1}} {E_0(p)^{m(p)-1}} \times (E_\text{beam})^{m(p)-m(\pi)}
\label{eq:parameterization ratio_Fh(p)_over_Fh(pi)}
\end{equation}

%\FloatBarrier
%------------
%\FloatBarrier
% ------------
Fits of Eq.~(\ref{eq:parameterization ratio_Fh(K)_over_Fh(pi)}) to the histograms of Figure~\ref{fig:FK/Fpi and Fp/Fpi} (a) and of Eq.~(\ref{eq:parameterization ratio_Fh(p)_over_Fh(pi)}) to the histograms of Figure~\ref{fig:FK/Fpi and Fp/Fpi} (b) allow a determination of 
\begin{equation}
B_{K/\pi} = \frac {E_0(\pi)^{m(\pi)-1}}{E_0(K)^{m(K)-1}},
\label{eq:A'_k}
\end{equation}
\begin{equation}
C_{K/\pi} = m(K)-m(\pi)
\label{eq:B'_k}
\end{equation}

and

\begin{equation}
B_{p/\pi} = \frac {E_0(\pi)^{m(\pi)-1}}{E_0(p)^{m(p)-1}},
\label{eq:A'_p}
\end{equation}
\begin{equation}
C_{p/\pi} = m(p)-m(\pi)
\label{eq:B'_p}
\end{equation}
respectively.
The fit curves to the experimental and simulated data are show in the figure. The strips display the correlated systematic uncertainties $\Delta_\text{syst.} F_\text{h} (K)/F_\text{h} (\pi)$ (Figure~\ref{fig:FK/Fpi and Fp/Fpi}~(a)) and $\Delta_\text{syst.} F_\text{h} (p)/F_\text{h} (\pi)$ (Fig.~\ref{fig:FK/Fpi and Fp/Fpi}~(b)). They are defined by the curves obtained fitting Eq.~(\ref{eq:parameterization ratio_Fh(K)_over_Fh(pi)}) (Eq.~(\ref{eq:parameterization ratio_Fh(p)_over_Fh(pi)})) to the points \\ $F_\text{h} (K)/F_\text{h} (\pi)\pm\Delta_\text{syst.} F_\text{h} (K)/F_\text{h} (\pi)$ \\ ($F_\text{h} (p)/F_\text{h} (\pi)\pm\Delta_\text{syst.} F_\text{h} (p)/F_\text{h} (\pi)$). All the fits were performed using as uncertainties the statistical uncertainties of the determinations. The values of the parameters obtained in the fits are reported in Table~\ref{tab:Fh ratios parametrization}. The first uncertainty value is the statistical uncertainty. It corresponds to the square root of the corresponding diagonal term of the fit error matrix. The systematic uncertainty (second uncertainty value) is  equal to half of the differences of the determinations obtained fitting Eq.~(\ref{eq:parameterization ratio_Fh(K)_over_Fh(pi)}) to the points \\ $F_\text{h} (K)/F_\text{h} (\pi)\pm\Delta_\text{syst.} F_\text{h} (K)/F_\text{h} (\pi)$ and Eq.~(\ref{eq:parameterization ratio_Fh(p)_over_Fh(pi)}) to the points $F_\text{h} (p)/F_\text{h} (\pi)\pm\Delta_\text{syst.} F_\text{h} (p)/F_\text{h} (\pi)$. In the table, the $\chi^2$ probability values of the fits performed to the central points are reported. In the case of the fits to simulated data the probabilities are very small.

% ---------
\begin{table}[ht]
\centering
\caption{Values of the parameters $B_{K/\pi}$ (Eq.~(\ref{eq:A'_k})) and $C_{K/\pi}$ (Eq.~(\ref{eq:B'_k})) obtained fitting Eq.~(\ref{eq:parameterization ratio_Fh(K)_over_Fh(pi)}) to the experimental and simulated values of $F_\text{h} (K)/F_\text{h} (\pi)$ as a function of $E_\text{beam}$ shown Figure~\ref{fig:FK/Fpi and Fp/Fpi} (a). Values of the parameters $B_{p/\pi}$ (Eq.~(\ref{eq:A'_p})) and $C_{p/\pi}$ (Eq.~(\ref{eq:B'_p})) obtained fitting Eq.~(\ref{eq:parameterization ratio_Fh(p)_over_Fh(pi)}) to the experimental and simulated values of $F_\text{h} (p)/F_\text{h} (\pi)$ as a function of $E_\text{beam}$ shown Figure~\ref{fig:FK/Fpi and Fp/Fpi} (b). In the case of experimental data results, statistical and systematic uncertainties are reported. Only statistical uncertainties appear in the case of simulated data results. The $\chi^2$ probability values of the fits are reported.
%Previous pion results reported in Ref.~\cite{production_modules} are also shown.
}
\label{tab:Fh ratios parametrization}       % Give a unique label
% For LaTeX tables you can use
\begin{tabular}{	|c|c|c|}
\hline
& Experimental data & Simulated data  \\ \hline
$B_{K/\pi}$ & 0.936$\pm$0.065$\pm$0.083 & 1.014$\pm$0.004 \\ \hline
$C_{K/\pi}$ & 0.057$\pm$0.021$\pm$0.023 & 0.0182$\pm$0.0014 \\ \hline
$\chi^2$ prob. & 0.944 &  $1.767\times10^{-7}$ \\ \hline
\hline
& Experimental data & Simulated data  \\ \hline
$B_{p/\pi}$ & 1.975$\pm$0.023$\pm$0.192 & 1.735$\pm$0.0074 \\ \hline
$C_{p/\pi}$ & -0.137$\pm$0.004$\pm$0.027 & -0.090$\pm$0.0014 \\ \hline
$\chi^2$ prob. & 0.311 &  $8.17910\times10^{-7}$ \\ \hline
\end{tabular}
%\begin{tabular}{	|c|c|c|} 
%\hline 
%\multicolumn{3}{|c|}{Experimental data} \\ \hline
%$B_{K/\pi}$ & $C_{K/\pi}$ & %$\chi^2$ prob. \\ \hline
% 0.936$\pm$0.065$\pm$0.083 & 0.057$\pm$0.021$\pm$0.023 & 0.944  \\ \hline
% \hline
%$B_{p/\pi}$ & $C_{p/\pi}$ & $\chi^2$ prob. \\ \hline
%1.975$\pm$0.023$\pm$0.192 & -0.137$\pm$0.004$\pm$0.027 & 0.311 %\\ \hline
%\hline
%\multicolumn{3}{|c|}{Simulated data} \\ \hline
%$B_{K/\pi}$ & $C_{K/\pi}$ & $\chi^2$ prob. \\ \hline
%1.014$\pm$0.004 & 0.0182$\pm$0.0014 & $1.767\times10^{-7}$ \\ \hline
%\hline
%$B_{p/\pi}$ & $C_{p/\pi}$ & $\chi^2$ prob. \\ \hline
%1.735$\pm$0.0074 & -0.090$\pm$0.0014 & $8.17910\times10^{-7}$ \\ \hline
%\end{tabular}
%\end{center}
\end{table}

The values of $B_{K/\pi}$ and $C_{K/\pi}$ obtained using experimental and simulated data agree within two sigmas. The values of $B_{p/\pi}$ and $C_{p/\pi}$ obtained using experimental and simulated data differ significantly. 

Fits of Eq.~(\ref{eq:parametrization_results}) to the determinations of $R^{\langle E^\text{raw} \rangle}$ as a  function of $E_\text{beam}$ (see Figure~\ref{fig:Response means})  allow a determination~\cite{GROOM2007633} of $m$ and 

\begin{equation}
A = \frac{1} {(E_0)^{m-1}}[(\frac{e} {h})^{-1}-1] .
\label{eq:A}
\end{equation}

The fit curves
%Fig.~\ref{fig:Response means} 
%were obtained fitting Eq.~(\ref{eq:parametrization_results}) 
to the experimental and simulated determinations 
%of $R^{E^\text{raw}}$ as a function of $E_\text{beam}$
are reported in the figure. The strips display correlated systematic uncertainties $\Delta R_\text{syst.}^{E^\text{raw}}$. They are bounded by the curves obtained fitting Eq.~(\ref{eq:parametrization_results}) to the points $R^{E^\text{raw}}\pm\Delta R_\text{syst.}^{E^\text{raw}}$. All the fits were performed using as uncertainties the statistical uncertainties of the determinations. The obtained values of $A$ and $m$ are reported in Table~\ref{tab:Response parametrization}. The first uncertainty value is the statistical uncertainty. It corresponds to the square root of the diagonal term of the error matrix. The systematic uncertainty (second uncertainty value) is  equal to half of the differences of the determinations obtained fitting Eq.~(\ref{eq:parametrization_results}) to the points $R^{E^\text{raw}}\pm\Delta R_\text{syst.}^{E^\text{raw}}$. In the table the $\chi^2$ probability values of the fits performed to the central points are reported. In the case of the fits to kaon and proton simulated data the probabilities are very small. 
%The values of $A$ ($m$) obtained analyzing pions and kaons are consistent and equal to about -0.26 (0.92). The values obtained analyzing proton data are smaller. The results obtained analyzing simulated data are smaller and show the same features.  
%The values of $A$ ($m$) obtained analyzing experimental data are about 25\% (15\%) larger then the ones obtained analysing simulated data.

% ---------
\begin{table}[ht]
\centering
\caption{Values of the parameters $A$ (Eq.~(\ref{eq:A})) and $m$ obtained fitting Eq.~(\ref{eq:parametrization_results}) to the experimental and simulated energy response normalized to incident beam energy, $R^{E^\text{raw}}$, as a function of $E_\text{beam}$. The fit functions are overlapped to the determinations in Figure~\ref{fig:Response means}. In the case of experimental data results statistical and systematic uncertainties are reported. Only statistical uncertainties appear in the case of simulated data results. The $\chi^2$ probability values of the fits are reported. 
%Previous pion results reported in Ref.~\cite{production_modules} are also shown.
}
\label{tab:Response parametrization}       % Give a unique label
% For LaTeX tables you can use 
\begin{tabular}{	|c|c|c|}
\hline 
\multicolumn{3}{|c|}{Pions} \\ \hline
& Experimental data & Simulated data  \\ \hline
$A$ & -0.2612$\pm$0.0019$\pm$0.0165 & -0.3557$\pm$0.0052
\\ \hline
$m$ & 0.9187$\pm$0.0023$\pm$0.0041 & 0.8258$\pm$0.0048
\\ \hline
$\chi^2$ prob. & 0.004 & 0.238
\\ \hline
\hline
\multicolumn{3}{|c|}{Kaons} \\ \hline
& Experimental data & Simulated data  \\ \hline
$A$ & -0.2481$\pm$0.0171$\pm$0.0095 & -0.3620$\pm$0.0025
\\ \hline
$m$ & 0.9715$\pm$0.0210$\pm$0.0008 & 0.8428$\pm$0.0023
\\ \hline
$\chi^2$ prob. & 0.981 & $2.634\times10^{-14}$ 
\\ \hline
\hline
\multicolumn{3}{|c|}{Protons} \\ \hline
& Experimental data & Simulated data  \\ \hline
$A$ & -0.5041$\pm$0.0044$\pm$0.0005 & -0.6208$\pm$0.0031
\\ \hline
$m$ & 0.7885$\pm$0.0029$\pm$0.0004 & 0.7338$\pm$0.0016
\\ \hline
$\chi^2$ prob. & 0.632 & $4.295\times10^{-12}$
\\ \hline
\end{tabular}
%\begin{tabular}{	|c|c|c|c|} 
%\hline 
%& \multicolumn{3}{c|}{Experimental data} \\ \hline
%& $A$ & $m$ & $\chi^2$ prob. \\ \hline
%$\pi$ &  -0.2612$\pm$0.0019$\pm$0.0165 & 0.9187$\pm$0.0023$\pm$0.0041 & 0.004 \\
%&  $\pm$0.0165 & $\pm$0.0041 & 0.004 \\ \hline
%$K$ & -0.2481$\pm$0.0171$\pm$0.0095 & 0.9715$\pm$0.0210$\pm$0.0008 & 0.981 \\ \hline
%$p$ & -0.5041$\pm$0.0044$\pm$0.0005 & 0.7885$\pm$0.0029$\pm$0.0004 & 0.632 \\ \hline \hline
%& \multicolumn{3}{c|}{Simulated data} \\ \hline
%& $A$ & $m$ & $\chi^2$ prob. \\ \hline
%$\pi$ &  -0.3557$\pm$0.0052 & 0.8258$\pm$0.0048 & 0.238 \\ \hline
%$K$ &  -0.3620$\pm$0.0025 & 0.8428$\pm$0.0023 & 0.8428$\pm$0.0023\\ \hline
%$p$ &  -0.6208$\pm$0.0031 
%\textcolor {red} {1.53 $\pm$ 0.004} 
%& 0.7338$\pm$0.0016 & $4.295\times10^{-12}$\\
%\textcolor {red} {8.77 $\pm$ 0.2}\\ 
%\hline 
%\end{tabular}
%\end{center}
\end{table}
%\FloatBarrier
% ------------
The values of $m$ obtained using pions, kaons and protons data without making any assumption on the values of $e/h$ and $E_0$ are: 0.919 $\pm$ 0.005 (0.826 $\pm$ 0.005), 0.97 $\pm$ 0.02
(0.843 $\pm$ 0.002) and 0.789 $\pm$ 0.003 (0.734 $\pm$ 0.002) respectively. The values in parenthesis were obtained using simulated data.
According to Ref.~\cite{GROOM2007633} values of $m$ around 0.87 are expected. The determinations can be compared with previous pion measurements summarized in~\cite{GROOM2007633}. 

To compare the results discussed in this paper with the ones obtained previously using pions beams with energy in the range 10-350 GeV and incident in the TileCal modules at $\eta$ = 0.35~\cite{production_modules}, Eq.~(\ref{eq:parametrization_results}) was fitted to the pion determinations fixing $E_0$ = 1~GeV. The obtained values $e/h$~=~1.3535~$\pm$~0.0304 and $m$~=~0.9187~$\pm$~0.0047 agree with the previous determination 1.33~$\pm$~0.02 and 0.85~$\pm$~0.03 respectively~\cite{production_modules}. The uncertainties include statistical and systematic uncertainties combined in quadrature.
%~\cite{production_modules}. 
%The $F_\text{h} (K)/F_\text{h} (\pi)$ ($F_\text{h} (p)/F_\text{h} (\pi)$) results as a function of $E_\text{beam}$ are also shown in Fig.~\ref{fig:FK/Fpi and Fp/Fpi} (a) (Fig.~\ref{fig:FK/Fpi and Fp/Fpi} (b)). In the case of experimental results, the error bars were obtained combining in quadrature statistical and systematic uncertainties.

%\FloatBarrier
% ------------
\subsection{Parametrization of the energy resolution as a function of the beam energy}
\label{subsec:resolution_parametrization}

The resolution of the energy measurements as a function of the beam energy $E_\text{beam}$ can be parametrized according to

\begin{equation}
R^{\sigma^\text{raw}} = \frac{a}{\sqrt{E_\text{beam}}} \oplus b .
\label{eq:fractional_resolution}
\end{equation}

where the first term describes the fluctuations on the number of particle produced in the showers, the second term 
%where $a$ is the statistical term, the constant term $b$ 
describes the non-uniformity of the cell response and the symbol $\oplus$ indicates the sum in quadrature. In the considered beam energy range the noise contribution is negligible (see Section~\ref{subsec:shower_energy}). 

The curves in Figure~\ref{fig:Response sigmas} were obtained fitting Eq.~(\ref{eq:fractional_resolution}) to the experimental and simulated determinations of $R^{\sigma^\text{raw}}$ as a function of $1/\sqrt{E_\text{beam} [\text{GeV}]}$. The strips in the figure display correlated systematic uncertainties $\Delta R_\text{syst.}^{\sigma^\text{raw}}$. They are defined by the curves obtained fitting Eq.~(\ref{eq:fractional_resolution}) to the points $R^{\sigma^\text{raw}}\pm\Delta R_\text{syst.}^{\sigma^\text{raw}}$. All the fits were performed using as uncertainties the statistical uncertainties of the determinations. The resulting values of $a$ and $b$ are reported in Table~\ref{tab:Resolution parametrization}. The statistical uncertainty (first uncertainty value) is equal to the square root of the corresponding diagonal term of the fit error matrix. The systematic uncertainty (second uncertainty value) is  equal to half of the differences of the determinations obtained fitting Eq.~(\ref{eq:fractional_resolution}) to the points $R^{\sigma^\text{raw}} + \Delta R_\text{syst.}^{\sigma^\text{raw}}$ and $R^{\sigma^\text{raw}} - \Delta R_\text{syst.}^{\sigma^\text{raw}}$. In the table the $\chi^2$ probability values of the fits performed to the central values are reported. 

%The values of statistical terms $a$ obtained analyzing pions and kaons are consistent inside the large uncertainties of about 4\%. The value obtained using protons is 14\%  smaller. The constant term $b$ is about 5\% and equal for the three particle beams. The values of $a$ and $b$ obtained analyzing pion data are consistent within about 2.6 sigmas with a previous determination~\cite{production_modules}. Analyses of simulated events produced values of $a$ 10\% smaller than the ones obtained using experimental data. The determinations of the constant terms $b$ are 30\% larger.    

% ---------
\begin{table}[ht]
\centering
\caption{Values of the parameters $a$ and $b$ obtained fitting Eq.~(\ref{eq:fractional_resolution}) to the experimental and simulated fractional resolution values $R^{\sigma^\text{raw}}$ obtained using pions ($\pi$), kaons ($K$) and prtons ($p$) as a function of 1⁄$\sqrt{E_\text{beam} [\text{GeV}] }$ (see Figure~\ref{fig:Response sigmas}). In the case of experimental data results, statistical and systematic uncertainties are reported. Only statistical uncertainties appear in the case of simulated data results. The $\chi^2$ probability values of the fits are reported. Previous pion~\cite{production_modules} results are also shown ($\pi$ old).}

\label{tab:Resolution parametrization}       % Give a unique label
% For LaTeX tables you can use 
\begin{tabular}{	|c|c|c|c|} 
\hline 
\multicolumn{4}{|c|}{Experimental data} \\ \hline
& $a$ [$\%~\text{GeV}^{-1/2}$] & $b$ [\%] & $\chi^2$ prob. \\ \hline
$\pi$ & 46.68$\pm$0.30$\pm$2.22  & 4.99$\pm$0.11$\pm$0.58 & 0.941 \\ \hline
$K$ & 49.9$\pm$2.60$\pm$2.46 & 1.78$\pm$2.78$\pm$1.03 & 0.935 \\ \hline
$p$ & 40.28$\pm$0.38$\pm$0.08 & 4.79$\pm$0.15$\pm$1.44 & 0.007 \\ \hline
$\pi$ old & 52.9$\pm$0.9 & 5.7$\pm$0.2 & -\\ \hline
\hline 
\multicolumn{4}{|c|}{Simulated data} \\ \hline
& $a$ [$\%~\text{GeV}^{-1/2}$] & $b$ [\%] & $\chi^2$ prob.  \\ \hline
Pions & 42.25$\pm$1 & 6.2$\pm$0.4 & 0.0008\\ \hline
Kaons & 42.8$\pm$0.3 & 5.3$\pm$0.1 & 0.058\\ \hline
Protons & 38.05$\pm$0.23 & 5.12$\pm$0.08 & 0.392 \\ \hline 
\end{tabular}
%\end{center}
\end{table}
%\FloatBarrier
% -------------

The values of $a$ obtained analyzing pions and kaons are consistent inside the large uncertainties of about 4\%. The value obtained using protons is 14\%  smaller. The constant term $b$ is about 5\% and equal for the three particle beams. Analyses of simulated events produced values of $a$ 10\% smaller than the ones obtained using experimental data. The determinations of the constant terms $b$ are 30\% larger.

The values of $a$ and $b$ obtained analyzing pion data are consistent within about 2.6 sigmas with the results obtained in a previous study~\cite{production_modules}.

%-------------------------------------------------------------------------------
\section{Summary and conclusions}
\label{sec:conclusion}

The results described in this paper were obtained by exposing three modules of the ATLAS Tile Calorimeter to positive pion and kaon and proton beams with energies equal to 16, 18, 20 and 30 GeV and incident at the centre of the front face of a calorimeter module cell with an angle of 14 degrees from the normal. Two Cherenkov counters in the beam line made it possible to identify pions, kaons and protons. The effects of electrons contaminating the pion samples in reconstructing the pion energy were determined by exploiting the difference of electromagnetic and hadronic shower profiles in the detector.

The main purpose of the study is to compare the measured energy of the particles with the predictions of the Geant4-based simulation program used in ATLAS to simulated jets produced in proton-proton collisions at the Large Hadron Collider.

Eleven (Nine) determinations of the twelve energy responses (resolutions) normalized to incident beam energy have a total uncertainty smaller than 1.4\% (1.9\%). In the case of kaons with $E_\text{beam}$ = 16~GeV, due to the large statistical error, the uncertainty on the determination of $R^{\langle E^\text{raw} \rangle}$, is equal to 2.4\%. The uncertainty values of the determinations of $R^{\langle \sigma^\text{raw} \rangle}$ obtained in the case of 16 GeV pion and kaon and 18 GeV kaon beams are equal to 3.1\%, 20.3\% and 10.4\% respectively.

Determinations of all the energy responses and of the pion and kaon energy resolutions obtained using experimental and simulated data agree within the uncertainties. The average of the absolute values of the differences of all the energy response measurements was found to be 1.1\% with an average total uncertainty of 1.4\%. The average difference of all the resolution measurements was found to be 3.4\%. The average total uncertainty of pion and kaon (proton) resolution measurements is 5.6\% (0.6\%).

In the considered $E_\text{beam}$ range, the measured ratios of the kaon over pion energy responses is constant with a weighted average equal to 0.967 $\pm$ 0.002 (-0.014). In parenthesis are reported the differences with the determinations obtained using simulated data. The ratios of the energy responses of protons and pions range between 0.908 $\pm$ 0.008 (+0.009) at $E_\text{beam}$ = 16~GeV to 0.941 $\pm$ 0.001 (+0.010) at $E_\text{beam}$ = 30~GeV. The values of the ratios of the energy resolution determinations are constants. The weighted averages values are   $R^{\sigma^\text{raw}}$($K$)⁄$R^{\sigma^\text{raw}}$($\pi$) = 0.95 $\pm$ 0.01 (-0.011) and \\ $R^{\sigma^\text{raw}}$($p$)⁄$R^{\sigma^\text{raw}}$($\pi$)  = 0.888 $\pm$ 0.005 (+0.011).

The differences of pion, kaon and proton responses and resolutions are due to the different fraction of non-electromagnetic energy deposited by incident particles: $F_\text{h}(\pi)$, $F_\text{h}(K)$ and $F_\text{h} (p)$ and to the non-compensating nature of the detector.
Data show constant ratios \\ $F_\text{h} (K)/F_\text{h}(\pi)$. The weighted average numerical value is 1.13 $\pm$ 0.01 (1.072 $\pm$ 0.001). The values in parenthesis were obtained analysing simulated data. The ratio $F_\text{h} (p)/F_\text{h}(\pi)$ decreases from 1.351 $\pm$ 0.04  (1.361 $\pm$ 0.003) at $E_\text{beam}$~16~GeV to 1.24 $\pm$ 0.01 (1.281 $\pm$ 0.003) at $E_\text{beam}$~30~GeV. 

As discussed in Section \ref {subsec:response_parametrization} the fraction of non--electromagnetic energy deposited by incident particles can be expressed in terms of the parameters $m$ and $E_0$ [GeV]. The ratio between the responses to the purely EM and hadronic components of showers $e/h$ describes the non-compensation nature of the calorimeter..
The values of $m$ obtained using experimental (simulated) pions, kaons and protons data are 0.919 $\pm$ 0.005 (0.826 $\pm$ 0.005), 0.97 $\pm$ 0.02 (0.843 $\pm$ 0.002) and 0.789 $\pm$ 0.003 (0.734 $\pm$ 0.002) respectively. 
%The determinations were obtained without any assumption on the values of $e/h$ and $E_0$. Fixing $E_0$ = 1~GeV, the values $e/h$~=~1.3535~$\pm$~0.0304 and $m$~=~0.9187~$\pm$~0.0047 obtained analysing pion data agree with previous determinations 1.33~$\pm$~0.02 and 0.85~$\pm$~0.03 respectively obtained by ATLAS Collaboration.

The energy resolution as a function of the energy can be parametrized with a statistical term \\ $a$⁄$\sqrt{E_\text{beam} [\text{GeV}] }$ and a constant terms $b$ (see Section \ref{subsec:resolution_parametrization}). 
The values of $a~[\%~\text{GeV}^{-1/2}]$ obtained analysing pions, kaons and protons are 47 $\pm$ 2 (42 $\pm$ 1), 50 $\pm$ 3 (42.8 $\pm$ 0.3) and 40.3 $\pm$ 0.4 (38.1 $\pm$ 0.2) respectively. The values in parenthesis were obtained analysing simulated data. The corresponding $b$ [$\%$] values are 5.0 $\pm$ 0.6 (6.2 $\pm$ 0.4), 2 $\pm$ 3
(5.3 $\pm$ 0.1) and 5 $\pm$ 1 (5.12 $\pm$ 0.08). 
%The term $b$ is affected by incomplete shower containment. 

%\begin{acknowledgements}
%\end{acknowledgements}

\Urlmuskip=0mu plus 1mu\relax
\bibliographystyle{spphys}       % APS-like style for physics
\bibliography{TestbeamAnalysis2020-epjc}   % name your BibTeX data base
%\begin{thebibliography}{}
% etc
%\end{thebibliography}

\end{document}